\documentclass[preprint,aps,superscriptaddress,nofootinbib,showkeys,10pt]{revtex4}

\usepackage{graphicx}
\usepackage{amsmath}
\usepackage{amsfonts}
\usepackage{placeins}
\usepackage{mathtools}
\usepackage{todonotes}
\usepackage{physics}
\usepackage{amssymb}
\usepackage{bm}
\usepackage{csquotes}
\usepackage{epstopdf}
\usepackage{float}
\usepackage{multirow}
\usepackage{subfigure}
\usepackage{capt-of}
\allowdisplaybreaks
\usepackage{hyperref}

\hypersetup{
     colorlinks   = true,
     citecolor    = red,
     linkcolor    = blue,
     urlcolor     = blue,
}
\definecolor{deeppink}{rgb}{0.9, 0.17, 0.31}

\def\beq{\begin{equation}}
\def\eeq{\end{equation}}
\def\bea{\begin{align}}
\def\eea{\end{align}}

\def\pitt{\Pi_{ij}^{\rm TT}}
\def\mpl{{ M}_{\rm pl}}
\def\l{\left}
\def\r{\right}
\def\vk{\bm k}
\def\kbar{\bar{k}}
\def\vq{\bm q}
\def\vx{\bm x}
\def\mGk{\mathcal{G}_k}
\def\Pt{\mathcal{P}_{\rm T}}

\def\wre{w_{\phi }}
\def\xe{x_{\rm e}}
\def\kf{k_{\rm end}}
\def\ee{\eta_{\rm end}}
\def\ae{a_{\rm end}}
\def\HI{H_{\rm end}}
\def\xre{x_{\rm re}}
\def\mI{\mathcal{I}}
\def\nn{\nonumber}
\def\umin{u_{\rm min}}
\def\umax{u_{\rm max}}
\def\kre{k_{\rm re}}
\def\mF{\mathcal{F}}
\def\mE{\mathcal{E}}
\def\ere{\eta_{\rm re}}
\def\ke{k_{\rm end}}
\def\He{H_{\rm end}}

\def\MP{{ M}_{\rm pl}}
\def\kmin{k_{\rm min}}
\def\Pts{\mathcal{P}_{\rm T}^{\rm sec}}

\def\Tre{T_{\rm re}}

\def\As{A_{\rm s}}
\def\Nre{N_{\rm re}}
\def\gre{g_{\rm re}}

\def\Gk{\mathcal{G}_{k}}
\def\Gkre{\mathcal{G}_{k}^{\rm re}}
\def\cH{\mathcal{H}}
\def\rhogw{\rho_{\rm gw}}
\def\ogw{\Omega_{\rm gw}}
\def\Ptra{\mathcal{P}_{\rm T}^{\rm rad}}

\def\ogwp{\Omega_{\rm gw}^{\rm pri}}
\def\nw{n_w}

\def\ogws{\Omega_{\rm gw}^{\rm sec}}
\def\ogwh{\Omega_{\rm gw}h^2}
\def\pijlm{P_{ij}^{lm}}
\def\Tlm{T_{lm}}
\def\mPchi{\mathcal{P}_{\chi}}
\def\mJ_{\mathrm{J}}

\def\rhora{\rho_{\rm ra}}
\def\rhon{\rho_\nu}
\def\rhogw{\rho_{\rm gw}}
\def\neff{N_{\rm eff}}
\def\Ps{\mathcal{P}_{\mathcal{S}}}
\def\Pr{\mathcal{P}_{\mathcal{R}}}
\def\rhop{\rho_{\gamma}}
\def\dneff{\Delta N_{\rm eff}}
\def\dneffc{\Delta N_{\chi}}
\def\dneffgw{\Delta N_{\rm gw}}
\def\omegac{\Omega_\chi}
\def\omegap{\Omega_\gamma}
\def\rhoc{\rho_{\rm c}}
\def\vx{\textbf{x}}

\graphicspath{{./figs/}}


\begin{document}

 \title{ Probing a nonminimal coupling through superhorizon instability and secondary gravitational waves
 %\\ 
%  Super-horizon instability and secondary gravitational waves:\\ Scalar dark radiation with non-minimal coupling
% \\Curvature coupling induced parametric instability of Massless Scalar Fields and its signature on Secondary Gravitational Waves}
}
\author{Ayan Chakraborty}
\email{E-mail:chakrabo@iitg.ac.in}
\affiliation{Department of Physics, Indian Institute of Technology, Guwahati, 
Assam, India}
 \author{Subhasis Maiti}
\email{E-mail: subhashish@iitg.ac.in}
\affiliation{Department of Physics, Indian Institute of Technology, Guwahati, 
Assam, India}
\author{Debaprasad Maity}
\email{E-mail: debu@iitg.ac.in}
\affiliation{Department of Physics, Indian Institute of Technology, Guwahati, 
Assam, India}

 \begin{abstract}
{ In this paper, we investigate the impact of scalar fluctuations ($\chi$) non-minimally coupled to gravity, $\xi\chi^2 R$, as a potential source of secondary gravitational waves (SGWs). 
Our study reveals that when reheating EoS $\wre < 1/3$ and $\xi \lesssim 1/6$ or $\wre > 1/3$ and $\xi \gtrsim 1/6$, the super-horizon modes of scalar field  experience a \textit{tachyonic instability} 
during the reheating phase. Such instability causes a substantial growth in the scalar field amplitude leading to pronounced production of SGWs in the low and intermediate-frequency ranges that are strong enough to be detected by \textit{Planck} and future gravitational wave detectors. Such growth in super-horizon modes of the scalar field and associated GW production may have a significant effect on the strength of the tensor fluctuation at the Cosmic Microwave Background (CMB) scales (parametrized by $r$) and the number of relativistic degrees of freedom (parametrized by $\dneff$) at the time of CMB decoupling. 
To prevent such overproduction, the \textit{Planck} constraints on tensor-to-scalar ratio $r \leq 0.036$ and $\dneff \leq 0.284$ yield a strong lower bound on $\xi$ for $\wre < 1/3$, and upper bound on the value of $\xi$ for $\wre > 1/3$. Taking into account all the observational constraints we found the value of $\xi$ should be $ \gtrsim 0.02$ for $\wre =0$, and $\lesssim 4.0$ for $\wre \geq 1/2$ for a wide range of reheating temperature within $10^{-2} \lesssim \Tre \lesssim 10^{14}$ GeV, and for a wide range of inflationary energy scales. Further, as one approaches $\wre$ towards $1/3$, the value of $\xi$ remains unconstrained. Finally, we identify the parameter regions in $(\Tre,\xi)$ plane which can be probed by the upcoming GW experiments namely BBO, DECIGO, LISA, and ET.

}
\end{abstract}
 
\maketitle

\section{Introduction}\label{intro}
Primordial Gravitational wave (GW) is one of the unique observable predictions of inflationary paradigm \cite{Guth:1980zm, Senatore:2016aui, Linde:1981mu, Albrecht:1982wi, Lemoine:2008zz, MUKHANOV1992203, Martin:2003bt, Martin:2004um,Linde:2014nna, RevModPhys.78.537, Sriramkumar:2009kg, Baumann:2008bn,Baumann:2018muz, Piattella:2018hvi}. Given the advent of a large number of existing \cite{LIGOScientific:2016aoc, LIGOScientific:2016emj, LIGOScientific:2017bnn, LIGOScientific:2016jlg, NANOGrav:2020bcs, NANOGrav:2023gor, NANOGrav:2023hde, Antoniadis:2023utw, EPTA:2023fyk, EPTA:2023xxk, Reardon:2023gzh, Zic:2023gta, Xu:2023wog} and upcoming \cite{Punturo:2010zz, Sathyaprakash:2012jk, Crowder:2005nr, Corbin:2005ny, Baker:2019pnp, Seto:2001qf, Kawamura:2011zz, Suemasa:2017ppd, Amaro-Seoane:2012aqc, Barausse:2020rsu, Janssen:2014dka} GWs detection experiments, inflationary framework proves to be an interesting playground to look for new physics at very high energy scales \cite{Arkani-Hamed:2015bza, Chen:2009we}. Exponential expansion leading to tachyonic growth of the super-horizon modes, and their subsequent evolution are known to be imprinted in the distribution of various cosmological relics such as Cosmic Microwave Background (CMB), Dark Matter (DM), GWs in the form of various correlations of scalar, vector and tensor fluctuations. Over the years enormous efforts have been put into estimating those correlations through the CMB anisotropy (\cite{Planck:2018jri} and references therein), dark and baryonic matter distribution \cite{DESI:2024mwx,eBOSS:2019dcv}, and placed tight constraints on the possible physics of inflation. Out of those different relics GW is said to be unique due to its extremely weak (Planck- suppressed) but universal coupling. Whereas weak coupling renders it an ideal probe of the very early universe, universal coupling on the other enables it to probe into the nature of all fundamental interactions in both the visible and dark sectors. Utilizing inflation as a mechanism and the subsequent reheating phase, in this paper we intend to probe the non-minimal gravitational coupling $\xi R \chi^2$ with a real scalar field $\chi$ through its imprints on the primordial GW spectrum. Such coupling is assumed to be inevitable in the low energy effective theory for the scalar field when coupled with gravity. Non-minimal gravitational coupling has been extensively explored in the context of inflation \cite{Faraoni_1996, PhysRevD.62.043512, Komatsu_1999, LUCCHIN1986163, SPOKOINY198439,PhysRevD.39.399, Shokri_2021,SalvatoreCapozziello_1994, NOZARI_2008,Gomes_2017, sarkar2023nonminimalinflationscalarcurvaturemixing, Pi:2017gih, Wang:2024vfv}, reheating \cite{Bassett:1997az, Tsujikawa:1999jh,Tsujikawa:1999iv,Ema_2017, Dimopoulos_2018,Figueroa:2021iwm,Opferkuch:2019zbd,Bettoni:2021zhq,Laverda:2023uqv, Figueroa:2024yja,Laverda:2024qjt}, DM \cite{Markkanen:2015xuw, Markkanen:2017edu, Fairbairn:2018bsw, Kainulainen:2022lzp, Lebedev:2022vwf, Kolb:2023ydq, Ema:2018ucl,Yu_2023,Cembranos:2023qph, Kolb:2020fwh, Capanelli:2024nkf, Capanelli:2024pzd}, and dark energy \cite{Setare:2010pfa, Sami:2012uh, Kase:2019veo,Ye:2024ywg}. In the present paper, however, we focus on exploring the dynamics of super-Hubble modes and associated induced GW spectrum. This particular aspect of the present study has been less explored in the literature. Depending on the strength of the non-minimal coupling, a certain range of super-Hubble modes of the scalar field realize tachyonic growth and may lead to potentially detectable secondary gravitational waves (SGW). Our analysis further reveals that the post-inflationary reheating phase plays an instrumental role. In the perturbative framework, the reheating phase is described by the equation of state $\wre$ of inflaton and reheating temperature $\Tre$. The instability that we pointed out for the super-Hubble modes turned out to be significant during the reheating phase, particularly for two distinct regions in the $(\wre,\xi)$ parameter space. Whereas for the equation of state $\wre < 1/3$, instability arises for $\xi <1/6$, for $\wre > 1/3$, on the other hand, $\xi$ should be greater than the conformal limit $1/6$. Such instability helps enhancing the overall growth of the amplitude of the scalar field modes after they enter into the reheating phase and depends upon the reheating parameters, namely $\wre$ and reheating temperature $\Tre$. Further, the cosmological background driven by matter fields with a stiff equation of state is known to amplify GW amplitude when propagating through such background \cite{Maiti:2024nhv}. In this article, we affirm that a combination of the above two non-trivial effects indeed leads to strong GW production for the stiff equation of state. 
For the $\wre <1/3$, on the other hand, the growth is weaker but GW acquires a non-trivial spectral behavior near the CMB scales that can have an appreciable impact on the tensor-to-scalar ratio. 
Taking into account CMB constraints on the inflationary tensor power spectrum, and BBN constraints on an effective number of degrees of freedom we demonstrate that SGW induced by non-minimal coupling put a tighter constraint on non-minimal coupling parameters $\xi$ as compared to the values generically assumed in the earlier studies \cite{Clery:2022wib, Barman:2022qgt, Ghoshal:2024gai} and also constraint recently reported considering primary gravitational waves \cite{Maity:2024cpq}. Maintaining all the existing observational constraints we finally estimate the range of non-minimal coupling against different reheating models that can be probed by the various upcoming GW experiments such as BBO, DECIGO, LISA, and ET.\\ 

The order of construction of this paper is as follows. In Section \ref{sec2}, we first give a brief overview of the non-perturbative framework of gravitational particle production. We then elaborately discuss the super-horizon instability dynamics(\textit{tachyonic instability}) of the scalar field in the presence of the non-minimal coupling with gravity $\xi\chi^2 R$ and we also compute the associated long-wavelength field solutions during reheating corresponding to three different ranges of non-minimal coupling strength, $0\leq \xi<3/16, \xi=3/16, \xi>3/16$ in the entire range of post-inflationary EoS $0\leq\wre\leq 1$. Further, we thoroughly investigate the behavior of the scalar power spectrum and explicitly point out the salient features of this spectrum for varying parameters ($\wre, ~ \xi$). We close this section by giving a short discussion on the model-independent definition of the reheating parameters ($N_{\rm re}, ~ T_{\rm re}$). In Section \ref{sec3}, we discuss the dynamics of the tensor fluctuations with (secondary gravitational waves, SGW) and without (primary gravitational waves, PGW) the anisotropy sourced by the gravitationally produced scalar field. In this section, our prime focus is on the effect of the parametric instability being caused by the non-minimal coupling of the scalar field on the gravitational wave spectrum(PGW+SGW) for both $\wre<1/3$ and $\wre>1/3$. Here we compute the primary as well as the secondary gravitational wave spectrum and show the nature of the full spectrum for varying parameters ($\wre, ~ T_{\rm re}, ~ \xi, ~ r_{0.05}$). Furthermore, we constrain the coupling strength $\xi$ in the light of the present-day tensor-to-scalar ratio at the CMB scale $r_{0.05}\leq 0.036$, the $\Delta N_{\rm eff}\leq 0.284$, and the present-day isocurvature bound as reported by the recent \textit{Planck} 2018 observation. We also constrain the negative $\xi$ values based on the observational bounds in the present scenario. We also provide a feasible parameter space of the parameter set ($T_{\rm re}, ~ \xi$) that future GW detectors like LISA, DECIGO, BBO, ET, etc can detect. Section \ref{sec4} concludes this paper by clearly highlighting the main outcomes of this work and shedding light on some possible directions of the present work that are left for future endeavors. In Appendix \ref{appenc}, we have studied the behavior of the suppression factor $\l(1-\frac{a^2\xi\langle\chi^2\rangle}{M^2_{\rm pl}}\r)^{-2}$ that we encounter in the dynamics of the secondary gravitational waves for different reheating parameters($\wre,~\Tre,~\He,~\xi$). In Appendix \ref{appenA}, we illustrate the computation of the secondary GW spectrum for $\wre<1/3$ and $\wre>1/3$, and in Appendix \ref{appenB}, we detail the computation of the total energy density of the produced scalar field.         

%%%%%%%%%%%%%%%%%%%%%%%%%%%

\section{Spectrum of Gravitationally Produced Massless Particles}\label{sec2}

We shall begin this section by briefly discussing the basic formalism of non-perturbative particle production. We consider the following general Lagrangian for inflaton ($\phi$) and a massive daughter field $(\chi)$ non-minimally coupled to gravity and inflaton as follows. %%%%%%%%%%%%%%%%%%%%%%%%%%%%
\begin{equation}\label{lagrangian1}
    \mathcal{L}_{[\phi,\chi]}= -\sqrt{-g}\bigg(\frac{1}{2}\partial_{\mu}\phi \partial^{\mu}\phi+V(\phi)+\frac{1}{2}\partial_{\mu}\chi \partial^{\mu}\chi+\frac{1}{2}(m_{\chi}^2+g^2\phi^2+\xi R)\chi^2 \bigg) .
\end{equation}
%%%%%%%%%%%%%%%%%%%%%%%%%%%
The background FLRW metric is expressed as $ds^2=a^2(\eta)\big(-d\eta^2+d\vec{x}^2\big)$ with the scale factor $a$ and $\sqrt{-g}=a^4(\eta)$. $V(\phi)$ is the inflaton potential, $m_{\chi}^2$ is the bare mass of the produced scalar particles. \enquote{$\xi$} is the dimensionless non-minimal coupling of $\chi$ field with gravity and $g$ is dimensionless coupling strength with the inflaton. Ricci scalar \enquote{$R$} generates a time-dependent effective mass for the $\chi$ field as, $m_{\text{eff}}^2(\eta)=\big(m_{\chi}^2+g^2\phi(\eta)^2+\xi R(\eta)\big)$. Since the background is set by the inflaton which is minimally coupled with gravity, we perform our computation in the Jordan frame for scalar field fluctuation $\chi$.   

From the background inflaton part of the Lagrangian (\ref{lagrangian1}), we get the inflaton dynamical equation as
%%%%%%%%%%%%%%%%%%%%%%%%%
\begin{equation}\label{inflatoneq}
  \phi^{\prime\prime} +2\mathcal{H} \phi^{\prime}+a^2\pdv{ V(\phi)}{\phi}=0 . 
\end{equation}
%%%%%%%%%%%%%%%%%%%%%%%%%%
With the Hubble scale
%%%%%%%%%%%%%%%%%%%%%%%%%%%
\begin{equation}\label{Hubble}
 \mathcal{H}=\sqrt{\frac{1}{3M_{\rm pl}^2}\bigg(\frac{1}{2} (\phi^{\prime})^2+a^2V(\phi)\bigg)}   ,
\end{equation}
%%%%%%%%%%%%%%%%%%%%%%%%%%%%%%%%
where $\phi^{\prime}=(d\phi/d\eta)$ and $M_{\rm pl}={1}/{\sqrt{8\pi G}}\approx 2.43\cross 10^{18} \text{GeV}$ is the reduced $Planck$ mass. Different dynamical features of inflaton in the Post-inflationary phase influence the Hubble scale which leaves a non-trivial impact on the particle production process.

Expressing the scalar field \enquote{$\chi$} field in terms of Fourier modes,
%%%%%%%%%%%%%%%%%%%%%%%%%%%%%%%%
\begin{equation}\label{fourier}
     \chi(\eta,{x})= \int\frac{d^3{k}}{(2\pi)^3}~\chi_k(\eta) ~e^{i {k}\cdot{x}},
 \end{equation}
%%%%%%%%%%%%%%%%%%%%%%%%%%%%%%
 and subject to the Lagrangian (\ref{lagrangian1}) we reach the following dynamical equation of mode function ($\chi_{{k}})$ as,
   %%%%%%%%%%%%%%%%%%%%%%%%%%%%%
   \begin{equation}\label{dynamical1}       \chi_{{k}}^{\prime\prime}+2\mathcal{H}\chi_{{k}}^{\prime}+\Big(k^2+a^2(\eta)\big(m_{\chi}^2+g^2\phi^2+\xi R\big)\Big)\chi_{{k}}=0 .
  \end{equation}
%%%%%%%%%%%%%%%%%%%%%%%%%%%%%%
  In the following dynamical Eq. (\ref{dynamical1}), note that there is a damping term, \enquote{$2\mathcal{H} \chi_{{k}}^{\prime}$} which is non-zero in expanding background.
Defining a new rescaled field $X_{{k}} = a(\eta)\chi_{{k}}(\eta)$, we obtain the following simplified equation of an oscillator with time-dependent frequency,
  %%%%%%%%%%%%%%%%%%%%%%%%%%
   \begin{equation}\label{dynamical2}       X_{{k}}^{\prime\prime}+\bigg[k^2+a^2\big(m_{\chi}^2+g^2\phi^2\big)+\frac{a^2 R}{6}(6\xi-1)\bigg]X_{{k}}=0 .
  \end{equation}
  %%%%%%%%%%%%%%%%%%%%%%%%%%
  The bracketed term in the above  Eq. (\ref{dynamical2})  can be identified as a time-dependent frequency \enquote{$\omega_k$} where,
  %%%%%%%%%%%%%%%%%%%%%%%%%%%%%%
  \begin{equation}\label{ffrequency}      \omega_k^2(\eta)=\bigg(k^2+a^2\big(m_{\chi}^2+g^2\phi^2\big)+\frac{a^2 R}{6}(6\xi-1)\bigg) .
  \end{equation}
  %%%%%%%%%%%%%%%%%%%%%%%%%%
To solve the Eq. (\ref{dynamical2}) we choose the positive frequency Bunch-Davies vacuum,
%%%%%%%%%%%%%%%%%%%%%%%%%
\begin{equation}\label{Bunchdavies}
   X_{k}(\eta_0)=\frac{1}{\sqrt{2\omega_k}}e^{-i\omega_k\eta_0}, \quad X_{k}^{\prime}(\eta_0)=-i\sqrt{\frac{\omega_k}{2}}e^{-i\omega_k\eta_0} .
\end{equation}
%%%%%%%%%%%%%%%%%%%%%%%%
Where $\eta_0$ is some initial time at which positive-frequency Bunch-Davies vacuum solution is satisfied. The particle occupation number or number density power spectrum for the scalar field is usually expressed as \cite{Kofman:1997yn}, 
 %%%%%%%%%%%%%%%%%%%%%%%%%%%
 \begin{equation}\label{fnoden}
     n_k=\frac{1}{2\omega_k}|\omega_kX_k-iX_k^{\prime}|^2 .
 \end{equation}
%%%%%%%%%%%%%%%%%%%%%%%%%%%%
Integrating Eq.(\ref{fnoden}) over all the momentum modes, we get the total number and UV convergent energy density as \cite{PhysRevD.9.341, PhysRevD.36.2963}
%%%%%%%%%%%%%%%%%%%%%%%%%%%%%
\begin{align}\label{fnoEden}
 n_{\chi}&=\frac{1}{(2\pi)^3a^3} \int d^3k n_k, \nonumber\\
 \rho_{\chi}&=\frac{1}{(2\pi)^3a^4} \int d^3k\omega_k n_k .
\end{align}
%%%%%%%%%%%%%%%%%%%%%%%%%%
The general formalism we constructed in this section is the main foundation of the non-perturbative study of particle production. To this end, we point out that our goal is to analyze the impact of instability due to non-minimal coupling on those modes that remain outside the horizon just after the inflation namely $k < \ae \He$. Where $(\ae, \He)$ are the scale factor and Hubble scale at the inflation end respectively. On the other hand, from the expression of the effective frequency $\omega_k$ one notes that if the $k \gg \ae m_\chi$ condition is satisfied, the bare mass of the $\chi$ field can be ignored.  Therefore, for the dynamics of the super-horizon modes, the $m_\chi \ll \He$ condition is equivalent to the massless limit of the scalar field. This is precisely the limit to which we will perform our computation. We further assume that the $\chi$ field is non-interacting with the standard model and hence it can be assumed as either dark radiation or dark matter. We have numerically checked that the massless limit is perfectly consistent with our present results as long as the scalar field mass $m_\chi \lesssim 10^{-5} \He$. Masses beyond this specified limit give rise to a possible mass-breaking effect(departure from the massless spectrum) in the number and energy density spectrum, hence they are likely to modify the GW spectrum as well as the constraint on the non-minimal coupling parameter $\xi$.  However, the detailed computation for the entire mass range including the super-Hubble mass($m_{\chi}>\He$) we defer for our future studies. 

As just pointed out we consider the massless non-minimally coupled fluctuation which has no other interaction except with gravity. As per these considerations, Lagrangian (\ref{lagrangian1}) becomes
%%%%%%%%%%%%%%%%%%%%%%%%%%%
\begin{equation}\label{lagrangian2}
  \mathcal{L}_{[\phi,\chi]}=- \sqrt{-g}\bigg(\frac{1}{2}\partial_{\mu}\phi \partial^{\mu}\phi+V(\phi)+\frac{1}{2}\partial_{\mu}\chi \partial^{\mu}\chi +\frac{1}{2}\xi R\chi^2\bigg)  .
\end{equation}
%%%%%%%%%%%%%%%%%%%%%%%%%%%
The scalar field mode equation (\ref{dynamical2}) becomes,
%%%%%%%%%%%%%%%%%%%%%%%
\begin{align}\label{dynamical3}  
%X_{\vec{k}}^{\prime\prime}+\bigg[k^2-\frac{a^2 R}{6}(1-6\xi)\bigg]X_{\vec{k}}=0 \nonumber\\
%& \Rightarrow 
X_{\vec{k}}^{\prime\prime}+\bigg[k^2-\frac{a^{\prime\prime}}{a}(1-6\xi)\bigg]X_{\vec{k}}=0 ,
\end{align}
%%%%%%%%%%%%%%%%%%%%%%%%%%
where in conformal coordinate, Ricci scalar has been expressed as $R=\left(6 a^{\prime\prime}/a^3\right)$. In order to calculate the general solution of the above equation (\ref{dynamical3}), we first need to calculate the \textit{adiabatic vacuum} solution of (\ref{dynamical3}) and any general solution of the equation can then be expressed as a linear combination of that vacuum solution with the knowledge of the \textit{Bogoliubov coefficients} $\alpha_k$ and $\beta_k$ what we shall compute now.

In the present context, we are interested in the spectrum associated with those modes that left the horizon during inflation and again reenter at some point during reheating and later. These long-wavelength modes experience an instability called \textit{tachyonic instability} after getting out of the horizon during inflation. To take into account the enhancement of field amplitude owing to this instability, we study their dynamics from the early inflationary era to the late reheating phase when all the modes are well inside the horizon. The evolution of scale factor during inflation and reheating with any general EoS can be represented as, 
%%%%%%%%%%%%%%%%%%%%%%%%%%
\begin{align}\label{scalefactor}
     a(\eta)=
     \begin{cases}
        -\frac{1}{H_{\text{end}}\eta}\quad \quad \quad -\infty<\eta\leq \eta_{\text{end}} & \\
a_{\text{end}}\Big(\frac{1+3\wre}{2|\eta_{\rm end}|}\Big)^{\frac{2}{1+3\wre}}\bigg(\eta-\eta_{\text{end}}+\frac{2 |\eta_{\rm end}|}{1+3\wre}\bigg)^{\frac{2 }{1+3\wre}}  \eta\geq\eta_{\text{end}} & 
\end{cases} .
\end{align}
%%%%%%%%%%%%%%%%%%%%%%%%%%%%%%%%%%
Considering pure de Sitter inflation, we assume the Hubble scale at the end of inflation as %$H_{\text{ds}}=H_{\text{end}}$. 
$H_{\text{end}}$.
It is straightforward to check that during the transition from inflation to reheating, the scale factor and its first derivative change continuously at the junction point, that is at the end of inflation, $\eta=\eta_{\text{end}}=-(1/a_{\rm end}H_{\rm end})$. Here $a_{\rm end}$ is the scale factor at $\eta=\eta_{\rm end}$ and $\wre$ is the background inflaton EoS during reheating.

Associated Hubble scale behaves as
%%%%%%%%%%%%%%%%%%%%%%%%%%
\begin{equation}\label{hubble}
 \mathcal{H}(\eta\geq \eta_{\rm end})=\frac{a^{\prime}(\eta)}{a(\eta)}=\frac{2 (a_{\rm end}H_{\rm end})}{(1+3\wre)}\bigg((\eta a_{\rm end} H_{\rm end})+\frac{3(1+\wre)}{(1+3\wre)}\bigg)^{-1}   \end{equation}
%%%%%%%%%%%%%%%%%%%%%%%%%%%

It is well-known that the violation of the adiabaticity condition owing to the changing background geometry in the abrupt transition from de Sitter vacuum to a post-inflationary vacuum state causes particle production associated with long-wavelength modes. In particular, one defines a dimensionless factor $|\omega^{\prime}_k/\omega_k^2|$ to study the departure from the adiabatic limit and in the process of transition,  this adiabaticity condition gets violated($|\omega^{\prime}_k/\omega_k^2|>>1$) at some intermediate point giving a burst of particles in long-wavelength regime.

Now let us suppose $X_k^{(\rm inf)}(\eta)$ is the adiabatic vacuum solution of (\ref{dynamical3}) during the de Sitter phase in the time interval $-\infty<\eta\leq\eta_{\text{end}}$ and $X_k^{(\rm reh)}(\eta)$ is the adiabatic vacuum solution during reheating phase for $\eta\geq\eta_{\text{end}}$. 
Any general field solution during reheating can thus be expressed as
%%%%%%%%%%%%%%%%%%%%%%%%%%%%%%%
\begin{equation}\label{Xgen}
    X_k(\eta)=\alpha_k X_k^{(\rm reh)}+\beta_k X_k^{*(\rm reh )},
\end{equation}
%%%%%%%%%%%%%%%%%%%%%%%%%%%%%%
where, $(\alpha_k, \beta_k)$ can be identified as Bogoliubov coefficients.
Making these solutions and their first derivatives continuous at the junction $\eta=\eta_{\text{end}}$, we compute the Bogoliubov coefficients as follows:  \cite{deGarciaMaia:1993ck}
%%%%%%%%%%%%%%%%%%%%%%%%%%%%%%%%%
\begin{align}\label{bogo}
  &\alpha_k= i\left({X_k^{(\rm inf)}}'(\eta_{\rm end}){X_k^{(\rm reh)}}^{*}(\eta_{\rm end})-{X_k^{(\rm inf)}}(\eta_{\rm end}){X_k^{(\rm reh)}}^{*\prime}(\eta_{\rm end})\right) ,\nonumber\\
  & \beta_k=-i\left({X_k^{(\rm inf)}}'(\eta_{\rm end}) X_k^{(\rm reh)}(\eta_{\rm end}) - {X_k^{(\rm reh)}}'(\eta_{\rm end}) X_k^{(\rm inf)}(\eta_{\rm end})\right) ,
  \end{align}
%%%%%%%%%%%%%%%%%%%%%%%%%%%%%%%
where ($'$) denotes the derivative with respect to conformal time. 

Now, our goal would be to obtain the adiabatic vacuum solutions in both phases. Using the scale factor (\ref{scalefactor}) in the  equation (\ref{dynamical3}), we obtain the form of the dynamical equation during de Sitter inflation  ($\eta\leq\eta_{\text{end}}$) as, %\cite{Baumann:2018muz}
%%%%%%%%%%%%%%%%%%%%%%%%%%%%%
\begin{equation}\label{dynamicalinf}
   X_k^{\prime\prime}+\underbrace{\Bigg[k^2-\frac{2(1-6\xi)}{\eta^2}\Bigg]}_{\omega_k^2}X_k=0  .
   \end{equation}
%%%%%%%%%%%%%%%%%%%%%%%%%%%%%%%%
From the above equation it can be noted that long wavelength modes after their horizon crossing becomes tachyonic ($\omega_k^2<0$)  for $0\leq\xi<1/6$. As $\xi$ exceeds the conformal limit $\xi=1/6$, this instability ceases to exist. 

The general solution of this equation is
\begin{align}\label{Xinf1}
   &X_k=C_1\sqrt{|\eta|}J_{\nu_1}(k|\eta|)+C_2\sqrt{|\eta|}Y_{\nu_1}(k|\eta|) .
   %\frac{e^{-i k\eta}}{\sqrt{2k}}\left(1-\frac{i}{k\eta}\right)
  %&{X_k^{(\rm inf)}}^{\prime}=\frac{d X_k^{(\rm inf)}(\eta)}{d \eta}\simeq i\sqrt{\frac{k}{2}}\frac{e^{-ik\eta}}{k^2\eta^2} 
  \end{align}
%%%%%%%%%%%%%%%%%%%%%%%%%%%
With the order of the Bessel functions $\nu_1={\sqrt{9-48\xi}}/{2}$ and $C_1, C_2$ are integration constants.
%%%%%%%%%%%%%%%%%%%%%%%%
%\begin{equation}\label{nu1}
%\nu_1=\frac{\sqrt{9-48\xi}}{2} 
%\end{equation}
%%%%%%%%%%%%%%%%%%%%%%%%%%
To evaluate $\alpha_k$ and $\beta_k$,  we need to define first the vacuum solution $X_k^{(\rm inf)}$ during inflation.
To compute the de Sitter vacuum solution, we use the Bunch-Davies vacuum condition at the beginning of inflation. In this limit $k|\eta|>>1$, the mode solution (\ref{Xinf1}) becomes,
%%%%%%%%%%%%%%%%%%%%%%%%%%%%%%%%%
\begin{equation}\label{Xinf2}
    X_k(\eta)\sim\frac{1}{\sqrt{2\pi k}}\bigg[(C_1-i C_2)e^{-i(k\eta+\pi/4+\pi\nu_1/2)}+(C_1+i C_2)e^{i(k\eta+\pi/4+\pi\nu_1/2)}\bigg] .
\end{equation}
%%%%%%%%%%%%%%%%%%%%%%%%%%%%%%%%%
In the asymptotic in-vacuum limit, the positive-frequency outgoing mode function behaves as 
%%%%%%%%%%%%%%%%%%
\begin{equation}\label{adiabaticVinf}
    X_k(\eta)\xrightarrow{{\eta}\rightarrow -\infty} \frac{e^{-ik\eta}}{\sqrt{2k}}.
\end{equation}
%%%%%%%%%%%%%%%%%%%%%%
Comparing (\ref{Xinf2}) with (\ref{adiabaticVinf}) we have
%%%%%%%%%%%%%%%%%%%%%%%%%
\begin{equation}\label{c1c2}
  C_1= \frac{\sqrt{\pi}}{2}e^{i(\pi/4+\pi\nu_1/2)}, \quad C_2= \frac{i\sqrt{\pi}}{2}e^{i(\pi/4+\pi\nu_1/2)} .
\end{equation}
%%%%%%%%%%%%%%%%%%%%%%%%%%
Therefore, the adiabatic vacuum solution during de Sitter inflation is
%%%%%%%%%%%%%%%%%%%%%%%%%%%%%%
\begin{equation}\label{Xinf3}
    X_k^{(\rm inf)}=\frac{\sqrt{-\pi \eta}}{2}e^{i(\pi/4+\pi\nu_1/2)}H^{(1)}_{\nu_1}(-k\eta) .
\end{equation}
%%%%%%%%%%%%%%%%%%%%%%%%%%%%%%%
Similarly the dynamical equation for general post-inflationary($\eta>\eta_{\rm end}$) EoS \enquote{$\wre$} is
%%%%%%%%%%%%%%%%%%%%%%%%%%%
  \begin{equation}\label{dynamicalreh}
    X_k^{\prime\prime}+\underbrace{\Bigg[k^2-\frac{2(1-3\wre)(1-6\xi)}{(1+3\wre)^2\Big(\eta+\frac{3(1+\wre)}{a_{\rm end}H_{\text{end}}(1+3\wre)}\Big)^2}\Bigg]}_{\omega_k^2}X_k=0  ,
  \end{equation}
  %%%%%%%%%%%%%%%%%%%%%%%%%
From the above equation, it can be again noted that the modes which were stable during inflation for $\xi>1/6$ become tachyonic ($\omega_k^2<0$) during reheating for $\wre>1/3$  \cite{Fairbairn:2018bsw}. We will see that this will play a significant role in our subsequent studies.  
The general solution of this equation is
%%%%%%%%%%%%%%%%%%%%%%%%%
\begin{equation}\label{Xreh1}
  X_k(\eta)=C_34^{\nu_2}\Gamma(\nu_2+1)\sqrt{2 i k\bar{\eta}} I_{\nu_2}(i k \bar{\eta})+C_4\sqrt{\frac{2 i k\bar{\eta}}{\pi}}K_{\nu_2}(i k\bar{\eta}) .
\end{equation}
%%%%%%%%%%%%%%%%%%%%%%%%%
Where we use the symbol $\bar{\eta} = (\eta + {3\mu}/a_{\rm end}H_{\text{end}})$. $I_{\nu_2}$ and $K_{\nu_2}$ are modified Bessel functions of order $\nu_2$ with
%%%%%%%%%%%%%%%%%%%%%%%%
\begin{align}\label{symbol}
   &\mu=\frac{(1+\wre)}{(1+3\wre)}, \quad 
\nu_2=\frac{\sqrt{3(1+\wre)\Big(3(1-\wre)^2+16\xi(3\wre-1)\Big)}}{2\sqrt{1+3\wre}\sqrt{1+4\wre+3\wre^2}}  .
\end{align}
%%%%%%%%%%%%%%%%%%%%
and $C_3, C_4$ are the integration constants.

We now seek the solution of (\ref{dynamicalreh}) compatible with the requirements of an adiabatic vacuum. If spacetime changes very slowly or equivalently particle momentum is so large that it hardly feels the background dynamics, the mode function can be safely assumed to behave as a positive frequency mode in Minkowski space in its asymptotic limit. 
For this we assume the ($k\eta>>1$) limit, and the mode solution (\ref{Xreh1}) transforms into,
%%%%%%%%%%%%%%%%%%%%%%%%%%
\begin{equation}\label{Xreh2}
X_k(\eta)\sim \Bigg[C_3\frac{2^{2\nu_2}\Gamma(\nu_2+1)}{\sqrt{\pi}}e^{ik \bar{\eta} } +C_4e^{-ik \bar{\eta}}\Bigg] .
\end{equation}
%%%%%%%%%%%%%%%%%%%%%%%%%
In the adiabatic out-vacuum limit that is for $\eta>>1$ or equivalently $a(\eta)\rightarrow \infty$ mode function behaves as a positive frequency state
%%%%%%%%%%%%%%%%%%
\begin{equation}\label{adiabaticVreh}
    X_k(\eta)\xrightarrow{{\eta}\rightarrow \infty} \frac{e^{-ik\eta}}{\sqrt{2k}},
\end{equation}
%%%%%%%%%%%%%%%%%%%%%%
Comparing (\ref{Xreh2}) with (\ref{adiabaticVreh}) we have
%%%%%%%%%%%%%%%%%%%%%%%%%%%
\begin{equation}\label{c3c4}
     C_3=0, \quad
    C_4= \frac{1}{\sqrt{2k}}\text{exp}\bigg[\frac{3ik\mu}{a_{\rm end} H_{\text{end}}}\bigg] .
\end{equation}
%%%%%%%%%%%%%%%%%%%%%%%%%%%%%%
Therefore, the adiabatic vacuum solution of massless particles for general reheating EoS \enquote{$\wre$} becomes
%%%%%%%%%%%%%%%%%%%%%%%%%%%%
\begin{equation}\label{Xreh3}
 X_k^{(\rm reh)}(\eta)=\sqrt{\frac{\bar{\eta}}{\pi}}\text{exp}\bigg[\frac{3ik\mu}{a_{\rm end}H_{\text{end}}}+\frac{i\pi}{4}\bigg]K_{\nu_2}(i k \bar{\eta}) .
\end{equation}
%%%%%%%%%%%%%%%%%%%%%%%%%%%%%%%%%%%%%%%%%%%%%
It is important to note that depending upon the value of the non-minimal coupling constant $\xi$, the order of the inflationary vacuum solution $\nu_1$ becomes positive for $0\leq\xi<3/16$, zero for $\xi=3/16$, and imaginary for $\xi>3/16$. In addition to this, the index of post-inflationary vacuum solution $\nu_2$ also becomes imaginary in the range $\xi>3/16$ for EoS $0\leq\wre<1/3$ and it becomes real positive for $1/3\leq\wre\leq 1$. This varying nature of the indices $\nu_1, \nu_2$ depending upon different ranges of non-minimal coupling strength and post-inflationary EoS greatly influences the nature of the post-inflationary field solution. Now we shall compute the field solution during reheating for general EoS in three specified ranges of the non-minimal coupling strength $\xi$.

\subsection{Field solution at large scale for $0\leq\wre<1/3$  }
%%%%%%%%%%%%%%%%%%%%%%%%%%%%%%5
\subsubsection{For $0\leq\xi<3/16$ :}
%%%%%%%%%%%%%%%%%%%%%%%%%%%
As mentioned earlier, the general field solution in a particular phase can be expressed as a linear combination of the respective vacuum solution with the help of Bogoliubov coefficients $\alpha_k$ and $\beta_k$ (See Eq.(\ref{Xgen}). So, our main task is to compute $\alpha_k$ and $\beta_k$ using the relations in (\ref{bogo}). In this specified range of $\xi$ we get both $\nu_1$ and $\nu_2$ to be positive definite. Substituting these vacuum solutions (\ref{Xinf3}) and (\ref{Xreh3}) into (\ref{bogo}), in long-wavelength limit $k/k_{\rm end}<<1$, we compute the Bogoliubov coefficients as
%%%%%%%%%%%%%%%%%%%%%%%%%%%%%%%%%%
\begin{subequations}\label{bogolarge1}
\begin{align}
  &\alpha_k\approx \frac{\Gamma(\nu_1)\Gamma(\nu_2)2^{\nu_1}}{8\pi}\left(\frac{2}{3\mu-1}\right)^{\nu_2}\left(\frac{3\mu(1-2\nu_1)+2(\nu_1-\nu_2)}{\sqrt{(3\mu-1)}}\right) \left(\frac {1}{\kbar}\right)^{\nu_1+\nu_2}
  e^{i\left(\frac{\pi\nu_1}{2}+\frac{\pi\nu_2}{2}+\frac{\pi}{2}-3\mu \kbar\right)}, \nonumber \\
  & \beta_k\approx  \alpha_k  e^{i\left(-\pi\nu_2-\frac{\pi}{2}+6\mu \kbar\right)}
%  \frac{\Gamma(\nu_1)\Gamma(\nu_2)2^{\nu_1}}{8\pi}\left(\frac{2}{3\mu-1}\right)^{\nu_2}\left(\frac{3\mu(1-2\nu_1)+2(\nu_1-\nu_2)}{\sqrt{(3\mu-1)}}\right)\frac{{\rm exp}\left(i(\pi\nu_1/2-\pi\nu_2/2+3\mu k/k_{\rm end})\right)}{(k/k_{\rm end})^{(\nu_1+\nu_2)}}
\end{align}
\end{subequations}
%%%%%%%%%%%%%%%%%%%%%%%%%%%%%%%
For simplified expression we define a new symbol $\kbar = k/\ke$. We define the energy density of the produced particles at a time during reheating when the associated longest wavelength is well inside the horizon. In this sense, we always have $k\eta>>1$ for any mode which will contribute to the energy density. Using the long-wavelength approximated form of the coefficients $\alpha_k$ and $\beta_k$(See Eq.(\ref{bogolarge1})) in (\ref{Xgen}) we obtain the general long-wavelength solution of scalar field for general EoS in the specified $\xi$ range for $k\eta>>1$ as
%%%%%%%%%%%%%%%%%%%%%%%%%%%%%%%%
\begin{equation}\label{Xreh4}
   X^{\rm long}_k(\eta)\approx\frac{\Gamma(\nu_1)\Gamma(\nu_2)2^{\nu_1}}{4\pi\sqrt{k_{\rm end}}}\left(\frac{2}{3\mu-1}\right)^{\nu_2}\left(\frac{3\mu(1-2\nu_1)+2(\nu_1-\nu_2)}{\sqrt{2(3\mu-1)}}\right) \frac{  \text{cos}(k\eta)}{\kbar^{(\nu_1+\nu_2+1/2)}} 
\end{equation}
%%%%%%%%%%%%%%%%%%%%%%%%%%%
where $k_{\rm end}=a_{\rm end}H_{\rm end}$ is the scale that leaves the horizon at the end of inflation.

\subsubsection{For $\xi=3/16$ :}
For this particular value of the coupling strength $\xi$, $\nu_1$ vanishes. Following the same procedure, in the long-wavelength limit, $\alpha_k$ and $\beta_k$ can now be evaluated as
%%%%%%%%%%%%%%%%%%%%%%%%%%%%%%%%%
\begin{subequations}\label{bogolarge2}
\begin{align}
  &\alpha_k\approx -\frac{\Gamma(\nu_2)}{2}\left(\frac{2}{3\mu-1}\right)^{\nu_2}\left(\frac{3\mu-2\nu_2}{4\sqrt{(3\mu-1)}}+\frac{i\sqrt{3\mu-1}}{\pi}\right)\frac{{\rm exp}\left(i(\pi\nu_2/2-\pi/2-3\mu \kbar)\right)}{\kbar^{\nu_2}}\\
  & \beta_k\approx \frac{\Gamma(\nu_2)}{2}\left(\frac{2}{3\mu-1}\right)^{\nu_2}\left(\frac{3\mu-2\nu_2}{4\sqrt{(3\mu-1)}}+\frac{i\sqrt{3\mu-1}}{\pi}\right)\frac{{\rm exp}\left(i(-\pi\nu_2/2+3\mu \kbar)\right)}{\kbar^{\nu_2}}
\end{align}
\end{subequations}
%%%%%%%%%%%%%%%%%%%%%%%%%%%%%%%
Associated general field solution for $\xi=3/16$ in $k\eta>>1$ limit becomes
%%%%%%%%%%%%%%%%%%%%%%%%%%%%%
\begin{equation}\label{Xreh5}
 X^{\rm long}_k(\eta)\approx \frac{\Gamma(\nu_2)}{\sqrt{2k_{\rm end}}}\left(\frac{2}{3\mu-1}\right)^{\nu_2}\left(\frac{3\mu-2\nu_2}{4\sqrt{(3\mu-1)}}+\frac{i\sqrt{3\mu-1}}{\pi}\right)\frac{i \text{sin}(k\eta)}{\kbar^{\nu_2+1/2}}  
\end{equation}

%%%%%%%%%%%%%%%%%%%%%%%%%%%%
\subsubsection{For $\xi>3/16$ :}

In this range of $\xi$ values, $\nu_1$ and $\nu_2$ become imaginary. Long-wavelength approximated Bogoliubov coefficients are 
%%%%%%%%%%%%%%%%%%%%%%%%%%%%%%%
\begin{subequations}\label{bogolarge3}
\begin{align}
     \alpha_k\approx &
     \bigg(K_{\nu_2}\Big(i \kbar(3\mu-1)\Big)H^{(1)}_{\nu_1}(\kbar)\left(\frac{3\mu}{\sqrt{3\mu-1}}\right)+\kbar\sqrt{3\mu-1}\Big(H^{(1)}_{\nu_1-1}(\kbar)-H^{(1)}_{\nu_1+1}(\kbar)\Big)\bigg)\nonumber\\
     \begin{split}
     &\times \text{exp}(i(\pi/2)-3\mu \kbar)\frac{\text{exp}(-\pi \tilde{\nu}_1/2)}{4}
     \end{split}
     \\
   \beta_k\approx & \bigg(K_{\nu_2}\Big(i \kbar(3\mu-1)\Big)H^{(1)}_{\nu_1}(\kbar)\left(\frac{3\mu}{\sqrt{3\mu-1}}\right)+\kbar\sqrt{3\mu-1}\Big(H^{(1)}_{\nu_1-1}(\kbar)-H^{(1)}_{\nu_1+1}(\kbar)\Big)\bigg)\nonumber\\
  \begin{split}
& \times  \text{exp}(3\mu \kbar) \frac{\text{exp}(-\pi \tilde{\nu}_1/2)}{4}
  \end{split} 
\end{align}
\end{subequations}
%%%%%%%%%%%%%%%%%%%%%%%%%%%%%%%%%%%%%%%%
%\frac{\Gamma(\nu_2)}{4}\left(\frac{2}{3\mu-1}\right)^{\nu_2}\left(\frac{3\mu-2\nu_2}{\sqrt{(3\mu-1)}}\right)H^{(1)}_{\nu_1}(k/k_{\rm end})\frac{{\rm exp}\left(i(\pi\nu_1/2+\pi\nu_2/2+\pi/2-3\mu k/k_{\rm end})\right)}{(k/k_{\rm end})^{\nu_2}}\nonumber\\

 %\frac{\Gamma(\nu_2)}{4}\left(\frac{2}{3\mu-1}\right)^{\nu_2}\left(\frac{3\mu-2\nu_2}{\sqrt{(3\mu-1)}}\right)H^{(1)}_{\nu_1}(k/k_{\rm end})\frac{{\rm exp}\left(i(\pi\nu_1/2-\pi\nu_2/2+3\mu k/k_{\rm end})\right)}{(k/k_{\rm end})^{\nu_2}}
%%%%%%%%%%%%%%%%%%%%%%%%%%%%%%%%
General field solution for $\xi>3/16$ becomes
%%%%%%%%%%%%%%%%%%%%%%%%%%%%
\begin{align}\label{Xreh6}
    X^{\rm long}_k(\eta)\approx & 
    \frac{\text{exp}(-\pi \tilde{\nu}_1/2)}{4\sqrt{2 k_{\rm end}}}\bigg(K_{\nu_2}\Big(i \kbar(3\mu-1)\Big)H^{(1)}_{\nu_1}(\kbar) \left(\frac{3\mu}{\sqrt{3\mu-1}}\right)+\kbar\sqrt{3\mu-1}\nonumber\\ 
    &\times \Big(H^{(1)}_{\nu_1-1}(\kbar)-H^{(1)}_{\nu_1+1}(\kbar)\Big)\bigg)\frac{\text{cos}(k\eta)}{\kbar^{1/2}}
    %\frac{\Gamma(\nu_2)}{2\sqrt{k_{\rm end}}}\left(\frac{2}{3\mu-1}\right)^{\nu_2}\left(\frac{3\mu-2\nu_2}{\sqrt{2(3\mu-1)}}\right)H^{(1)}_{\nu_1}(k/k_{\rm end})\frac{ \text{cos}(k\eta)}{(k/k_{\rm end})^{\nu_2+1/2}}
\end{align}

%%%%%%%%%%%%%%%%%%%%%%%%%%%%%%%%
where $\tilde{\nu}_1=(\sqrt{48\xi-9})/2$.
\subsection{Field solution at large scale for $1/3\leq\wre\leq 1$ }

\subsubsection{For $0\leq\xi<3/16$ :}
Likewise in the previous case, the indices $\nu_1, \nu_2$ are also real positive in this EoS range. In the limit $k/k_{\rm end}<<1$, using equations (\ref{Xinf3}),(\ref{Xreh3}), and (\ref{bogo}), we obtain Bogoliubov coefficients as
%%%%%%%%%%%%%%%%%%%%%%%%%%%%%%%%%%%%%%%%%%%%%%%%%
\begin{subequations}\label{bogolarge4}
\begin{align}
     &\alpha_k\approx -\frac{\Gamma(\nu_1)\Gamma(\nu_2)2^{\nu_1}}{8\pi}\left(\frac{2}{3\mu-1}\right)^{\nu_2}\left(\frac{3\mu(1-2\nu_1)+2(\nu_1-\nu_2)}{\sqrt{(3\mu-1)}}\right)\frac{{\rm exp}\left(i(\pi\nu_1/2+\pi\nu_2/2-\pi/2-3\mu \kbar)\right)}{\kbar^{(\nu_1+\nu_2)}}\\
  & \beta_k\approx \frac{\Gamma(\nu_1)\Gamma(\nu_2)2^{\nu_1}}{8\pi}\left(\frac{2}{3\mu-1}\right)^{\nu_2}\left(\frac{3\mu(1-2\nu_1)+2(\nu_1-\nu_2)}{\sqrt{(3\mu-1)}}\right)\frac{{\rm exp}\left(i(\pi\nu_1/2-\pi\nu_2/2+3\mu \kbar)\right)}{\kbar^{(\nu_1+\nu_2)}}
\end{align}
\end{subequations}
%%%%%%%%%%%%%%%%%%%%%%%%%%%%%%%%%%%%%%%%%%%%%
Subject to the following coefficients, the general field solution in $k\eta>>1$ limit becomes
%%%%%%%%%%%%%%%%%%%%%%%%%%%%%%%%%%%%%

\begin{equation}\label{Xreh7}
  X^{\rm long}_k(\eta)\approx\frac{\Gamma(\nu_1)\Gamma(\nu_2)2^{\nu_1}}{4\pi\sqrt{k_{\rm end}}}\left(\frac{2}{3\mu-1}\right)^{\nu_2}\left(\frac{3\mu(1-2\nu_1)+2(\nu_1-\nu_2)}{\sqrt{2(3\mu-1)}}\right) \frac{ i \text{sin}(k\eta)}{\kbar^{(\nu_1+\nu_2+1/2)}}  
\end{equation}

%%%%%%%%%%%%%%%%%%%%%%%%%%%%%%%%%%%%%%%%%
\subsubsection{For $\xi=3/16$ :}
For this particular value of $\xi$, the expression of Bogoliubov coefficients as well as general field solution will remain the same for this EoS range $1/3\leq \wre\leq1$ also. Here also we obtain $\alpha_k$ and $\beta_k$ as
%%%%%%%%%%%%%%%%%%%%%%%%%%%%%%%%%
\begin{subequations}\label{bogolarge5}
\begin{align}
  &\alpha_k\approx -\frac{\Gamma(\nu_2)}{2}\left(\frac{2}{3\mu-1}\right)^{\nu_2}\left(\frac{3\mu-2\nu_2}{4\sqrt{(3\mu-1)}}+\frac{i\sqrt{3\mu-1}}{\pi}\right)\frac{{\rm exp}\left(i(\pi\nu_2/2-\pi/2-3\mu \kbar)\right)}{\kbar^{\nu_2}}\\
  & \beta_k\approx \frac{\Gamma(\nu_2)}{2}\left(\frac{2}{3\mu-1}\right)^{\nu_2}\left(\frac{3\mu-2\nu_2}{4\sqrt{(3\mu-1)}}+\frac{i\sqrt{3\mu-1}}{\pi}\right)\frac{{\rm exp}\left(i(-\pi\nu_2/2+3\mu \kbar)\right)}{\kbar^{\nu_2}}
\end{align}
\end{subequations}
%%%%%%%%%%%%%%%%%%%%%%%%%%%%%%%
Associated general field solution for $\xi=3/16$ in $k\eta>>1$ limit will be
%%%%%%%%%%%%%%%%%%%%%%%%%%%%%

\begin{equation}\label{Xreh8}
 X^{\rm long}_k(\eta)\approx \frac{\Gamma(\nu_2)}{\sqrt{2k_{\rm end}}}\left(\frac{2}{3\mu-1}\right)^{\nu_2}\left(\frac{3\mu-2\nu_2}{4\sqrt{(3\mu-1)}}+\frac{i\sqrt{3\mu-1}}{\pi}\right)\frac{i \text{sin}(k\eta)}{\kbar^{\nu_2+1/2}}  
\end{equation}

%%%%%%%%%%%%%%%%%%%%%%%%%%%%

\subsubsection{For $\xi>3/16$ :}
A significant difference in terms of spectral behavior will appear between two given EoS ranges in this particular case $\xi>3/16$. In this case, we get one index $\nu_1$ to be imaginary as expected but another index $\nu_2$ to be real positive which differs from the previous case for $0\leq \wre<1/3$. This causes a noticeable change in the spectral behavior as we see soon. Long-wavelength approximated coefficients are evaluated to be
%%%%%%%%%%%%%%%%%%%%%%%%%%%%%%%%%%%%%%%%%%
\begin{subequations}
\begin{align}\label{bogolarge6}
  \alpha_k\approx &\frac{\Gamma(\nu_2)\text{exp}(-\pi \tilde{\nu}_1/2)}{8}\left(\frac{3\mu-2\nu_2}{\sqrt{(3\mu-1)}}H^{(1)}_{\nu_1}(\kbar)+\kbar\sqrt{3\mu-1}\Big(H^{(1)}_{\nu_1-1}(\kbar)-H^{(1)}_{\nu_1+1}(\kbar)\Big)\right)\nn\\
  &\times\frac{{\rm exp}\left(i(\pi\nu_2/2+\pi/2-3\mu \kbar)\right)}{\kbar^{\nu_2}} \left(\frac{2}{3\mu-1}\right)^{\nu_2} \\
  \beta_k\approx &\frac{\Gamma(\nu_2)\text{exp}(-\pi \tilde{\nu}_1/2)}{8}\left(\frac{3\mu-2\nu_2}{\sqrt{(3\mu-1)}}H^{(1)}_{\nu_1}(\kbar)+\kbar\sqrt{3\mu-1}\Big(H^{(1)}_{\nu_1-1}(\kbar)-H^{(1)}_{\nu_1+1}(\kbar)\Big)\right)\nn\\
  &\times\frac{{\rm exp}\left(i(3\mu \kbar-\pi\nu_2/2)\right)}{\kbar^{\nu_2}} \left(\frac{2}{3\mu-1}\right)^{\nu_2}
\end{align}
\end{subequations}
%%%%%%%%%%%%%%%%%%%%%%%%%%%%%%%%%%%%%%%%%%%%
Therefore, the associated general field solution takes the following form.
%%%%%%%%%%%%%%%%%%%%%%%%%%%%%%%%%%%%%%%%
\begin{align}\label{Xreh9}
   X^{\rm long}_k(\eta)\approx & \frac{\Gamma(\nu_2)\text{exp}(-\pi \tilde{\nu}_1/2)}{4\sqrt{2k_{\rm end}}}\left(\frac{3\mu-2\nu_2}{\sqrt{(3\mu-1)}}H^{(1)}_{\nu_1}(\kbar)+\kbar\sqrt{3\mu-1}\Big(H^{(1)}_{\nu_1-1}(\kbar)-H^{(1)}_{\nu_1+1}(\kbar)\Big)\right)\nonumber\\
  &\times\frac{ \text{cos}(k\eta)}{\kbar^{\nu_2+1/2}}\left(\frac{2}{3\mu-1}\right)^{\nu_2}
\end{align}

\begin{figure}[t]
     \begin{center}
\includegraphics[scale=0.29]{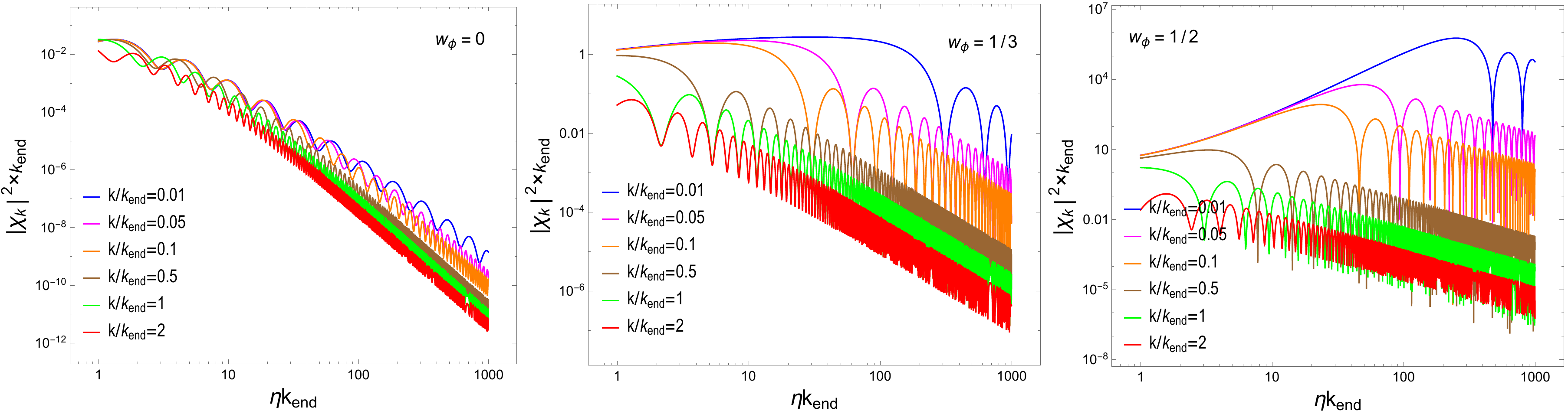}
\caption{\textit{Figure represents the time-evolution of dimensionless field amplitude square $(|\chi_k|^2\times\ke)$ associated with some large scale($k<\ke$) and small scale($k>\ke$) modes for three different EoS. In all three plots, we choose the non-minimal coupling strength to be $\xi=3$. It is clearly seen that for $\xi>1/6$, the post-inflationary instability effect associated with long-wavelength modes is effective for $\wre>1/3$ as discussed earlier. }}
\label{fieldevolutionfig}
\end{center}
\end{figure}

%%%%%%%%%%%%%%%%%%%%%%%%%%%%%%%%%%%%
%\subsection{Comoving number density spectrum ($|\beta_k|^2$)}
From our analysis so far it is revealed that the long-wavelength scalar field modes gets amplified through tachyonic instability during and after inflation depending upon the value of $(\xi, \wre)$. To this end let us reiterate again that for $0\leq\xi<1/6$ long wavelength modes after their horizon crossing during inflation, get amplified due to the tachyonic instability effect (see Eq.(\ref{dynamicalinf})). 
As $\xi$ exceeds the conformal limit $\xi=1/6$, the inflationary instability diminishes, whereas new tachyonic instability develops during reheating, particularly for stiff equation of state $\wre>1/3$ (see Eq.(\ref{dynamicalreh}). Enhancement of the scalar field modes due to those instabilities can indeed be observed in Fig.\ref{fieldevolutionfig}. 
In the figure, we have plotted the time evolution of different field modes assuming $\xi = 3$ for three different equations of state $\wre = (0, 1/3, 1/2)$. It can indeed be seen that for $\xi >1/6, \wre > 1/3$, the amplitude of the original field modes $\chi_k$ increases appreciably.  
%and the authors of \cite{Fairbairn:2018bsw} also hint at this possibility. 

%Taking these inflationary and post-inflationary instabilities into account, we study the behavior of the number density spectrum $|\beta_k|^2$ of the massless fluctuations for EoS $0\leq\wre\leq1$ $\big(\text{See Equations}~ 
%(\ref{no.den1})~ \text{and}~ (\ref{no.den2})\big)$. Depending upon different post-inflationary EoS, the large-scale(IR modes) number density spectrum follows different power-law behavior upon horizon reentry during reheating. 

The excitation of these large-scale modes due to the instability effect during and after inflation for different parameter regions of $\xi$ and $\wre$ motivates us to investigate the induced GWs. In the subsequent study of the generation of secondary GWs sourced by the anisotropy, we pay attention to the long-wavelength modes of the source, that lie in the range $k_{\ast}<k<k_{\rm end}$ where $\l(k_{\ast }/a_0\r)=0.05 \text{Mpc}^{-1}$ is the present-day CMB pivot scale.
%\subsubsection{For $0\leq\wre<1/3$ :}
%%%%%%%%%%%%%%%%%%%%%%%%%%%%%
%\begin{align}\label{no.den1}
%   |\beta_k|^2\propto
%   \begin{cases}
%       (k/k_{\rm end})^{-2(\nu_1+\nu_2)} \quad &\text{for} \quad 0\leq\xi<3/16 \\
%       (k/k_{\rm end})^{-2\nu_2} \quad &\text{for} \quad \xi=3/16 \\
       %(k/k_{\rm end})^{-2\nu_2} \quad &\text{for} \quad \xi>3/16 
%        \end{cases} 
%\end{align}
%%%%%%%%%%%%%%%%%%%%%%%%%%%%%

%\subsubsection{For $1/3\leq\wre\leq1$ :}

%%%%%%%%%%%%%%%%%%%%%%%%%%%%%
%\begin{align}\label{no.den2}
%   |\beta_k|^2\propto
%   \begin{cases}
%       (k/k_{\rm end})^{-2(\nu_1+\nu_2)} \quad &\text{for} \quad 0\leq\xi<3/16 \\
%       (k/k_{\rm end})^{-2\nu_2} \quad &\text{for} \quad \xi=3/16 \\
%       (k/k_{\rm end})^{-2\nu_2} \quad &\text{for} \quad \xi>3/16 
%        \end{cases} 
%\end{align}
%%%%%%%%%%%%%%%%%%%%%%%%%%%%%

%\subsection{Time evolution of the amplitude square of field mode ($|\chi_k|^2$)}

%%%%%%%%%%%%%%%%%%%%%%
%%%%%%%%%%%%%%%%%%%%%%%%%%%%%%%%%%%%%%%%%%%%%%%%%
\subsection{Behavior of scalar field energy density spectrum \enquote{$\rho_{\chi_k}(\eta)$} }

While defining the anisotropic stress tensor later, we require the nature of the power spectrum of the source field($\chi$). However, we shall soon see in the subsequent section that the gravitational wave energy density spectrum follows the spectral behavior of the field energy density spectrum($\rho_{\chi_k}$) in the long-wavelength regime $k<<\kre$. 
%That is why it is imperative to study the behavior of field power spectrum $\Pchi(k,\eta)$ as well as the field energy density spectrum $\rho_{\chi_k}(\eta)$ simultaneously to understand the nature of GW energy spectrum in all scales. 
According to the standard definition given in \cite{Baumann:2018muz}, we define the field
power spectrum of the produced fluctuations corresponding to the original field mode $\chi_k$ as $\mPchi(k,\eta)=\frac{k^3}{2\pi^2 a^2}|X_k|^2$.
Excluding the non-minimal coupling term, the standard  expression of energy density in terms of rescaled field mode($X_k$) and its derivative($X_k^{\prime}$) is $\rho_{\chi}=\frac{1}{4\pi^2 a^4}\int d(\text{ln}k)k^3(|X_k^{\prime}|^2+k^2|X_k|^2\big)$\cite{Ema:2018ucl, Kolb:2023ydq}. 
Hence associated energy density spectrum can be expressed as $\rho_{\chi_k}(\eta)=\frac{k^3}{4\pi^2 a^4}(|X_k^{\prime}|^2+k^2|X_k|^2\big) = \l(k^2/a^2\r)\mPchi(k,\eta)$. 
%Using the general post-inflationary field solution (\ref{Xgen}) and it's conformal time-derivative, we finally associate the energy density spectrum $\rho_{\chi_k}(\eta)$ with field power spectrum $\mPchi(k,\eta)$ at any time $\eta$ during reheating for any reheating EoS, $0\leq\wre\leq 1$, as $\rho_{\chi_k}(\eta)=\l(k^2/a^2\r)\mPchi(k,\eta)$. 
Now we shall discuss the spectral features of $\rho_{\chi_k}$in the entire post-inflationary EoS range $0\leq \wre\leq 1$ for three different $\xi$ ranges.
%%%%%%%%%%%%%%%%%%%%%%
\subsubsection{For $0\leq\wre<1/3$ :}
\begin{align}\label{powerspec1}
    \rho_{\chi_{k}}(\eta>\ee) = \frac{k^2}{a^2} \mPchi(k,\eta >\eta_{\rm end})
    %=\frac{k^3}{2\pi^2}|\chi^{\rm long}_k|^2
    \propto 
    \begin{cases}
        (\text{cos}^2(k\eta)/a(\eta)^4)(k/k_{\rm end})^{2(2-\nu_1-\nu_2)} & \text{for} \quad 0\leq\xi<3/16 \\
(\text{sin}^2(k\eta)/a(\eta)^4)(k/k_{\rm end})^{2(2-\nu_2)} & \text{for} \quad \xi=3/16\\
%(\text{cos}^2(k\eta)
(1/a(\eta)^4)(k/k_{\rm end})^4 & \text{for} \quad \xi>3/16
\end{cases}
\end{align}
%%%%%%%%%%%%%%%%%%%%%%%%%%%%%
\subsubsection{ For $1/3\leq\wre\leq1$ : }
%%%%%%%%%%%%%%%%%%%%%%%%%%%%%%%
\begin{align}\label{powerspec2}
    \rho_{\chi_k}(\eta>\ee)
    %=\frac{k^3}{2\pi^2}|\chi^{\rm long}_k|^2
    \propto 
    \begin{cases}
        (\text{sin}^2(k\eta)/a(\eta)^4)(k/k_{\rm end})^{2(2-\nu_1-\nu_2)} & \text{for} \quad 0\leq\xi<3/16 \\
(\text{sin}^2(k\eta)/a(\eta)^4)(k/k_{\rm end})^{2(2-\nu_2)} & \text{for} \quad \xi=3/16\\
(\text{cos}^2(k\eta)/a(\eta)^4)(k/k_{\rm end})^{2(2-\nu_2)} & \text{for} \quad \xi>3/16
\end{cases}
\end{align}
%%%%%%%%%%%%%%%%%%%%%%%%%%%%%
Depending upon EoS in three ranges $0\leq\wre<1/3$, $\wre=1/3$, and $1/3\leq\wre\leq 1$, the scalar field energy density spectrum $\rho_{\chi_k}(\eta)$ has interesting spectral behavior with the variation of coupling strength $\xi$. We first illustrate those important features of the $\rho_{\chi_k}(\eta)$ spectrum at a fixed time(and this is true for any time) for varying EoS $\wre$ in different ranges of $\xi$ values. We shall next discuss the behavior of the spectrum at varying instants of time for a fixed coupling strength.  
%%%%%%%%%%%%%%%%%%%%%%%%%%%%%%%%%%%%%%%
\subsubsection{\underline{Spectral behavior of \enquote{$\rho_{\chi_k}(\eta)$} at a fixed time }:}
\begin{figure}[t]
     \begin{center}
\includegraphics[scale=0.2]{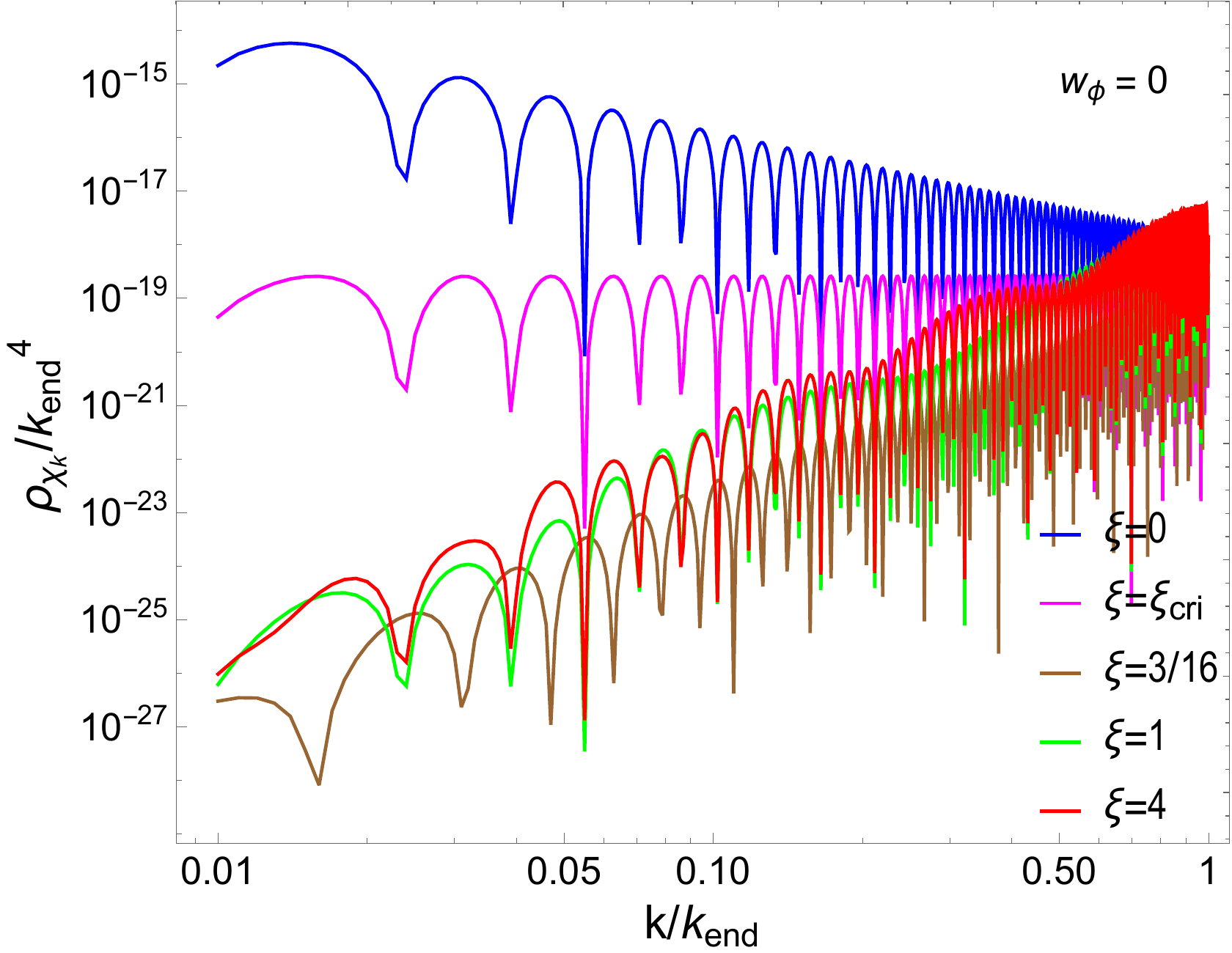}
\includegraphics[scale=0.2]{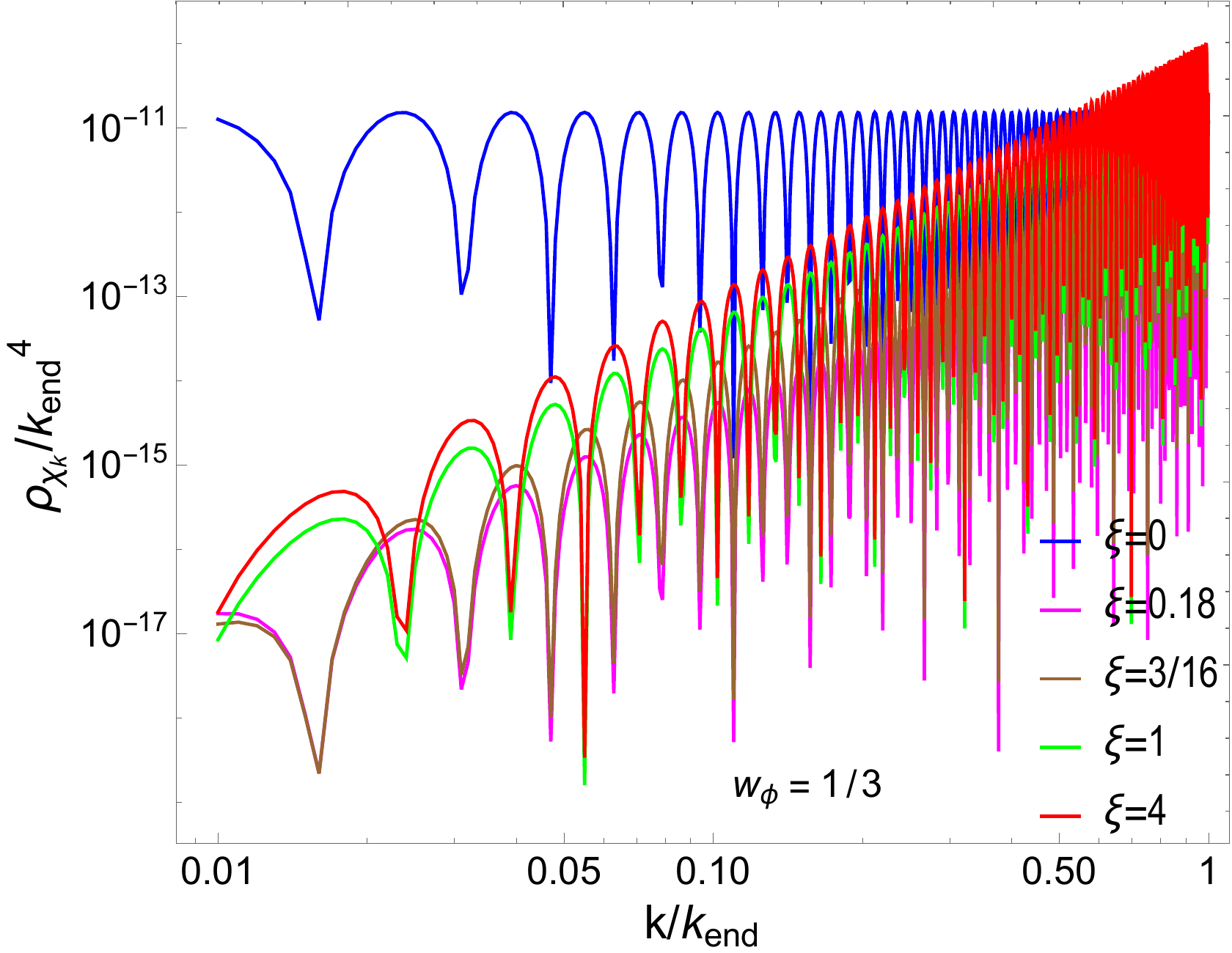}
\includegraphics[scale=0.2]{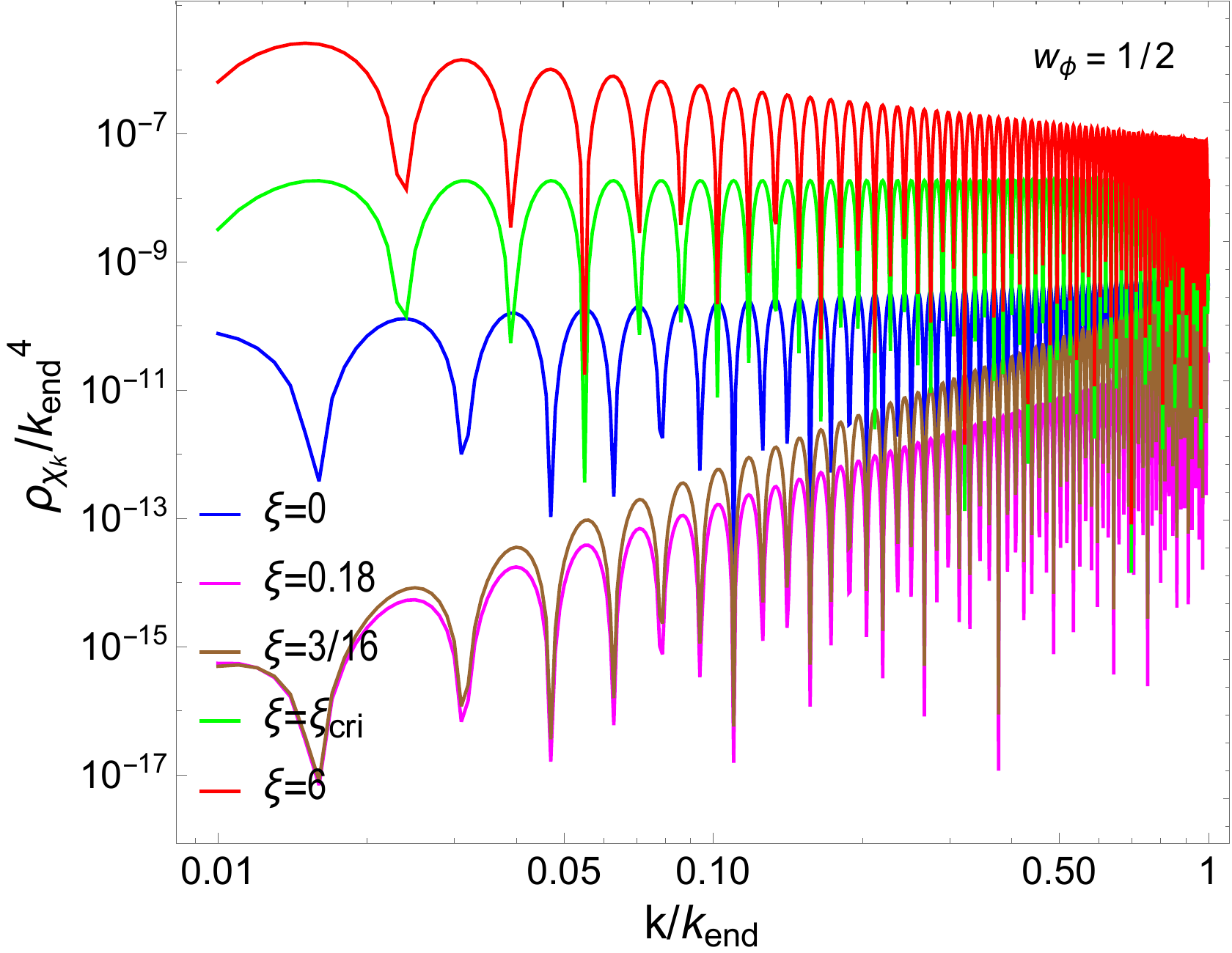}
\caption{\textit{Figure represents the variation of dimensionless scalar field energy density spectrum $\rho_{\chi_k}(\eta)/k_{\rm end}^4$ with non-minimal coupling strength $\xi$ for three reheating EoS, $\wre=0,~1/3,~1/2$. For $\wre=0$, the critical coupling $\xi_{\rm cri}=5/48$, and for $\wre=1/2$, critical coupling $\xi_{\rm cri}\approx 4.073$. }}
\label{powerspectrumfig}
\end{center}
\end{figure}

%%%%%%%%%%%%%%%%%%%%%%%%%%%%%
We first study the nature of the energy density spectrum for varying coupling strength at a fixed time during reheating. 
%%%%%%%%%%%%%%%%%%%%%%%%%%%%%%%%
\begin{itemize}
\item \underline{For $0\leq\wre<1/3$ :}~~
Examining the spectrum as given in (\ref{powerspec1}) for $0\leq\wre<1/3$, we find that for $0\leq\xi<1/6$, the spectrum is always red-tilted or IR divergent $\rho_{\chi_k}(\eta)\propto k^{2(2-\nu_1-\nu_2)}$ with $(2-\nu_1-\nu_2)<0$. However, the amount of red tilt depends upon the choice of EoS $\wre$ through the following relations,
%%%%%%%%%%%%%%%%%%%%%%%%%
\begin{equation}\label{indices}
\nu_1=\frac{\sqrt{9-48\xi}}{2}~~~;~~~\nu_2=\frac{\sqrt{3(1+\wre)\Big(3(1-\wre)^2+16\xi(3\wre-1)\Big)}}{2\sqrt{1+3\wre}\sqrt{1+4\wre+3\wre^2}} .
\end{equation}
%%%%%%%%%%%%%%%%%%%%%%
With the above mentioned parameter ranges $\nu_2$ should lie within $\sqrt{9-48\xi}/2 > \nu_2 > 1/2$. For example, as $\wre$ approaches zero the spectrum becomes maximally red-tilted $\rho_{\chi_k}(\eta)\propto k^{2(2-2\nu_1)}$ for given $\xi < 1/6$ up to the sinusoidal function of $k$. Such red tilt can indeed be observed in the blue curve in the left panel of Fig.\ref{powerspectrumfig}. Furthermore, there exists a critical coupling strength $\xi_{\rm 
 cri}$ lying in this range $0<\xi_{\rm cri}<1/6$, at which energy spectrum becomes scale-invariant giving $(4-2(\nu_1+\nu_2))=0$ (see the magenta line in the left panel of Fig.\ref{powerspectrumfig}). Therefore, for $0<\xi_{\rm cri}<1/6$, energy spectrum remains IR divergent in the range $0\leq\xi<\xi_{\rm cri}$ and it turns blue-tilted $(4-2(\nu_1+\nu_2))>0$ for $\xi>\xi_{\rm cri}$. Due to this red-tilted behavior of the energy density spectrum, the gravitational wave amplitude would be very large at the CMB scale. The \textit{Planck} constraint on tensor to scalar ratio $r_{0.05} < 0.036$ will, therefore be shown to set a lower limit on the value of $\xi$.    
Once $\xi$ exceeds conformal limit $\xi>1/6$, the spectrum remains blue-tilted till one reaches the $\xi = 3/16$.  
As $\xi$ exceeds 3/16, both the indices $\nu_1,\nu_2$ being imaginary results in the energy density spectrum to be insensitive to the non-minimal coupling strength with $\rho_{\chi_k}(\eta)\propto k^4$.
In the left panel of Fig.\ref{powerspectrumfig}
we can indeed see the blue titled spectrum for $\xi = (3/16,~ 1,~4)$ in brown, green and red respectively. In summary, in the entire range of $\xi>\xi_{\rm cri}$, the spectrum being blue-tilted draws the maximum contribution to the amplitude of the scalar power spectrum amplitude for those modes which left the horizon at the inflation end, that is $k_{\rm end}$. 
%%%%%%%%%%%%%%%%%%%%%%%%%%%%%%%%%%%%%
\item \underline{For $\wre=1/3$ :}~~
For this particular value of the equation of state, $\nu_2 =1/2$, irrespective of the choice of $\xi$, and in the range $0\leq\xi<1/6$, $\nu_1$ lies within $[3/2,1/2)$. Consequently, we obtain a scale-invariant energy density spectrum(see the blue line in the middle panel of Fig.\ref{powerspectrumfig}) for $\xi=0$, and a blue-tilted spectrum in the range $0<\xi<1/6$. As $\xi$ exceeds the conformal limit, the index $\nu_1$ gradually decreases with the increase of $\xi$ up to $\xi=3/16$, and in this range, $\nu_1$ lies within $(0.5,0] $. According to the spectral index given in Eq.(\ref{powerspec2}), spectrum in the range $1/6 <\xi\leq 3/16$ becomes blue-tilted as shown in magenta and brown color for $\xi=(0.18, 3/16)$ respectively in the middle panel of Fig. \ref{powerspectrumfig}. For $\xi>3/16$, $\nu_1$ becomes imaginary as is obvious from (\ref{indices}). With the further increase of $\xi$, the independence of $\nu_2$ makes the slope of the spectrum completely insensitive to the coupling strength in the entire range $\xi>3/16$ although our numerical analysis shows very slow growth of the amplitude with increasing $\xi$ (see the green and red lines in the middle panel of Fig.\ref{powerspectrumfig} for $\xi=(1,~4)$ respectively). From the spectral index given in Eq.(\ref{powerspec2}), in the range $\xi\geq 3/16$, the power spectrum behaves as $\rho_{\chi_k}(\eta)\propto k^3$. 
%%%%%%%%%%%%%%%%%%%%%%%%%%%%%%%%%
\item \underline{For $1/3<\wre\leq 1$ :}~~

%In the range $0\leq\xi<1/6$, again we obtain a red-tilted (IR divergent) power spectrum

 Contrary to the previous case, the energy density spectrum for $1/3< \wre\leq 1$, is blue-tilted in the range $0\leq\xi<3/16$ as obvious in the right panel of Fig.\ref{powerspectrumfig} for $\xi=0$ (blue line, also see the magenta and brown lines in the right panel of Fig.\ref{powerspectrumfig}). However, for $\xi>3/16$, the energy spectrum has some noticeable features. For this case $\nu_2$ should lie within the range $(1/2, \sqrt{3\xi/2})$. Given an EoS $\wre>1/3$, there exists a particular coupling strength say, $\xi_{\rm cri} = \frac{(9 \wre +7) (15 \wre+1)}{48 (3 \wre-1)}$, at which the spectrum becomes perfectly scale-invariant giving $(2-\nu_2)=0$ (see Eq.(\ref{powerspec2})). For $3/16<\xi<\xi_{\rm cri}$, spectrum $\rho_{\chi_k}$ remains blue-tilted, $(2-\nu_2)>0$ and turns into red-tilted or IR divergent, $(2-\nu_2)<0$ for $\xi > \xi_{\rm cri}$. In the right panel of Fig.\ref{powerspectrumfig}, we notice the scale-invariant and red-tilted spectrum for $\xi = (\xi_{\rm cri},~ 6)$ in green and red lines respectively.

All these important characteristics of the energy density spectrum are shown in Fig.\ref{powerspectrumfig} for three EoS $\wre=0, \wre= 1/3, \text{and}~\wre=1/2$ for different $\xi$ values at some point of time during reheating. Variations of spectral tilt with the variation of $\xi$ are believed to leave a discernible imprint on the induced GWs spectrum, which we intend to investigate in the subsequent section.
\end{itemize}

%%%%%%%%%%%%%%%%%%%%%%
%%%%%%%%%%%%%%%%%%%%%%%%%%%%%%%%%%%%%%%%%%%%%%%%%

%%%%%%%%%%%%%%%%%%%%%%%%%%%%%%%%%%
\subsubsection{\underline{Spectral behavior of \enquote{$\rho_{\chi_k}(\eta)$} for varying time }:}

\begin{figure}[t]
     \begin{center}
\includegraphics[scale=0.3]{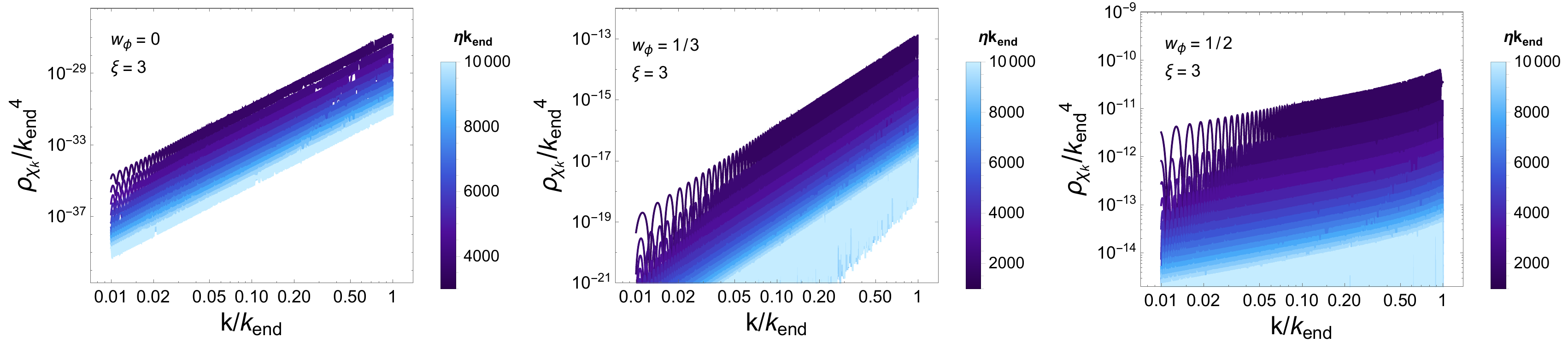}
\caption{\textit{Figure represents the behavior of dimensionless scalar field energy density spectrum $\rho_{\chi_k}(\eta)/k_{\rm end}^4$ for three EoS and a specific coupling strength $\xi=3$ with the variation of dimensionless time-variable $\eta\ke$ indicated by the color bar where the color gradient is indicating this time-evolution in the post-inflationary phase.}}
\label{powerspectrumgradfig}
\end{center}
\end{figure}

Here we show for a certain coupling strength how the spectral shape will change over time during the reheating phase. We notice that as we go deep into the reheating phase $\eta>>\eta_{\rm end}$, the amplitude of the spectrum gets diminished for a given $k$-mode as obvious in Fig.\ref{powerspectrumgradfig}, and this is because of the decaying nature of every mode after horizon reentry during reheating. For the chosen value of coupling strength $\xi=3$, the blue-tilted nature of the energy density spectra for $\wre=0,~1/3,~ 1/2$ are also consistent with the expressions given in Eqs. (\ref{powerspec1}) and (\ref{powerspec2}).  

%For a given reheating EoS $\wre>1/3$, the spectral tilt of $\mathcal{P}_{\chi}(k,\eta)$ changes noticeably in three different ranges of $\xi$ values as given in Eq.(\ref{powerspec2}). For $0\leq\xi<3/16$, the power spectrum $\mathcal{P}_{\chi}(k,\eta)$ is always red-tilted, for $\xi=3/16$, it changes to blue-tilted, and for $\xi>3/16$, it first goes towards a scale-invariant spectrum with the increase of the strength $\xi$, and for a specific EoS, there exists a particular value of $\xi$, for which spectrum becomes completely scale-invariant. After that, it again starts behaving like a red-tilted spectrum with the further increase of the non-minimal strength $\xi$. The variation of this spectral nature with $\xi$ is obvious in Fig.\ref{powerspectrumfig}. As the variation of non-minimal coupling strength is nicely captured in the behavior of scalar power spectrum $\mathcal{P}_{\chi}(k,\eta)$, this is also expected to have the imprint of this interesting slope-changing feature in secondary gravitational wave spectrum. 
%%%%%%%%%%%%%%%%%%%%%%%%%%%
%%%%%%%%%%%%%%%%%%%%%%%%%%%%%%%%%%%%%%%%%%%%%%%%%

\iffalse
where we have defined $\mathcal{A}$ as
%%%%%%%%%%%%%%%%%%%%%%%%%%%%%%%%
\begin{align}
    \mathcal{A}=\frac{\Gamma^2(\nu)}{8\pi}\l(\frac{2}{3\mu-1}\r)^{2\nu}\l(\frac{3(2\mu-1)}{\sqrt{2(3\mu-1)}}\r)^2
\end{align}
%%%%%%%%%%%%%%%%%%%%%%%%%%%%%%%%%
we recall that $\mu~\&~\nu$ are defined in Eq.(\ref{symbol}).\\
%%%%%%%%%%%%%%%%%%%%%%%%%%%%
%\begin{align}\label{power}
  %\mathcal{P}_{\chi_k}&= \frac{k^3}{2\pi^2}|\chi^{\rm long}_k|^2\nonumber\\
 % &=\frac{H^2_{\rm end}\bigg(\eta H_{\rm end}+\frac{3(1+w_{\rm re})}{(1+3w_{\rm re})}\bigg)^{2\nu+1}}{ 8\pi\times  4^{\nu}\nu^2\Gamma^2(\nu)}\Big(\frac{k}{H_{\rm end}}\Big)^{2\nu+3} \simeq \frac{k^2}{8\pi\times4^{\nu}\nu^2\Gamma^2(\nu)}(k\eta)^{2\nu+1}\quad \text{for}~~\eta>>1
 % \end{align}
%%%%%%%%%%%%%%%%%%%%%%%%%%%
We shall require this power spectrum while computing the tensor power spectrum later. Clearly, this is a red-tilted spectrum as the spectral index $(2\nu+1)$ is always positive for any EoS $0 \leq \wre\lesssim 1$. \\
\fi
%%%%%%%%%%%%%%%%%%%%%%%%%%%%%%%%
\subsection{Defining the Reheating Parameters ($N_{\rm re}$, $T_{\rm re}$) : }
%%%%%%%%%%%%%%%%%%%%%%%%%%%%%%%%%%
In this section we introduce pivotal inflationary parameters namely the inflationary energy scale ($\HI$), the duration of the inflationary period denoted by e-folding number $N_\ast$, and define a crucial reheating parameter, namely the \textit{Reheating Temperature} ($\Tre$). From the CMB pivot scale, we have $(k_\ast/a_0) = 0.05 \mathrm{Mpc}^{-1}$ where $a_0$ is the present-day scale factor, and $\ke$ denotes the wave number that crosses the Hubble radius at the end of inflation, and connecting these two scales, we express $\ke=k_\ast \mathrm{e}^{N_\ast}$.

The \textit{Planck} data imposes constraints on the amplitude of scalar perturbation and tensor-to-scalar ratio, setting them at $\As=2.1\times 10^{-9}$ and $r_{0.05} \leq 0.036$~\cite{Planck:2018jri, Planck:2018vyg}.
With $\HI = \pi \mpl\sqrt{rA_s/2}$, the constraint on $r$ implies $\HI \lesssim 10^{-5}\mpl$, where $\mpl=1/\sqrt{8\pi G}$ is the reduced \textit{Planck} mass as mentioned earlier.

Assuming the reheating dynamics are characterized by an average inflaton equation of state $\wre$, the evolution of the inflaton energy density during this period is described by the simple expression $\rho_{\phi} = \rho_{\phi}^{\rm end}(\ae/a)^{3(1+\wre)}$, where $\rho_{\phi}^{\rm end} = 3\HI^2\mpl^2$ is the total inflaton energy density at the end of inflation.

The reheating temperature is conventionally defined at the end of the reheating period, where the radiation energy density equals the inflaton energy density, i.e., $\rho_{\rm R}(\ere) = \rho_{\phi}(\ere)$, with $\ere$ being the conformal time defined at the end of the reheating period. Employing this condition, the \textit{Reheating Temperature} $(\Tre)$ can be expressed as \cite{PhysRevLett.113.041302, Maiti:2024nhv}
\begin{align}
\Tre = \left(\frac{90\HI^2\MP^2}{\pi^2\gre}\right)^{1/4} \exp\left[-\frac{3\Nre}{4}(1+\wre)\right].\label{eq:tre}
\end{align}
Alternatively, the duration of the reheating period can be expressed as \cite{PhysRevLett.113.041302, Afzal_2023}
%%%%%%%%%%%%%%%%%%%
\begin{align}
\Nre = \frac{1}{3(1+\wre)}\ln{\left(\frac{90\HI^2\MP^2}{\pi^2\gre\Tre^4}\right)}\label{eq:nre}.
\end{align}
%%%%%%%%%%%%%%%%%%%%%%%%%%%
Here, $g_{\text{re}} = 106.7$ represents the number of relativistic degrees of freedom at the beginning of the radiation epoch.

Assuming negligible entropy production after reheating, leading to the conservation of comoving entropy density ($a^3(\eta\geq\ere)s=\text{const}$), we can establish a connection between the lowest possible mode re-entering the horizon at the end of reheating and the reheating temperature as 
%%%%%%%%%%%%%%%%%%%%%%%%%
\begin{align}\label{kre}
    \l(\kre/a_0\r)=\left(\frac{43 }{11\gre}\right)^{1/3}\left(\frac{\pi^2\gre}{90}\right)^{1/2}\l(\frac{\Tre T_0}{M_{\rm pl}}\r)\simeq 1.82\times 10^5~\l(\frac{\Tre}{10^{-2}~\text{GeV}}\r)~\text{Mpc}^{-1} .
\end{align}
%%%%%%%%%%%%%%%%%%%%%%%%%

By utilizing Eq. (\ref{eq:nre}), we can define the largest mode that left the horizon at the end of inflation as
%%%%%%%%%%%%%%%%%%%%%%%%
\begin{align}
    \l(\ke/a_0\r)=\left(\frac{43 }{11\gre}\right)^{1/3}\left(\frac{\pi^2\gre}{90}\right)^{\alpha}\frac{\HI^{1-2\alpha}\Tre^{4\alpha-1}T_0}{\MP^{2\alpha}},
\end{align}
where $\alpha=1/3(1+\wre)$ and $T_0=2.725$ K is the present-day CMB temperature.

\section{Production of Gravitational Waves}\label{sec3}
%%%%%%%%%%%%%%%%%%%%%%%%%
In this section, our primary interest is to investigate the effect of the produced scalar fluctuations in the presence of non-minimal coupling as a possible source of anisotropy on gravitational waves. 
The perturbed FLRW metric can be written as
%%%%%%%%%%%%%%%%%%%%%%%%%%%
\begin{align}
ds^2=a^2(\eta)\left[-d\eta^2+(\delta_{ij}+h_{ij})dx^idx^j\right],
\end{align}
%%%%%%%%%%%%%%%%%%%%%%%%%%%%%
where \enquote{$\eta$} is the conformal time and $h_{ij}$ is the traceless tensor, i.e. $\partial^ih_{ij}=h^i_i=0$.
% Now to find the dynamical equation of tensor perturbation we shall treat `$h_{ij}(\eta,\textbf{x})$' as a quantum field in an unperturbed FRW background metric, and if we keep up to quadratic order in `$h_{ij}$',
The tensor perturbations in the presence of anisotropic stress are governed by the following action up to quadratic order \cite{Boyle:2005se}
 %%%%%%%%%%%%%%%%%%%%%%%%%%%
\begin{align}\label{eq2}
    S_{GW}=\int dx^4\sqrt{-g}\left[-\frac{g^{\mu\nu}}{64\pi G}\partial_{\mu}h_{ij}\partial_{\nu}h_{ij}+\frac{1}{2}\Pi_{ij}h_{ij}\right],
\end{align}
%%%%%%%%%%%%%%%%%%%%%%%%%%%%
where `$\Pi_{ij}$' is the anisotropic stress, defined as \cite{Boyle:2005se} $\Pi^i_j=T^i_j-p\delta^i_j$. Here `$\Pi_{ij}$' also satisfies the transverse $\left(\partial^i\Pi_{ij}=0\right)$ and traceless $(\Pi^i_i=0))$ conditions. Here $\Pi_{ij}$ coupled with the tensor-perturbations $h_{ij}$ acting like an external source. From the expression of the stress-energy tensor of a massless scalar field having non-minimal gravity coupling $\xi\chi^2 R$ \cite{Birrell_Davies_1982, Ema:2018ucl,Kolb:2023ydq}, we write the expression of anisotropic stress tensor as
%%%%%%%%%%%%%%%%%%%%%%%%%%%%%
\begin{equation}\label{anisotropicstress}
\Pi_{ij}\sim (1-2\xi)\partial_{i}\chi\partial_{j}\chi-2\xi\chi\partial_{i}\partial_{j}\chi+\xi\chi^2 G_{ij}
\end{equation}
%%%%%%%%%%%%%%%%%%%%%%%%%%%
By varying $h_{ij}$ in action (\ref{eq2}), we obtain the equation of motions of $h_{ij}$ \cite{Fu:2017ero,Arapoglu:2022vbf}
\begin{align}\label{eq3}
    h''_{ij}(\eta,\vx)+2\mathcal{H}h'_{ij}(\eta,\vx)-\nabla^2h_{ij}(\eta,\vx)=16\pi Ga^2\left(1-\frac{a^2\xi\langle\chi^2\rangle}{M_{pl}^2}\right)^{-1}\Big((1-2\xi)\partial_{i}\chi\partial_{j}\chi-2\xi\chi\partial_{i}\partial_{j}\chi\Big) ,
\end{align}
%%%%%%%%%%%%%%%%%%%%%%%%%%%%%%%%%%
where the vacuum expectation value of fluctuations square can be expressed in terms of Fourier modes as
%%%%%%%%%%%%%%%%%%%%%%%%%%%%%%%%
\begin{equation}\label{chisquarevev}
  \langle\chi^2\rangle= \frac{1}{a^2}\int_{\kre}^{\ke}\frac{k^3}{2\pi^2}|X_k|^2~ d(\text{ln}(k)) .
\end{equation}
%%%%%%%%%%%%%%%%%%%%%%%%%%%%%%%%%%%%%%
For our specific choices of $\xi$, $a^2\xi\langle\chi^2\rangle<M_{pl}^2$ should be always satisfied. Therefore, $\left(1-a^2\xi\langle\chi^2\rangle/M_{pl}^2\right)\approx 1$ is well justified(See the details in the Appendix \ref{appenc}). Now it is good to write the above Eq(\ref{eq3}) in the following fashion \cite{ Sorbo:2011rz, Caprini:2014mja, Ito:2016fqp, Sharma:2019jtb, Okano:2020uyr} 
%%%%%%%%%%%%%%%%%%%%%%%%%
\begin{align}\label{eq:hij_2}
     h''_{ij}(\eta,\vx)+2\mathcal{H}h'_{ij}(\eta,\vx)-\nabla^2h_{ij}(\eta,\vx)=\frac{2}{\mpl^2}\pijlm \Tlm(\eta,\vx)
\end{align}
%%%%%%%%%%%%%%%%%%%%%%%
where $\pijlm=P^l_iP^m_j-P_{ij}P^{lm}/2$ is the transverse traceless projector with $P_{ij}=\delta_{ij}-\partial_i\partial_j/\Delta$ and for massless non-minimally coupled system $\Tlm(\eta,\vx)= (1-2\xi)\partial_l \chi \partial_m \chi+\left(2\xi-\frac{1}{2}\right)g_{lm}(\partial_{\alpha} \chi\partial^{\alpha}\chi)+2\xi\left(g_{lm}\chi\Box\chi-\chi\nabla_l\partial_m\chi\right)+\xi G_{lm}\chi^2$\cite{Ema:2018ucl,Kolb:2023ydq} represents the spatial part of the stress-energy momentum tensor of the scalar field $\chi$.

Recall that the tensor perturbations~$h_{ij}(\eta,{\bm x})$ evolving in a 
Friedmann universe can be decomposed in terms of the Fourier modes, say, 
${h}_{\bm k}^{\lambda}(\eta)$, as follows:
%%%%%%%%%%%%%%%%%%%%%%%%%%%%
\begin{equation}
h_{ij}(\eta,{\bm x})
=\sum_{\lambda=(+,\times)}\int \frac{d^3{\vk}}{(2\pi)^{3/2}} 
e_{ij}^{\lambda}({\vk}) h_{\vk}^{\lambda}(\eta) 
\mathrm{e}^{i\vk\cdot{\vx}},
\end{equation}
%%%%%%%%%%%%%%%%%%%%%%%%%%%%
where $e^{\lambda}_{ij}({\bm k})$ is the polarization tensor corresponding
to the mode with wave vector~${\bm k}$ and the index~$\lambda$ represents the 
two types of polarization of the GWs.
Note that $e^{\lambda}_{ij}({\bm k})$ is assumed to be real in the linear polarization basis and 
implying $h_{-\bm k}^{\lambda}(\eta)
= {h}_{\bm k}^{\lambda *}(\eta)$, and the mode functions $h_{\bm k}^\lambda(\eta)$ 
satisfy the following inhomogeneous equation~\cite{Sorbo:2011rz,
Caprini:2014mja, Sharma:2019jtb}:
%%%%%%%%%%%%%%%%%%%%%%%%%%%%%%%%
\begin{equation}\label{gweq}
h_{\vk}^{\lambda{}''}+2\frac{a'}{a}h_{\vk}^{\lambda}{}'
+k^2h_{\vk}^{\lambda}=\frac{2}{\mpl^2} e^{ij}_{\lambda}({\vk})\pijlm(\hat{k})T_{lm}(\vk, \eta) .
\end{equation}
%%%%%%%%%%%%%%%%%%%%%%%%%%%%%%%%%%%
 For this type of source, we chose the linear polarization basis where both polarization modes contribute equally to the total energy density of the produced gravitational waves (GWs). Henceforth, We drop the polarization index to simplify the notation and incorporate its information into the tensor power spectrum. The tensor power spectrum is defined as
\begin{align}
    \Pt(k,\eta) = \delta^{(3)}(\vk - \vk_1) \frac{k^3}{2\pi^2} \sum_{\lambda = +,\times} \langle h^{\lambda}_{\vk}(\eta) h^{\lambda *}_{\vk_1}(\eta) \rangle = 4 \frac{k^3}{2\pi^2} |h_{\vk}(\eta)|^2 \label{eq:pt}
\end{align}
Utilizing the Green's function method, the solution for tensor perturbation can be expressed as\cite{PhysRevD.85.023534}
%%%%%%%%%%%%%%%%%%%%%%%%%%
\begin{align}
h_{\vk}(\eta)=h_{\vk}^{\rm vac}+\frac{2e^{ij}(\vk)}{\mpl^2}\int d\eta_1\mGk(\eta,\eta_1)\pitt(\vk,\eta_1).\label{eq:hij_sol}
\end{align}
%%%%%%%%%%%%%%%%%%%%%%%%%
 Here $h_{\vk}^{\rm vac}$ is the homogeneous contributions of the tensor fluctuations and $\mGk(\eta,\eta_1)$ is the retarded propagator solving the Eq.(\ref{eq:hij_2}) with delta function source. In the above Eq.(\ref{eq:hij_sol}) we define $\pitt(\vk,\eta_1)=\pijlm(\vk)\Tlm(\eta,\vk)$ with $\pijlm(\vk)$  and $\Tlm(\eta,\vk)$ being the Fourier transformation of the transverse-traceless projector and the energy-momentum tensor respectively. 
 %Now we can define the tensor power spectrum as\cite{2024arXiv240101864M}
 %%%%%%%%%%%%%%%%%%%%%%%%%%%%%
%\begin{align}
%\langle \hij(\vk,\eta)h^{ *ij}(\vk',\eta)\rangle=\frac{2\pi^2}{k^3}\Pt(k,\eta)\delta^{(3)}(\vk+\vk')\label{eq:pt}
%\end{align}
%%%%%%%%%%%%%%%%%%%%%%%%
%where tensor power spectrum $\Pt(k,\eta)$ is expressed in terms of the Fourier modes $\hij(\vk,\eta)$ as $\Pt(k,\eta)=4\frac{k^3}{2\pi^2}|\hij(\vk,\eta)|^2$.

By substituting Eq. (\ref{eq:hij_sol}) into Eq. (\ref{eq:pt}), we derive the secondary production of the tensor power spectrum, which is defined as
%%%%%%%%%%%%%%%%%%%%%%%%%
\begin{align}
    \Pts(k,\eta) &= 4 \frac{k^3}{2\pi^2} \frac{4}{\mpl^4} \int_{\eta_i}^{\eta_f} d\eta_1 \mGk(\eta,\eta_1) \int_{\eta_i}^{\eta_f} d\eta_2 \mGk(\eta,\eta_2) \Pi^2(k,\eta_1,\eta_2) . \label{eq:TP}
\end{align}
%%%%%%%%%%%%%%%%%%%%%%%%
Here, \(\eta_i\) and \(\eta_f\) represent the initial and final times when the source was active to produce the tensor fluctuations. The term \(\Pi^2\) on the right-hand side is defined as the correlator of the source \(\Pi_{ij}^{\rm TT}\), given by \cite{Figueroa:2017vfa}:
  \begin{align}
      \langle 0|\Pi_{ij}^{\rm TT}(\vk,\eta_1)\Pi_{ij}^{\rm TT *}(\vk_1,\eta_2)|0\rangle=\delta^3(\vk-\vk_1)\Pi^2(k,\eta_1,\eta_2) .
  \end{align}
Utilizing Eq.(\ref{fourier}) and promoting it so quantum field $a(t) \chi_k = \hat{a}_{\vq}X_{\vq}(\eta)+\hat{a}_{-\vq}^{\dagger}X_{\vq}^{*}(\eta) $ \cite{Figueroa:2017vfa} we can write the Fourier expression of the source term as
  \begin{align}
      \Pi_{ij}^{\rm TT}(\vk,\eta)=&\frac{(2\xi-1)\pijlm(\hat{k})}{(2\pi)^3a^2(\eta)}\int d^3 q q_l(k-q)_m \l ( \hat{a}_{\vq}X_{\vq}(\eta)+\hat{a}_{-\vq}^{\dagger}X_{\vq}^{*}(\eta) \r)\l( \hat{a}_{\vk-\vq}X_{\vk-\vq}(\eta)+\hat{a}^{\dagger}_{-(\vk-\vq)}X^{*}_{\vk-\vq}(\eta) \r)\nonumber\\
      & +\frac{2\xi\pijlm(\hat{k})}{(2\pi)^3a^2(\eta)}\int d^3q q_lq_m \l ( \hat{a}_{\vq}X_{\vq}(\eta)+\hat{a}_{-\vq}^{\dagger}X_{\vq}^{*}(\eta) \r)\l( \hat{a}_{\vk-\vq}X_{\vk-\vq}(\eta)+\hat{a}^{\dagger}_{-(\vk-\vq)}X^{*}_{\vk-\vq}(\eta) \r).\label{eq:PiTT}
  \end{align}
The creation $(\hat{a}_{\vq})$ and annihilation $\hat{a}_{-\vq}^{\dagger}$ operators which contribute to the expectation value of Eq.(\ref{eq:PiTT}) are
%%%%%%%%%%%%%%%%%%%%%%%%%%%%%
\begin{subequations}\label{eq:pow_inf}
\begin{align}
&\langle 0|\hat{a}_{\vq}\hat{a}_{\vk-\vq}\hat{a}^{\dagger}_{\vq_1}\hat{a}^{\dagger}_{\vk_1-\vq_1}|0\rangle =(2\pi)^6\l[\delta^{(3)}(\vk-\vq-\vq_1)+\delta^{(3)}(\vq-\vq_1) \r]\delta^{(3)}(\vk-\vk_1),\label{eq:comu1}\\
&\langle 0|\hat{a}_{\vq}\hat{a}^{\dagger}_{-(\vk-\vq)}\hat{a}_{\vq_1}\hat{a}^{\dagger}_{-(\vk_1-\vq_1)}|0\rangle =(2\pi)^6\delta^{(3)}(\vk)\delta^{(3)}(\vk_1-\vk) ,\label{eq:comu2}
\end{align}
\end{subequations}
%%%%%%%%%%%%%%%%%%%%%%%
where we used the following commutation relation $[ \hat{a}_{\vk},\hat{a}^{\dagger}_{\vk_1}]=(2\pi)^3\delta^{(3)}(\vk-\vk_1)$. Since the second term Eq.(\ref{eq:comu2}) does not contribute to $\Pi^2(k,\eta,\eta_1)$ due to the finite momenta i.e. $k=k_1\neq 0$. The only term Eq.(\ref{eq:comu1}) contributes to the final expression of the correlator as %\cite{Figueroa:2017vfa, PhysRevD.85.023534}
%%%%%%%%%%%%%%%%%%%%%%%%%%%%%
\begin{align}
     \Pi^2(k,\eta_1,\eta_2)=\frac{1}{4\pi^2a^2(\eta_1)a^2(\eta_2)}\int dq q^6 \int d\gamma (1-\gamma^2)^2\langle X_{\vq}(\eta_1)X_{{\vk}-{\vq}}(\eta_1)X_{\vq}^*(\eta_2)X^*_{\vk-\vq}(\eta_2)\rangle\label{eq:Pi2}
  \end{align}
  %%%%%%%%%%%%%%%%%%%%%%%%
  where $\gamma=\hat{k}\cdot\hat{q}=\cos(\theta)$, where $\theta$ is the angle between $\vk$ and $\vq$. 
  %Note that at $\xi =1/4$ the leading order ${\cal O}(1/M_p^2)$ contribution to the energy momentum vanishes, 
  %d like to point out that the vanishing $\Pi^2(k,\eta_1,\eta_2)$ in the above Eq.(\ref{eq:Pi2}) for $\xi=1/4$ does not signify the complete nonexistence of the anisotropy in the present system. and sub leading  ${\cal O}(1/M_p^4)$ contribution will contribute.
  %It has already been argued in equations (\ref{eq3}) and (\ref{chisquarevev}) that a suppressed contribution comes from the term, $\l(1-\frac{a^2\xi\langle\chi^2\rangle}{M_{pl}^2}\r)^{-1}$ for any coupling strength $\xi$, as compared to the others. However for $\xi=1/4$, because of the disappearance of higher-order dominant source terms, this subdominant term will then generate anisotropy in the system that will be too mild to leave a visible imprint on the SGW spectrum. However, we will see that to achieve an appreciable strength of the SGW spectrum, $\xi$ should always remain in the domain greater than unity. 

Now utilizing Eq.(\ref{eq:Pi2}) in Eq.(\ref{eq:TP}) we have found the tensor power spectrum to be
\begin{align}
    \Pts(k,\eta) = &\frac{k^3}{2\pi^2}\frac{4}{\pi^2\mpl^4}  \times\int_{0}^{\infty}dq q^6 \int_{-1}^1 d\gamma(1-\gamma^2)^2 \nn\\
   & \times\int_{\ee}^{\eta}d\eta_1 \frac{\Gkre(\eta,\eta_1)}{a^2(\eta_1)}\int_{\ee}^{\eta}d\eta_2 \frac{\Gkre(\eta,\eta_2)}{a^2(\eta_2)} \langle X_{\vq}(\eta_1)X_{{\vk}-{\vq}}(\eta_1)X_{\vq}^*(\eta_2)X^*_{\vk-\vq}(\eta_2)\rangle \label{eq:ptra_chi}
\end{align}
 Here $\Pts(k,\eta)$ defines the secondary tensor power spectrum during reheating at conformal time $\eta$ induced due to massless scalar field $\chi$.
  
 % \begin{align}
 %     \Pts(k,\eta) &=\frac{2\mathcal{A}^2k^4}{\mpl^4}\l(\frac{k}{\ke}\r)^{2\alpha_1}\l( \int_{\xe}^{\xre}dx_1 \frac{\Gkre(\xre,x_1)}{a^2(x_1)}\mI(x_1)\r)^2\times \int_{\kmin}^{\ke}\frac{dq}{k}\int_{-1}^1d\gamma (1-\gamma^2)^2 (q/k)^{5+\alpha_1}|1-\vq/\vk|^{\alpha_1-1}\label{eq:def_ptre}
%  \end{align}
  \subsection{Evolution of Primordial Tensor Power spectrum during Reheating:}
This subsection provides a concise overview of the primary tensor power spectrum and its evolution resulting from quantum fluctuations during inflation.
%Inflation, a crucial mechanism addressing cosmological challenges such as flatness and horizon problems, offers a well-established framework for generating tensor fluctuations from the quantum vacuum. The tensor power spectrum, characterizing the distribution of gravitational waves across cosmic scales, provides valuable insights into the universe's inflationary phase and subsequent evolution.
Within the context of a simple slow-roll inflationary background, the primary tensor power spectrum resulting from quantum production can be approximated as \cite{Guzzetti:2016mkm, Maiti:2024nhv, Haque:2021dha}:
%%%%%%%%%%%%%%%%%%%%%%%%%%
\begin{equation}
\mathcal{P}_{\rm T}^{\rm pri}(k,\eta_{\rm end}) \approx \frac{2}{\pi^2} \left(\frac{\HI}{\MP}\right)^2 \left(1 + \frac{k^2}{\ke^2}\right)
\end{equation}
%%%%%%%%%%%%%%%%%%%%
Here, $\ke$ represents the highest momentum leaving the horizon at the end of inflation.

Following inflation, the reheating phase converts inflaton energy into radiation, leading to a radiation-dominated universe characterized by the equation of state ($\wre$) and reheating temperature ($\Tre$). The Hubble parameter evolves as $H^2=\HI^2(a/\ae)^{-3(1+\wre)}$, influencing the scale factor's evolution $a(\eta)\approx\ae(\eta/\ee)^{2/(1+3\wre)}$. This results in the evolution of tensor fluctuations $h_{\vk}$ during reheating without any source term:
%%%%%%%%%%%%%%%%%%%%%%%%%%
\begin{equation}
h_{\vk}^{\prime\prime}(x) + \frac{4}{1+3\wre}\frac{1}{x}h_{\vk}^{\prime}(x) + h_{\vk}(x) = 0 .
\end{equation}
%%%%%%%%%%%%%%%%%%%%%%%%%
Here, we introduce the dimensionless variable $x=k\eta$. The well-known solution to this equation is given by:
%%%%%%%%%%%%%%%%%%%%%%
\begin{equation}
h_{\vk}(x) = \mathcal{C}_1x^{l(\wre)}J_{l(\wre)}(x) + \mathcal{C}_2x^{l(\wre)}J_{-l(\wre)}(x)\label{eq:hkre} ,
\end{equation}
%%%%%%%%%%%%%%%%%%%%%%%%
where $J_l(x)$ is the Bessel function of order $l(\wre)=3(\wre-1)/2(1+3\wre)$ and the two integration constants $\mathcal{C}_1$ and $\mathcal{C}_2$ contain critical information regarding the origin of tensor fluctuations during inflation. Determining these constants involves satisfying continuity conditions for both tensor fluctuations and their first derivatives at $\eta = \ee$. Focusing on modes beyond the Hubble radius at the end of inflation, we can calculate $\mathcal{C}_1$ and $\mathcal{C}_2$ in the super-horizon limit ($\xe = k\ee \ll 1$). The expressions are as follows:
%%%%%%%%%%%%%%%%%%%%%%%%%%%%
\begin{eqnarray}
&&\mathcal{C}_1 = \frac{\pi}{2\sin(l\pi)}\left(\frac{k}{\ke}\right)^{2(1-l)}\left(\frac{1}{\Gamma(2-l)}+\frac{1}{\Gamma(1-l)}\right)h_{\vk}(k,\ee), \\
&&\mathcal{C}_2 = \frac{\pi}{2\sin(l\pi)}\left(\frac{1}{\Gamma(l)}-\frac{1}{\Gamma(1+l)}\left(\frac{k}{\ke}\right)^2\right)h_{\vk}(k,\ee)\label{eq:C2}
\end{eqnarray}
%%%%%%%%%%%%%%%%%%%%%%%%%
where $h_{\vk}(k,\ee)$ is the amplitude of tensor fluctuation at the inflation end \cite{Maiti:2024nhv, Haque:2021dha}.
In the general scenario, the parameter $\wre$ lies within $0 \leq \wre \leq 1$, causing $l(\wre)$ to take negative values consistently. Given our interest in scales beyond the horizon at the end of inflation (i.e., $k < \ke$, implying $k/\ke < 1$), we assert that $\mathcal{C}_2$ greatly dominates over $\mathcal{C}_1$.

Finally, utilizing Eqs.(\ref{eq:hkre}) and (\ref{eq:C2}) in Eq.(\ref{eq:pt}), we obtain the primary GW spectrum at the end of reheating as \cite{Haque:2021dha}:
%%%%%%%%%%%%%%%%%%%%%%%%
\begin{equation}
\mathcal{P}_{\rm T}^{\rm pri}(k,\eta_{\rm re}) = \frac{\pi^2}{4\sin^2(l\pi)\Gamma^2(l)}\left(1-\frac{1}{l}\left(\frac{k}{\ke}\right)^2\right)^2\mathcal{P}_{\rm T}^{\rm pri}(k,\eta_{\rm end}) .\label{eq:ptrev}
\end{equation}
%%%%%%%%%%%%%%%%%%%%%%%%%%%%%%%

  \subsection{Productions of Secondary Tensor Power spectrum during Reheating:}
As previously discussed in Section \ref{sec2}, due to non-minimal coupling %particularly in reheating scenarios with $\wre>1/3$, 
results in the growth of the scalar fluctuation due to instability.  
For a high value of $\xi> 1$, it is evident from the right panel of Fig.\ref{fieldevolutionfig} that for any EoS $0\leq \wre\leq1$, the effect of this instability is much stronger for large scales($k<\ke$), leading to growth of amplitude as depicted in Fig.\ref{fieldevolutionfig}. Such growth is prominent exclusively in reheating scenarios where $\wre>1/3$ in larger $\xi$ regimes. On the contrary, for $\wre<1/3$, the growth of the large-scale modes becomes significant in the lower $\xi$ regimes, particularly below the conformal limit $0\leq \xi<1/6$, which causes an enhanced energy density spectrum $\rho_{\chi_k}$ as shown in the left panel of Fig.\ref{powerspectrumfig}. 
%This very fact, in turn, ascertains that for $\wre<1/3$, significant production of secondary gravitational waves (SGWs) owing to the substantial generation of scalar field anisotropic stress will occur in the lower coupling regime, $\xi<1/6$ and for stiff post-inflationary EoS $\wre>1/3$, significant production of SGWS will occur in higher coupling regime, $\xi\gtrsim 1$. 
%In this subsection, we first give the expression of the secondary tensor power spectrum $\Pts(k,\eta)$ for $\wre>1/3$, then in a similar fashion, we give the respective expression of the tensor power spectrum for $\wre<1/3$. \\

%The resulting production of massless scalar fields is sufficient to induce notable anisotropy in the system, which in turn generates a significant amount of secondary gravitational waves (SGWs) during the reheating era.

The secondary tensor power spectrum $\Pts(k,\eta)$ corresponding to this period is detailed in Eq. (\ref{eq:ptra_chi}), where $\Gkre(\eta,\eta_1)$ represents the Green's function associated with Eq. (\ref{eq:hij_2}), satisfying the following differential equation \cite{Maiti:2024nhv}
%%%%%%%%%%%%%%%%%%%%%%%%%%%
\begin{align}
\Gk''(\eta,\eta_1) + 2\cH \Gk'(\eta,\eta_1) + k^2 \Gk(\eta,\eta_1) = \delta(\eta-\eta_1).
\end{align}
%%%%%%%%%%%%%%%%%%%%%%%%%%%%%%%%%
The Green's function during the reheating epoch is expressed as \cite{Maiti:2024nhv}
%%%%%%%%%%%%%%%%%%%%%%%%%%%%%
\begin{align}
\Gkre(\eta,\eta_1) &= \Theta(\eta-\eta_1)\frac{\pi \eta^{l} \eta_1^{1-l}}{2\mathrm{sin}(l\pi)}\left[J_l(k\eta)J_{-l}(k\eta_1)-J_{-l}(k\eta)J_l(k\eta_1) .\right]\label{eq:gre}
\end{align}
%%%%%%%%%%%%%%%%%%%%%%%%%%%%%%
Utilizing Eq. (\ref{Xreh9}) and (\ref{eq:gre}) in Eq. (\ref{eq:ptra_chi}), we have obtained the general expressio of the secondary tensor power spectrum due to the massless scalar field $\chi$ at the end of the reheating era $\eta=\ere$ as follows,
%%%%%%%%%%%%%%%%%%%%%%%%%%%
\begin{align}
\Pts(k,\ere) &= \frac{2\mathcal{A}^2\HI^4}{\pi^4\mpl^4}\left( \frac{k}{\ke} \right)^{4+2\delta-4\nu_2} \left( \int_{\xe}^{\xre}dx_1 x_1^{-\delta} \Gkre(\xre,x_1) \mI(x_1) \right)^2 \mathcal{F}(k) \label{eq:pt1},
\end{align}
%%%%%%%%%%%%%%%%%%%%%%%
where $\delta(\wre)=4/(1+3\wre)$. This equation introduces one dimensionless variable: $x=k\eta$ or $x_1=k\eta_1$. 
The time integral limits range from $\xe=k\ee$ to $\xre=k\ere$. The momentum integral part $\mathcal{F}(k)$ is detailed in the Appendix \ref{appenA}. 

Upon evaluating all the integrals, the resulting secondary tensor power spectrum at the end of reheating for $\wre>1/3$ and $\xi>3/16$ turns out to be
%%%%%%%%%%%%%%%%%%%%%%%%%%
\begin{align}
\Pts(k, \ere) &= \frac{2\mathcal{A}^2\HI^4}{\pi^4\mpl^4}\left( \frac{k}{\ke} \right)^{4+2\delta-4\nu_2} \mI_t^2(\xre,\xe) \mathcal{F}(k) .\label{eq:ptres}
\end{align}
%%%%%%%%%%%%%%%%%%%%%%%%%%%
Here, $\mathcal{A}$ is defined as
%%%%%%%%%%%%%%%%%%%%%%%%%%%%
\begin{align}
\mathcal{A}\approx\left(\frac{\Gamma(\nu_2)\text{exp}(-\pi \tilde{\nu}_1/2)}{4}\left(\frac{2}{3\mu-1}\right)^{\nu_2}\sqrt{3\mu-1}\left|\frac{\left(\pi+i\text{cosh}(\pi\tilde{\nu_1})\Gamma(1-i\tilde{\nu_1})\Gamma(i\tilde{\nu_1})\right)}{\pi\Gamma(i\tilde{\nu_1})}\right|\right)^2 ,\label{eq:A}
\end{align}
%%%%%%%%%%%%%%%%%%%%%%%%%
whereas the time integral part $\mI_t$ is defined as (see the details in Appendix \ref{appenA})
\begin{align}
      \mI_t(\xre,\xe)=\int_{\xe}^{\xre} dx_1 x_1^{-\delta}\mI(x_1)\Gkre(\xre,x_1) ,\label{eq:time_int}
\end{align}
%\begin{align}
%    \mI_t(\xre,\xe)=\int_{\xe}^{\xre} dx_1 x_1^{-\delta}\mI(x_1)\Gkre(\xre,x_1)=\frac{\pi}{2\sin(l\pi)} \frac{x^lx_1^{2-2l-\delta}}{2^{1+l}}\l( -x_1^{2l}J_{-l}(x)\Gamma\l[1-\frac{\delta}{2} \r]\r.\nonumber\\
%    \l.\text{HypergeometricPFQRegularized}\l[\l\{1-\frac{\delta}{2}\r\},\l\{1+l,2-\frac{\delta}{2}\r\},-\frac{x_1^2}{4} \r]\r.\nonumber\\
 %  \l. \l.+4^l J_l(x)\Gamma\l[1-l-\frac{\delta}{2} \r]\text{HypergeometricPFQRegularized}\l[ \l\{1-l-\frac{\delta}{2} \r\},\l\{1-l,2-l-\frac{\delta}{2}\r\},-\frac{x_1^2}{4} \r] \r)\r|_{\xe}^{\xre}\label{eq:mI_t}
%\end{align}
and
%%%%%%%%%%%%%%%%%%%%%%%%%%%%%%
\begin{align}
      \mF(k)\simeq \frac{16}{15} \l\{ \frac{1}{6-2\nu_2}\l( 1-\l(\frac{\kmin}{k}\r)^{6-2\nu_2}\r) +\frac{1}{5-4\nu_2}\l( \l(\frac{\ke}{k}\r)^{5-4\nu_2}-1\r)\r\} .
\label{eq:fk}
\end{align}
%%%%%%%%%%%%%%%%%%%%%%%%%%%%

Proceeding in the same way, utilizing the Eq.(\ref{powerspec1}) we obtain the secondary tensor power spectrum for $\wre<1/3$ and $\xi<3/16$ as
%%%%%%%%%%%%%%%%%%%%%%%%%%%%
\begin{align}
\Pts(k, \ere) &= \frac{2\mathcal{B}^2\HI^4}{\pi^4\mpl^4}\left( \frac{k}{\ke} \right)^{4+2\delta-4(\nu_1+\nu_2)} \mI_t^2(\xre,\xe) \mathcal{E}(k) .\label{eq:ptres1}
\end{align}
%%%%%%%%%%%%%%%%%%%%%%%%%%%
where $\mathcal{B}$ is defined as
%%%%%%%%%%%%%%%%%%%%%%%%%%%%
\begin{align}
\mathcal{B}\approx\left(\frac{\Gamma(\nu_1)\Gamma(\nu_2)2^{\nu_1}}{8\pi}\left(\frac{2}{3\mu-1}\right)^{\nu_2}\left(\frac{3\mu(1-2\nu_1)+2(\nu_1-\nu_2)}{\sqrt{(3\mu-1)}}\right)\right)^2 .\label{eq:B}
\end{align}
%%%%%%%%%%%%%%%%%%%%%%%%%
Here the expression of the time-integral 
$\mI_t(\xre,\xe)$ will remain unchanged and momentum-integral will be modified as
%%%%%%%%%%%%%%%%%%%%%%%%%%%%%%%
\begin{align}
      \mE(k)\simeq \frac{16}{15} \l\{ \frac{1}{6-2(\nu_1+\nu_2)}\l( 1-\l(\frac{\kmin}{k}\r)^{6-2(\nu_1+\nu_2)}\r) +\frac{1}{5-4(\nu_1+\nu_2)}\l( \l(\frac{\ke}{k}\r)^{5-4(\nu_1+\nu_2)}-1\r)\r\} .
\label{eq:ek}
\end{align}
%%%%%%%%%%%%%%%%%%%%%%%%%%%%

Now we shall define the dimensionless energy density of the gravitational waves $\ogw(k)h^2$ for $\textit{today}$ in the following subsection. 

\subsection{GW spectrum for today :}
During the reheating and radiation-dominated epochs, perturbation modes progressively re-enter the Hubble radius, producing a stochastic gravitational wave (GW) signal. When produced in the early universe, this GW signal is assumed to possess statistical homogeneity, isotropy, and Gaussianity, inheriting properties from the FLRW universe, whether during inflation or the thermal era.

Due to the weak interaction of gravity with matter, GWs are decoupled from the rest of the universe at the Planck scale. Neglecting interactions with ordinary matter and self-interactions, we assume that sub-Hubble GWs propagate freely in space after their production or re-entry into the Hubble radius. The GW energy density decays with the expansion of the universe, mimicking the behavior of radiation, i.e., $\rhogw \propto a^{-4}$. Meanwhile, the physical wave number of GWs evolves as $k/a$. Deep inside the radiation-dominated universe, given the spectrum, the normalized GW energy density parameter at any time $\eta$ is defined as
%%%%%%%%%%%%%%%%%%%%%%%%%%%
\begin{align}\label{eq:ogwdef}
\ogw(k,\eta) = \frac{\rhogw(k,\eta)}{\rho_c(\eta)} = \frac{k^2 \Ptra(k,\eta)}{12a^2(\eta) H^2(\eta)} .
\end{align}
%%%%%%%%%%%%%%%%%%%%%%%%%%
Here, the critical energy density $\rho_c(\eta) = 3H^2(\eta)\MP^2$, and $\MP \simeq 2.43 \times 10^{18}~\text{GeV}$ represents the reduced \textit{Planck} mass.
The tensor power spectrum $\Ptra(k,\eta)$ during radiation-dominated era can further be expressed in terms of the spectrum at the end of reheating as \cite{Maiti:2024nhv}.%%%%%%%%%%%%%%%%%%%%%%%%%%%
\begin{align}
  \Ptra(k,\eta)=  \left(1 + \frac{k^2}{\kre^2}\right) \frac{\mathcal{P}_{\rm T}(k,\ere)}{2k^2 \eta^2} .
\end{align}
%%%%%%%%%%%%%%%%%%%%%%%%%%%%
%The spectrum of GW energy density per logarithmic frequency interval can be normalized by the critical energy density $\rho_c(\eta)$ as follows:
%\begin{align}\label{eq:ogwdef}
%\ogw(k,\eta) = \frac{\rhogw(k,\eta)}{\rho_c(\eta)} = \frac{1}{12} \frac{k^2 \Ptra(k,\eta)}{a^2(\eta) H^2(\eta)}
%\end{align}
%Here, $\rho_c(\eta) = 3H^2(\eta)\MP^2$, and $\MP \simeq 2.43 \times 10^{18}~\text{GeV}$ represents the reduced Planck mass.
%%%%%%%%%%%%%%%%%%%%%%%%%%
\begin{figure*}
\includegraphics[width=0.48\linewidth]{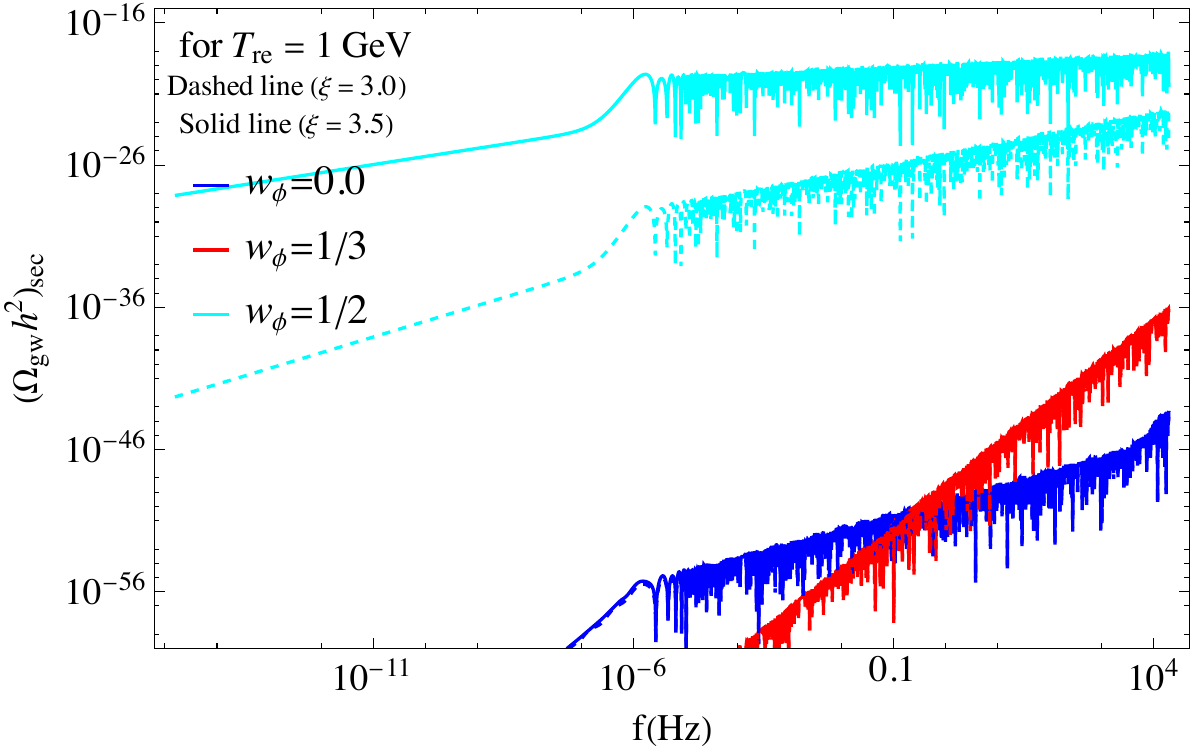}
\includegraphics[width=0.48\linewidth]{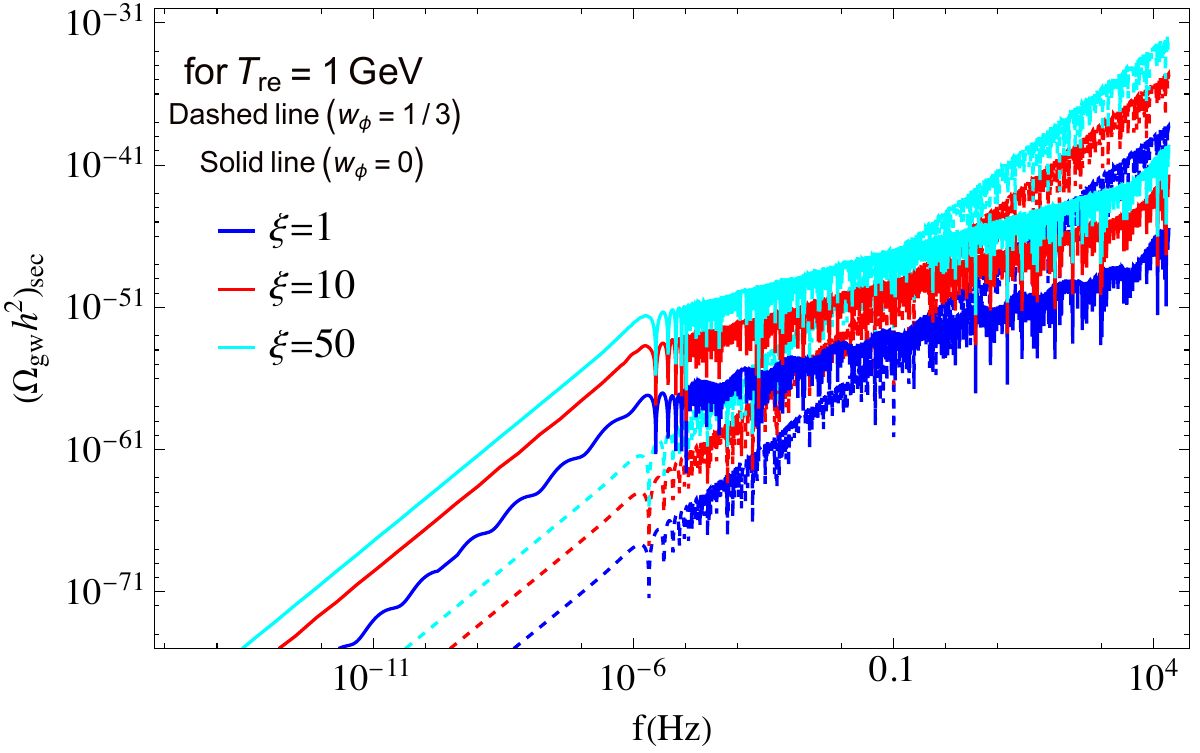}
\caption{\textit{In this figure, we have shown the present-day dimensionless gravitational waves energy density $(\ogw h^2)_{\rm sec}$ (induced by the scalar field only) as a function of frequency $f$ (Hz). The left panel illustrates the dependency of the EoS $\wre = 0, 1/3,~\&~1/2$ for two specific values of the coupling constant $\xi = 3.5$ (indicated by solid lines) and $\xi = 3.0$ (indicated by dashed lines). In the right panel, we have shown how the spectral energy density evolves for the coupling constants $\xi = 1, 10,~\&~50$, with fixed EoS values $\wre = 0$ (indicated by solid lines) and $\wre = 1/3$ (indicated by dashed lines). In both panels, we have considered the reheating temperature of our universe to be $\Tre = 1$ GeV. }}\label{fig:gws1}
\end{figure*}
%%%%%%%%%%%%%%%%%%%%%%%%%%%%
As is well-established, the energy density of gravitational waves (GWs) exhibits a behavior akin to radiation, scaling as $\rhogw \propto a^{-4}$. Our focus lies on modes that are well within the Hubble radius at a later time, particularly in proximity to the radiation-matter equality epoch during radiation domination. We express the dimensionless energy density parameter $\ogw(k)$ today in terms of $\ogw(k,\eta)$ as follows, as described in \cite{Haque:2021dha,Maiti:2024nhv}:
%%%%%%%%%%%%%%%%%%%%%%%%%%
\begin{align}
\ogw(k)h^2 &= \left(\frac{g_{r,eq}}{g_{r,0}}\right)\left(\frac{g_{s,0}}{g_{s,eq}}\right)^{4/3} \Omega_R h^2 \ogw(k,\eta) 
\approx \left(\frac{g_{r,0}}{g_{r,eq}}\right)^{1/3} \Omega_R h^2 \ogw(k,\eta) .\label{eq:ogw_today}
\end{align}
%%%%%%%%%%%%%%%%%%%%%%%%%%%%%
Here, $\Omega_R h^2 = 4.3 \times 10^{-5}$, representing the dimensionless energy density of radiation at the present epoch. $g_{r,eq}$ and $g_{r,0}$ represent the number of relativistic degrees of freedom at radiation-matter equality and in the present era, respectively, whereas $g_{s,eq}$ and $g_{s,0}$ signify the number of such degrees of freedom contributing to entropy at these respective epochs. In our calculation we adopt the value %$g_{r,eq} \simeq g_{s,eq} =  106.7$
$g_{r,0} \simeq g_{s,0}=3.35$.\\

\paragraph{\underline{\textit{Primary GWs spectrum today(PGWs):}}} After the reheating epoch, the subsequent era is predominantly characterized by radiation. Now utilizing Eqs (\ref{eq:ptrev}) and (\ref{eq:ogwdef}) in Eq.(\ref{eq:ogw_today}) we have found that the PGWs \textit{today} can be estimated as \cite{Maiti:2024nhv}
%%%%%%%%%%%%%%%%%%%%%%%
\begin{align}
    \ogwp(k)h^2\simeq 1.12\cdot 10^{-17} \l(\frac{\Omega_Rh^2}{4.3\times 10^{-5}}\r)\l(\frac{\HI}{10^{-5}\MP}\r)^2\times\l\{\begin{matrix}
        \mathcal{D}_1 & &\mbox{ for}~~ k<\kre\\
       \mathcal{D}_2 (k/\kre)^{\nw} & &\mbox{ for}~~ \kre<k<\ke
    \end{matrix}\r. .\label{eq:gws_pri}
\end{align}
%%%%%%%%%%%%%%%%%%%%%%%%%%
Here, we introduce the parameter $\nw = 2(3\wre-1)/(1+3\wre)$, and define $\mathcal{D}_1=\pi^22^{2 l}/4\sin^2(l\pi)\Gamma^2(l)\Gamma^2(1-l)\simeq \mathcal{O}(1)$ and $\mathcal{D}_2\simeq \l(\pi^2/2\sin^2(l\pi)\Gamma^2(l)\r)\cos^2\l(-k/\kre+(1-2 l)\pi/4\r)\simeq\mathcal{O}(1)$.\\

\paragraph{\underline{\textit{Secondary GWs spectrum today(SGWs):}}}

Similarly, using Eqs. (\ref{eq:ptres}),~(\ref{eq:ptres1}), and (\ref{eq:ogwdef}) in Eq.(\ref{eq:ogw_today}) we have found that the secondary GWs spectrum for the modes $k_{\ast}\leq k\leq\ke$ can be estimated as
%%%%%%%%%%%%%%%%%%%%%%%%%%%%
\begin{align}
\ogws(k)h^2\simeq~ &2.22\times 10^{-26}\l(\frac{\Omega_Rh^2}{4.3\times 10^{-5}}\r)\l(\frac{\HI}{10^{-5}\MP}\r)^4\l(1+\frac{k^2}{\kre^2}\r)\mathcal{I}_t^2(k,\ere)\nonumber\\
&\times
\begin{cases}
{\mathcal{B}^2}\l(\frac{k}{\ke}\r)^{4+2\delta-4(\nu_1+\nu_2)} \mE(k)& \text{for}~~ \wre<1/3,~\xi<3/16\\
{\mathcal{A}^2}\l(\frac{k}{\ke}\r)^{4+2\delta-4\nu_2} \mF(k)& \text{for} ~~ \wre>1/3,~\xi>3/16
\end{cases}
\label{eq:gws}
\end{align}
%%%%%%%%%%%%%%%%%%%%%%%%%%

In this context, $\mathcal{I}_t(k,\ere)$ represents the time integral as defined in Eq. (\ref{eq:time_int}), while $\mathcal{A}$ and $\mathcal{B}$ denote the constant coefficients consisting of the EoS parameter and coupling strength ($\wre, ~ \xi)$, as outlined in Eq.(\ref{eq:A}), and (\ref{eq:B}). 

%\textcolor{blue}{Now in the proceeding discussions we shall mainly concentrate on the behavior of SGWs for higher EoS $\wre>1/3$ in the large non-minimal coupling regime $\xi>3/16$. At the end of this discussion, we shall give a brief account of the behavior of the SGW spectrum for $\wre<1/3$ and $\xi<3/16$.} 

\begin{itemize}
\item \underline{For $\wre>1/3$,~ $\xi>3/16$ :}~~
This the region of parameter space, for which the scalar field growth is prominent for the range of scales which has wider detection prospects of a large pool of ongoing and future GW experiments.  We consider two distinct regimes to analyze the spectral behavior of secondary gravitational waves: $k \ll \kre$ and $\kre \ll k < \ke$. In the super-horizon limit, the SGW spectrum follows the relation $\ogws(k \ll \kre)h^2 \propto f^{4(2-\nu_2)}$. In contrast, in the sub-horizon limit, the spectrum exhibits a behavior of $\ogws(k \gg \kre)h^2 \propto f^{6+\delta-4\nu_2}$ (for further details, refer to the Appendix \ref{appenA}).

For an EoS with $\wre > 1/3$ and a coupling parameter $\xi > 1/6$, a post-inflationary parametric instability raises the overall production of the scalar field during the reheating era. 
On the contrary, for $\wre \leq 1/3$, no such post-inflationary instability occurs.
In Fig. (\ref{fig:gws1}), we showed how such growths are imprinted in SGWs depending upon $\wre$ and the coupling parameter $\xi$. In the left panel, we examine three EoS values: $\wre = 0$, $1/3$, and $1/2$, with a fixed reheating temperature of $\Tre = 1$ GeV. The secondary GW spectrum generically acquires a common factor $\ogws(k)h^2\propto (\ke/\kre)^{\frac{6\wre -2}{1+3\wre}}$ (see the detailed calculation in appendix \ref{appenA}). The factor immediately suggests that for $\wre <1/3$ $(\wre >1/3)$, the amplitude of the spectrum is suppressed (enhanced) as $\ke/\kre >> 1$. This suppression or enhancement can indeed be seen from the left panel of Fig. \ref{fig:gws1}. Detailed calculation further indicates that when $\wre \leq 1/3$, the secondary GW production is always overshadowed by the primary production for all scales as far as larger coupling values, $\xi>3/16$, are concerned. However, for $\wre = 1/2$ (i.e., $\wre > 1/3$) in the larger coupling regime, the production of SGWs due to the scalar field is enhanced in super-horizon modes, potentially surpassing the primary gravitational wave production.

In the top left panel of Fig. \ref{fig:gws}, we explore the dependency of the coupling parameter $\xi$ for two different EoS values: $\wre = 0$ and $1/3$. As neither of these EoS values induces instability in the system, no growth in the scalar field occurs during the reheating era (see Fig. \ref{fieldevolutionfig}). Consequently, the secondary production of gravitational waves is negligible, and the strength of the gravitational wave spectrum remains considerably lower than that of the primary gravitational waves, even when a very high coupling constant $\xi$ is considered.

%%%%%%%%%%%%%%%%%%%%%%%%%%%%%%
\begin{figure*}
\includegraphics[width=0.49\linewidth]{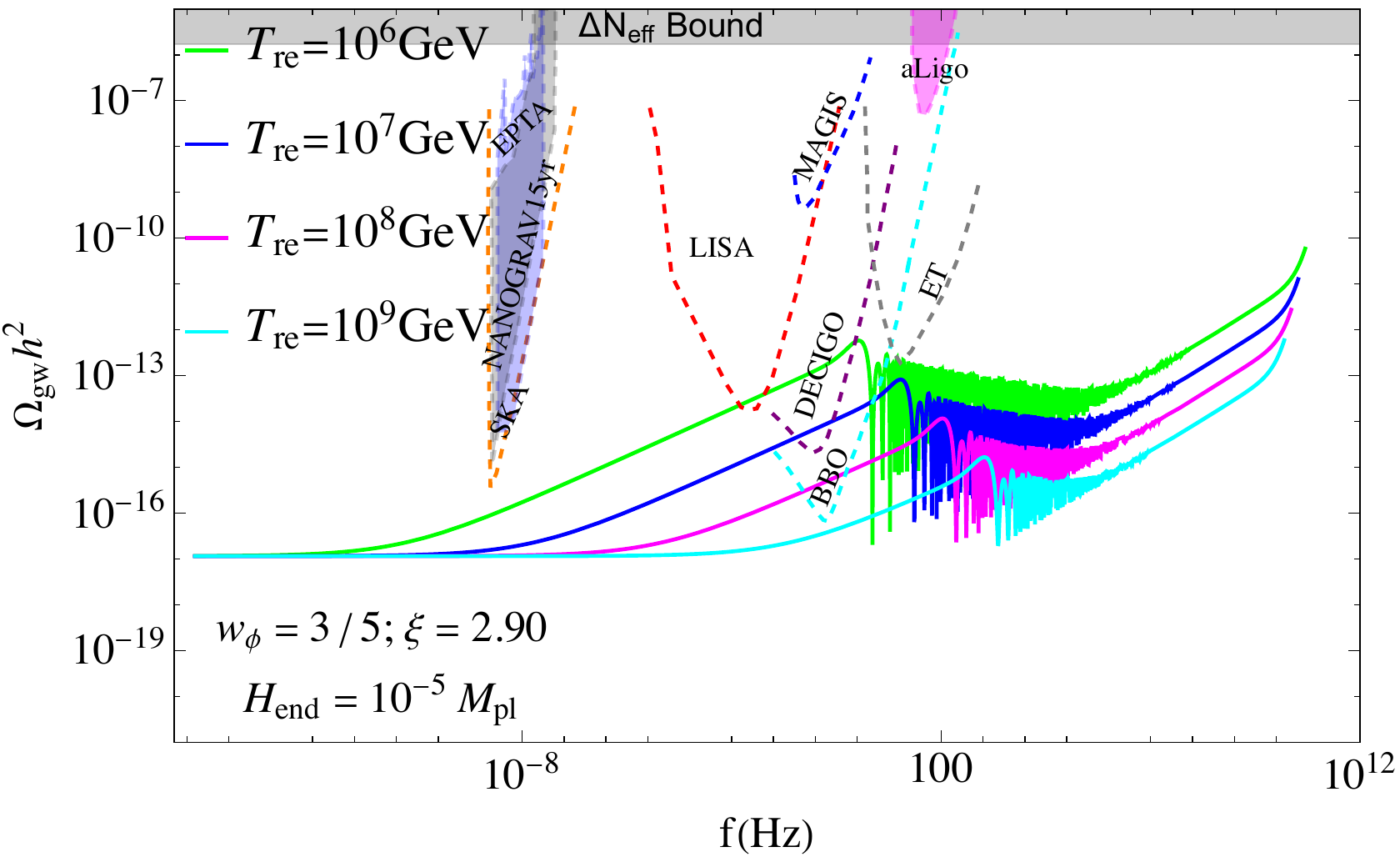}
\includegraphics[width=0.49\linewidth]{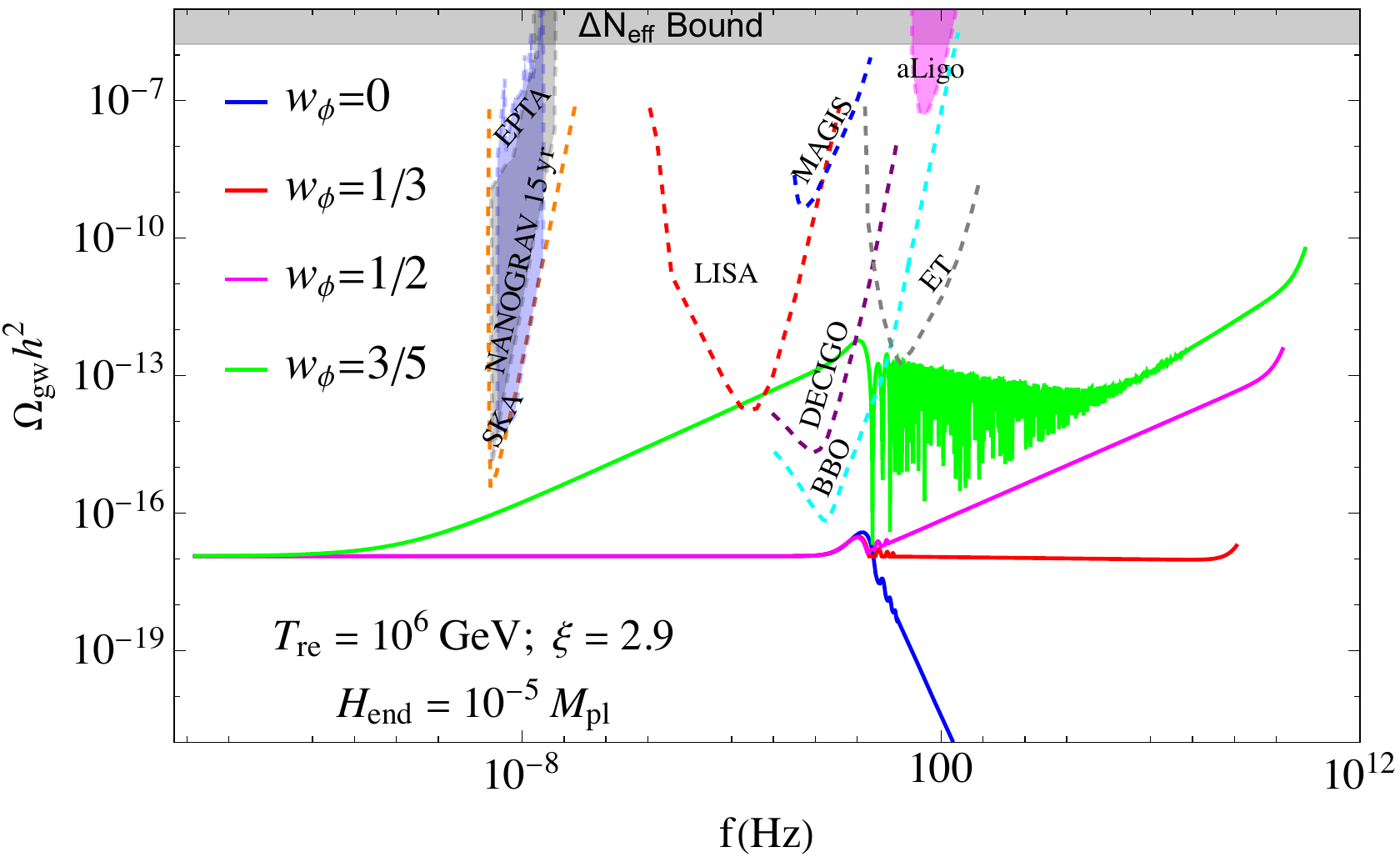}
\includegraphics[width=0.49\linewidth]{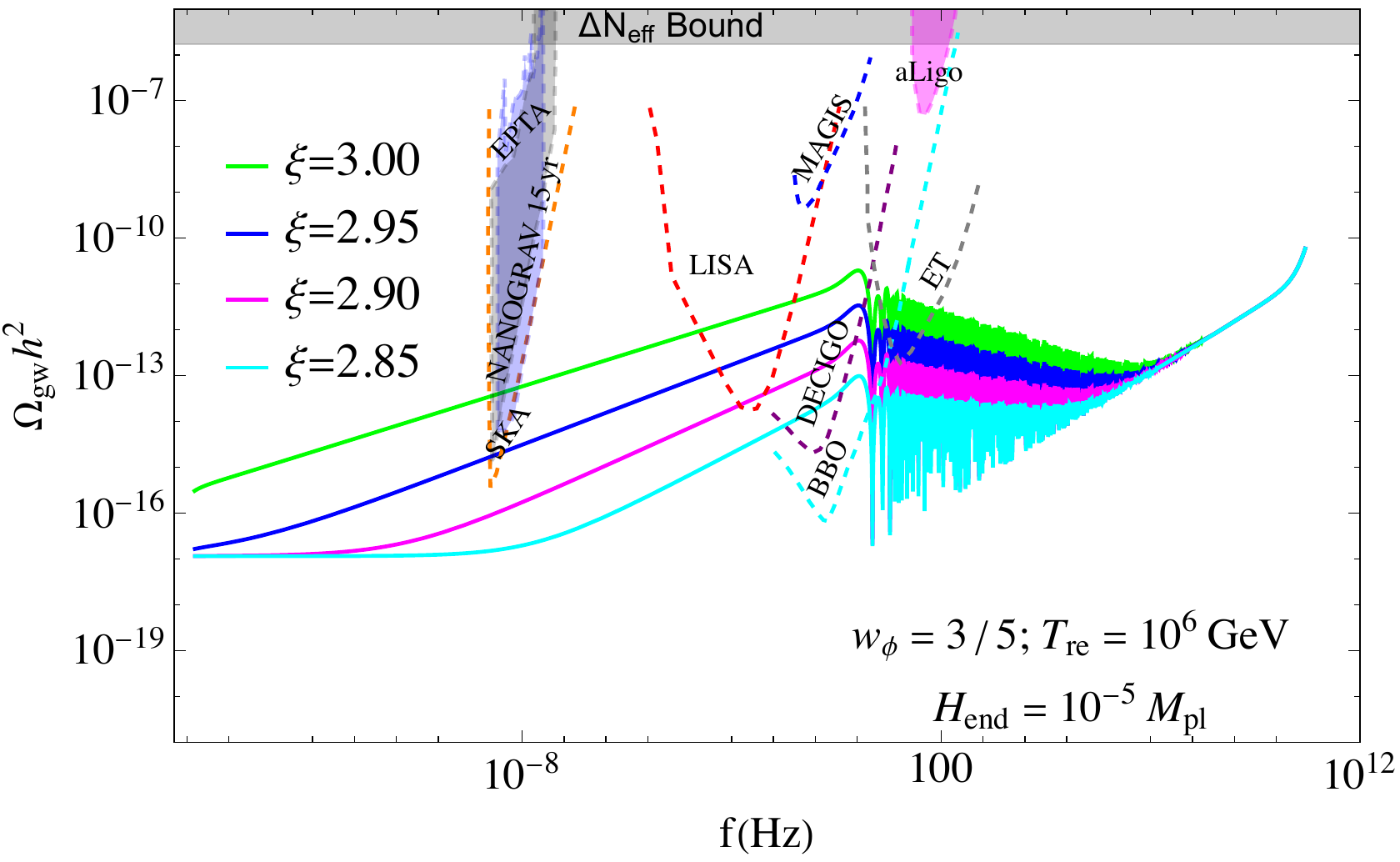}
\includegraphics[width=0.49\linewidth]{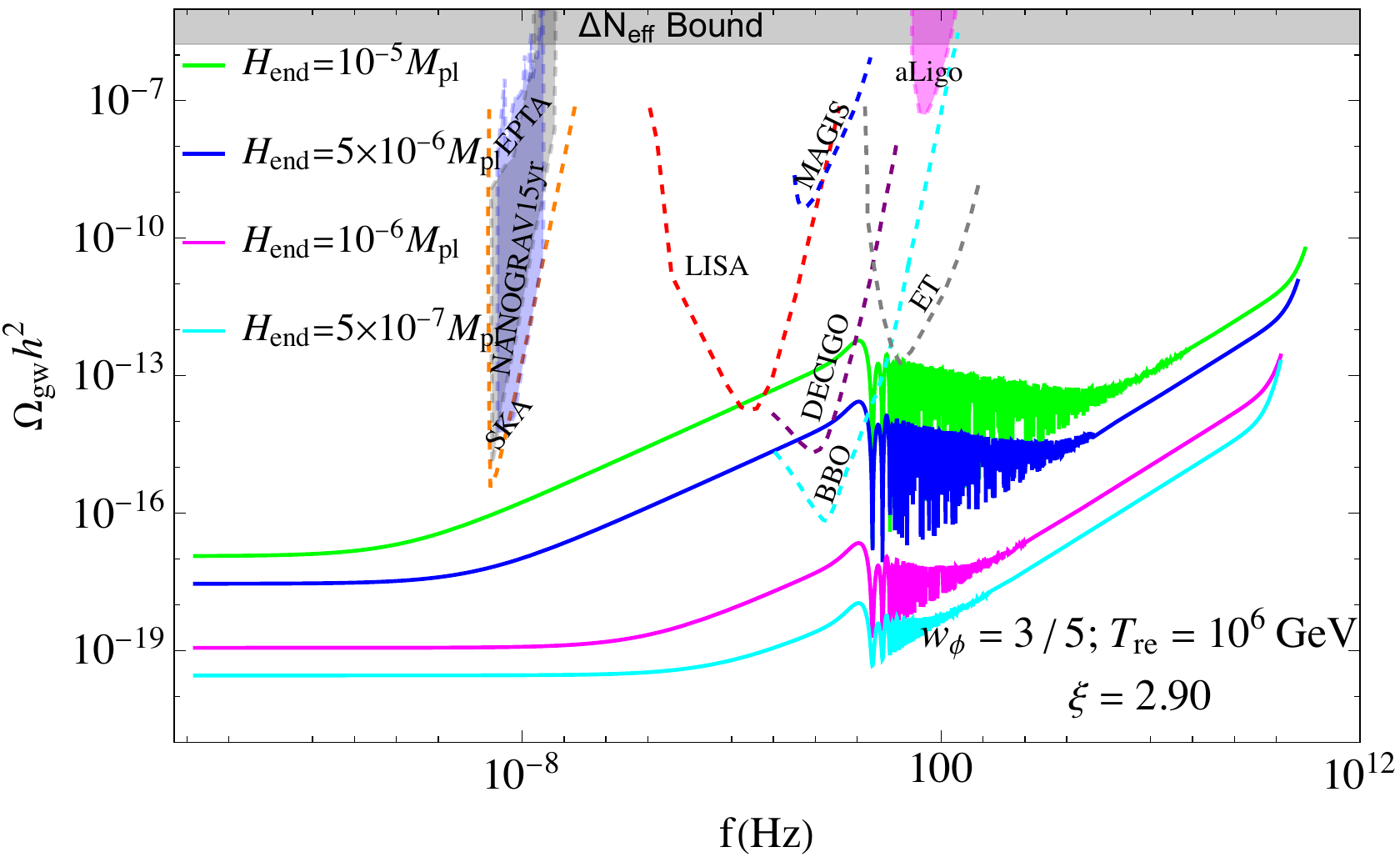}
\caption{\textit{
The figure presents the combined GW spectrum today for different theory parameters.
In the top-left panel, we illustrate the dependence of the GW spectrum on the reheating temperature $\Tre$ for a fixed EoS $\wre = 3/5$ with a coupling parameter $\xi = 3.0$. The top-right panel shows the GW spectra for four different EoS values: $\wre =0, 1/3$, $1/2$, and $3/5$, assuming a fixed reheating temperature of $\Tre = 10^6$ GeV and a coupling constant $\xi = 2.90$.
The bottom-left panel demonstrates the influence of the coupling constant $\xi$ on the GW spectrum, where we assume a fixed reheating scenario with $\Tre = 10^6$ GeV and $\wre = 3/5$. In all three of these plots, we have adopted the maximal allowed value of the tensor-to-scalar ratio, $r_{0.05} \simeq 0.036$, as constrained by the Planck observations.
Finally, in the bottom-right panel, we explore the impact of varying the tensor-to-scalar ratio on the GW spectrum for a fixed reheating scenario with $\Tre = 10^6$ GeV, $\wre = 3/5$, and $\xi =2.90$.}}
\label{fig:gws}
\end{figure*}
%%%%%%%%%%%%%%%%%%%%%%%%%%%%%%

\begin{figure*}
\includegraphics[width=0.49\linewidth]{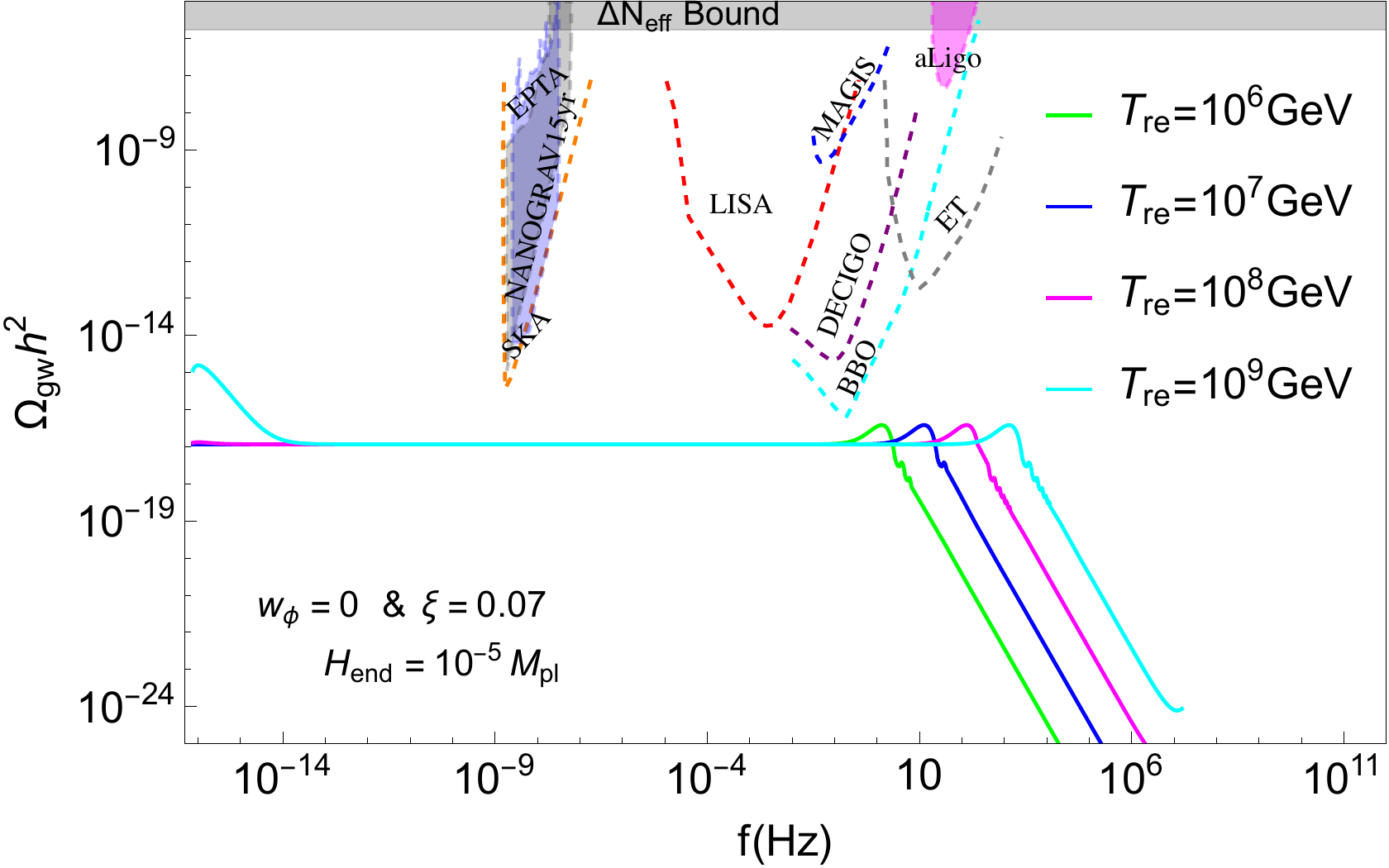}
\includegraphics[width=0.49\linewidth]{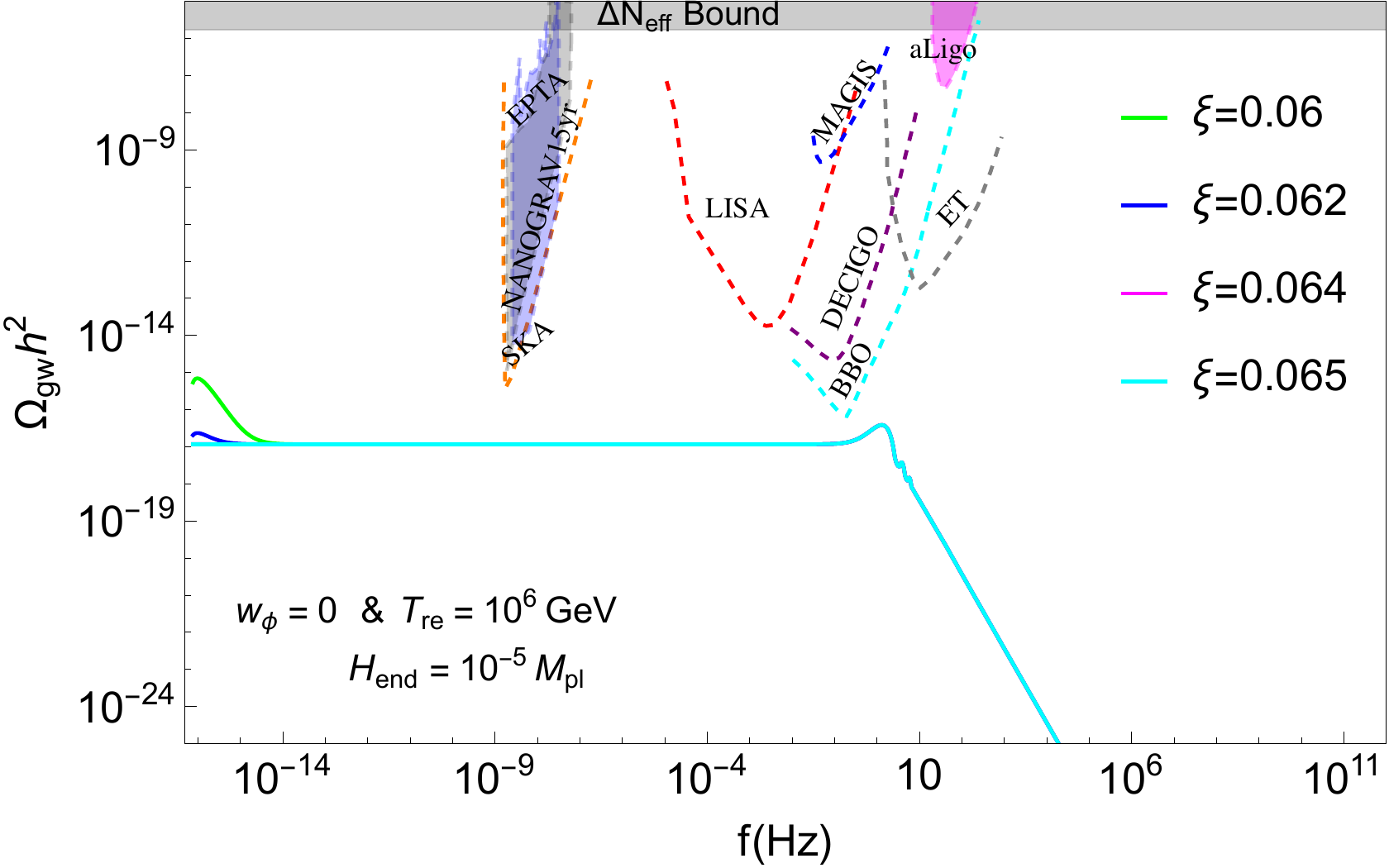}
\caption{\textit{
The figure presents the combined GW spectrum today for different theory parameters.
In the left panel of this figure, we show the variation of the GW spectrum for different reheating temperatures with a fixed coupling strength $\xi=0.07$ and reheating EoS $\wre=0$. In the right panel, we show the dependence of the spectrum on the coupling strength $\xi$ for a fixed reheating temperature $\Tre=10^6$ GeV and EoS $\wre=0$. In these two plots, we have taken the maximal value of the Hubble scale $\He=10^{-5}M_{\rm pl}$.}}
%Finally, in the right panel, we show the impact of varying tensor-to-scalar ratio on the GW spectrum for a fixed reheating scenario with $\Tre = 10^6$ GeV, $\wre = 0$, and $\xi =0.06$.  
\label{fig:gws_1}
\end{figure*}
%%%%%%%%%%%%%%%%%%%%%%%%%%%%

To this end, we would like to discuss the generic characteristics of the total GW spectrum (PGWs + SGWs) in terms of the reheating parameters, non-minimal coupling, and the inflationary energy scale. 
In the top left panel, we show the dependency of the reheating temperature on the GW spectrum, with a fixed equation of state $\wre = 3/5$ and a coupling constant $\xi = 2.90$. Evidently, in the intermediate frequency range, the secondary production of GWs can surpass the primary gravitational waves (PGWs). The decreasing temperature turns out to increase the overall amplitude of the spectrum through the factor $\ogw(k)h^2\propto T_{\rm re}^{\frac{8(1-3\wre)}{3(1+\wre)}}\propto T_{\rm re}^{~ -4/3} $ for $k<\kre$, $\wre=3/5$, and $\ogw(k)h^2 \propto T_{\rm re}^{\frac{4(1-3\wre)}{3(1+\wre)}}\propto T_{\rm re}^{~ -2/3}$ for $k>\kre$, $\wre=3/5$.
%with $\kre$ being directly proportional to $\Tre$ (see Appendix \ref{appenA} for detailed discussion). 
Beyond a certain reheating temperature, however, the large-scale spectrum can exceed the present bound on the tensor-to-scalar ratio $r_{0.05} \leq 0.036$ from the recent \textit{Planck}-2018 results \cite{Planck:2018vyg}, which we discuss in the subsequent section.
The spectrum can intersect with future GW sensitivity curves for specific parameter sets, such as LISA, DECIGO, and BBO. 

This enhancement of the SGWs with decreasing reheating temperature is effective only for $\wre > 1/3$ reheating scenarios due to post-inflationary instability to the scalar field mode. Further lowering the reheating temperature implies an increase in the duration of the reheating period. A longer reheating period implies that the scalar field experiences instability for a more extended period, resulting in greater growth.

Similarly, in the top right panel of Fig.\ref{fig:gws}, we show the evolution of the final GW spectrum for different EoS values, indicated by four different colors. We assume  $\xi = 2.90$ and $\Tre = 10^6$ GeV. For $\wre =0, 1/3, 1/2$ the PGW dominates the entire spectrum. Although for $\wre > 1/3$ and $\xi > 1$, there is instability (as seen in Fig.\ref{fieldevolutionfig}), this growth is insufficient to produce GWs of strengh which can surpass the PGWs generated during inflation. SGW turns out to dominate only when $\wre \geq 3/5$. For example assuming $\wre =3/5$  the spectrum transforms from scale-invariant to blue tilted in the small frequency(large scale) range $f^{8-4\nu_2}=f^{0.41}$, and from blue tilted to red-tiled $f^{6+\delta - 4\nu_2} = f^{-0.16}$ for the modes $k>\kre$(small scale) as indeed observed in green curve. Therefore, for a fixed reheating temperature and coupling constant $(\xi > 1)$, there exists a threshold EoS value above which the scalar field growth is sufficient to produce enough SGWs to overtake the PGWs.

In the bottom left panel of Fig. \ref{fig:gws}, we examine the effect of the coupling constant $\xi$, with a fixed equation of state $\wre = 3/5$ and reheating temperature $\Tre = 10^6$ GeV. The results show that increasing the coupling parameter enhances the scalar field's growth, thereby enhancing the GW spectrum.
The spectral behavior is the same as the previous one with blue tilt in the small frequency(large scale) range $f^{8-4\nu_2}$, and red-tilt $f^{6+\delta - 4\nu_2}$ for the modes $k>\kre$(small scale) as indeed observed in the green curve. For fixed reheating scenarios, a critical value of the coupling constant $\xi_{\rm max}$ exists above which the tensor fluctuations can be overproduced violating bound on  tensor-to-scalar ratio $r_{0.05} \leq 0.036$ set by the \textit{Planck}.

In the bottom right panel of Fig. \ref{fig:gws}, we demonstrate how the gravitational wave (GW) spectrum energy density varies with the inflationary energy scale $\He$ for a fixed reheating temperature of $\Tre=10^{6}$ GeV and a fixed EoS $\wre=3/5$, with the coupling parameter $\xi=2.90$. 
%The inflationary energy scale $\HI$ is directly related to the tensor-to-scalar ratio $r_{0.05}$ through the relation $\HI = \pi \mpl \sqrt{r_{0.05} A_s / 2}$, and
Since the GW spectral energy density follows, $\ogwh \propto \HI^4$, a decrease in $\He$ naturally leads to reduction of the total GWs strength. 

For a fixed reheating temperature, lowering the inflationary energy scale by reducing $\He$ leads to a shorter reheating period (see Eq.(\ref{eq:nre})). The duration of this period is intimately tied to the growth of the scalar field. When the inflationary energy scale $\He$ is reduced, it not only limits the growth of the scalar field during reheating but also affects its initial production during inflation since the scalar field’s amplitude depends on the inflationary energy scale. As a result, the overall amplitude of the gravitational wave (GW) spectrum decreases as $\He$ is lowered.

Considering these factors, we find that with $\He=10^{-5}M_{\rm pl}$,~$\wre = 3/5$, $\Tre = 10^6$ GeV, and $\xi = 2.9$, the tensor fluctuations are strong enough to be detected by future sensitivity curves like LISA, DECIGO, and BBO. However, with a lower value like $\He = 5\times 10^{-6}M_{\rm pl}$, the amplitude is suppressed, allowing detection by DECIGO and BBO but not by LISA. For $\He\leq 10^{-6}M_{\rm pl}$, the signal is too weak to be detected by upcoming experiments. Therefore, the inflationary energy scale significantly impacts the production of secondary GWs via the scalar field dynamics.

In all figures, at very high-frequency ranges near $f_{\rm end} = (\ke/2\pi)$, the spectrum follows the $\ogwh (f) \propto f^{\nw}$ \cite{Maiti:2024nhv, Haque:2021dha} behavior, which is due to PGWs contributions at these frequencies. At very high frequencies, PGWs dominate over SGWs. As seen in Fig.(\ref{powerspectrumgradfig}), when the modes becomes sub-horizon, the scalar field growth subsides because of adiabatic evolution leading to insufficient to produce significant anisotropy.\\

\item \underline{For $0\leq\wre<1/3$,~ $\xi<3/16$:}~~
In this region of parameter space the infrared scalar field growth turned out to be relevant around the CMB scales, \textit{Planck} is the only experiment that can place constraints. Likewise the previous case for $\wre>1/3$, we consider here also two distinct regimes of the SGW spectrum, $k \ll \kre$ and $\kre \ll k < \ke$. For the super-horizon modes, the SGW spectrum follows the relation $\ogws(k \ll \kre)h^2 \propto f^{4(2-\nu_1-\nu_2)}$. In contrast for sub-horizon modes, the spectrum exhibits a behavior of $\ogws(k \gg \kre)h^2 \propto f^{6+\delta-4(\nu_1+\nu_2)}$ (for detailed calculation see Appendix \ref{appenA}). Unlike $\wre>1/3$, for this case the GW spectrum exhibits a sharp enhancement $\big((2-\nu_1-\nu_2)<0\big)$ in the long- wavelength modes $k<<\kre$ particularly for $\xi<1/6$. We have presented the dependence of GW spectrum (PGW+SGW) on reheating temperature $\Tre$, and the coupling strength $\xi$ for EoS $\wre=0$ in the above Fig. \ref{fig:gws_1}. As stated earlier for lower EoS for the modes around CMB pivot scale $k_{\ast}$, the GW spectrum receives significant correction due to tachyonic instability. This makes the spectrum IR divergent with enhanced amplitude, and its value freezes at the end of inflation.  
Therefore, as the duration of reheating decreases or reheating temperature increases the GW amplitude enhances as $\ogw(k)h^2\propto (\ke/\kre)^{4-2\delta}\propto \Tre^{\frac{8(1-3\wre)}{3(1+\wre)}}\propto \Tre^{8/3}$ due to less dilution. For example if  $\wre=0$, $\ogw(k)h^2\propto \Tre^{8/3}$ for $\wre=0$ (see left panel of Fig. \ref{fig:gws_1}). 
%The above spectral behavior is calcualted for $\xi=0.07$ with $\wre=0$.
%According to the spectrum (\ref{gw11A}) in the low-frequency regime $k<<\kre$, $\ogw(k)h^2\propto (\ke/\kre)^{4-2\delta}\propto \Tre^{\frac{8(1-3\wre)}{3(1+\wre)}}\propto \Tre^{8/3}$ for $\wre=0$. 
The right panel depicts the GW spectral behavior for different non-minimal coupling strength $\xi$ with $\wre=0,~\Tre=10^6$ GeV. In the regime $k<<\kre$, it grows as $\ogw(k)h^2\propto (k/\ke)^{8-4(\nu_1+\nu_2)}$.
The spectral index $\big(4(\nu_1+\nu_2)-8\big)$ gradually increases with the decrease of $\xi$. This is a distinctive feature of the gravitational wave spectrum for $\wre<1/3$, which essentially puts a lower bound on $\xi$ set by CMB observation. 
% According to the Eq.(\ref{powerspec1}), the scalar field power spectrum is found to be red-tilted, $(1-\nu_1-\nu_2)<0$, for $0\leq\xi<1/6$. This behavior is also depicted in the SGW spectrum 

\end{itemize}

%From this figure, we conclude that in reheating dynamics with $\wre > 1/3$, the scalar field experiences tachyonic instability, affecting modes still outside the horizon at the end of reheating. As a result, even modes far outside the horizon show substantial growth, further producing tensor fluctuations at the CMB scale. There is a relationship between the reheating dynamics and the coupling constant $\xi$ so that for fixed reheating scenarios with $\wre > 1/3$, a cutoff value of the coupling constant prevents the overproduction of tensor fluctuations at the CMB scale. Therefore, we can place a strong constraint on the coupling parameter $\xi$, which we will discuss next.

\paragraph{\underline{Constraining the coupling strength $\xi$ from the tensor-to-scalar ratio bound:}}  

\begin{figure*}
\includegraphics[width=0.46\linewidth]{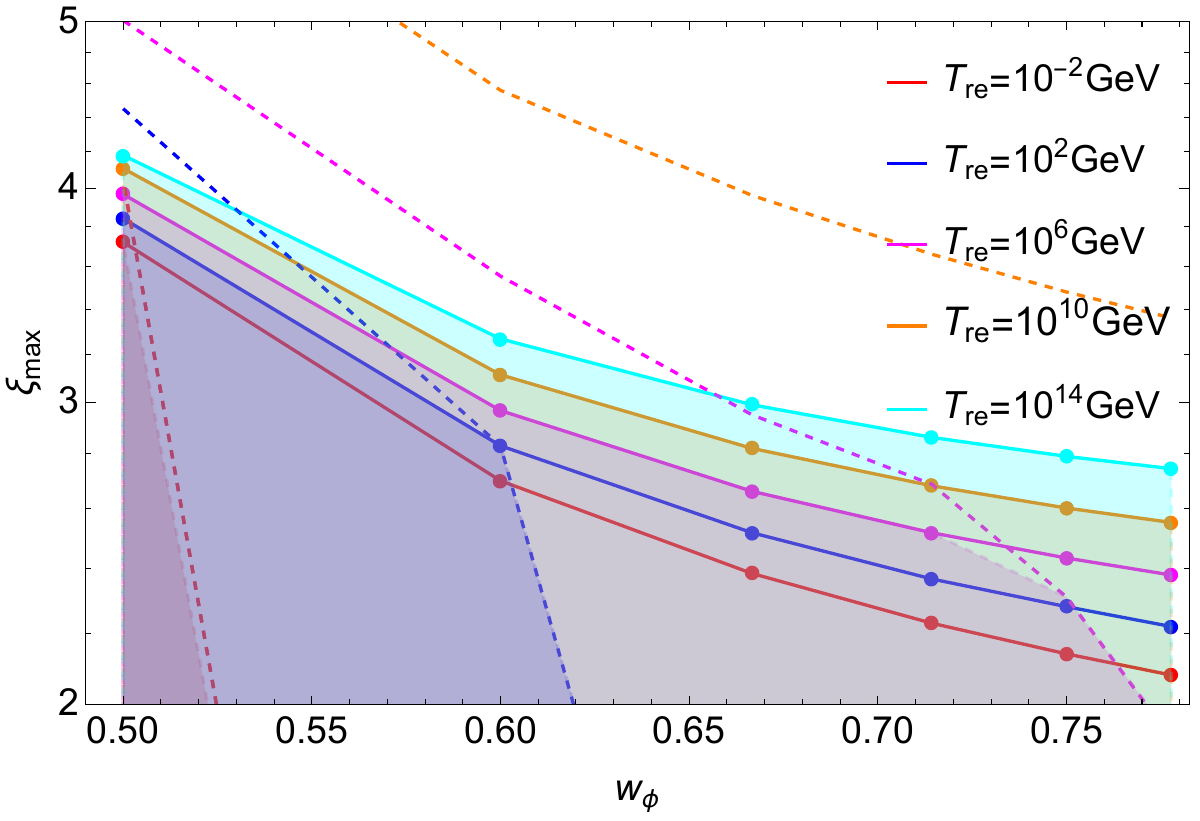}
\includegraphics[width=0.475\linewidth]{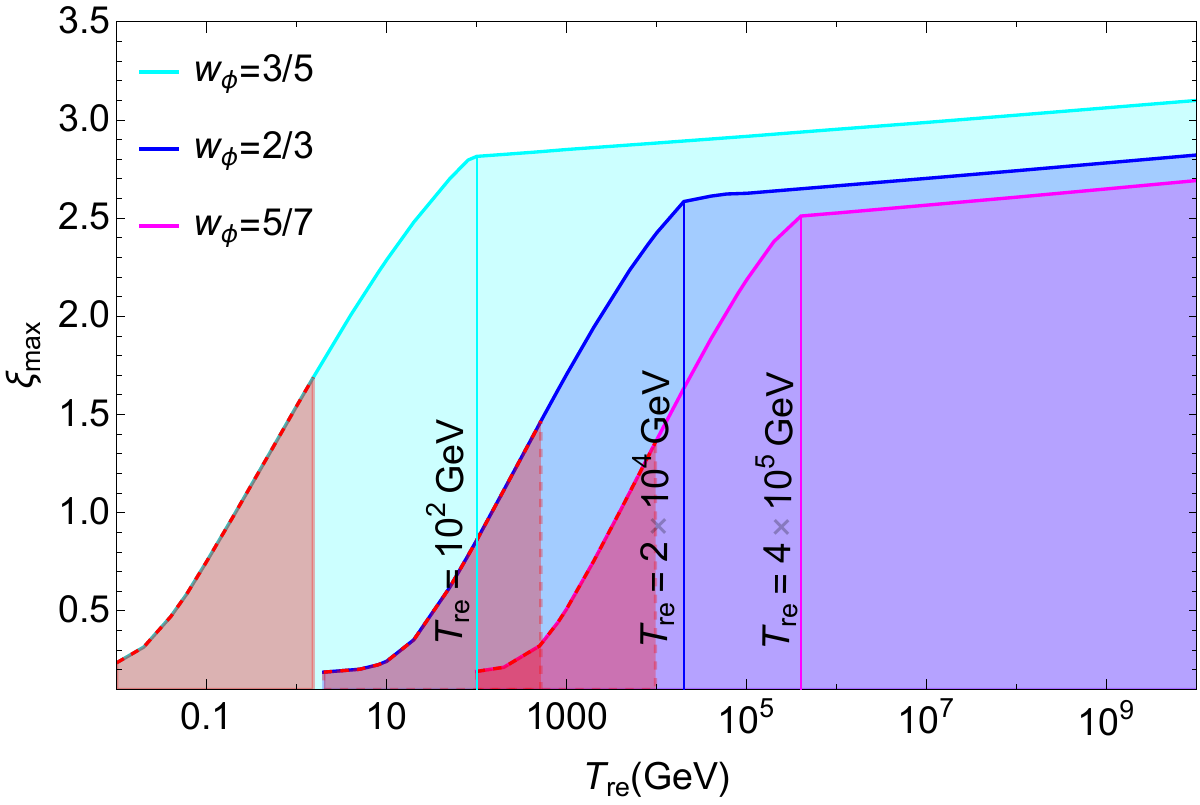}
\caption{\textit{In the left panel, we present the maximum allowed values of the coupling constant, $\xi_{\rm max}$, as a function of $\wre$. The colored dashed lines represent $\xi_{\rm max}$ values calculated from the $\dneff$ bound (see Eq. (\ref{eq:chi_neff})). In contrast, the solid colored lines indicate the limit derived from the tensor-to-scalar ratio (see Eq. (\ref{eq:r_con})). Different colors represent different $\Tre$ values. By combining both limits, we identify a common region allowed under both bounds, represented by the shaded areas corresponding to each color.
In the right panel, we plot $\xi_{\rm max}$ as a function of $\Tre$ for three discrete values of the equation of state: $\wre = 3/5$ (cyan), $\wre = 2/3$ (blue), and $\wre = 5/7$ (magenta). Combining both constraints, this plot shows the common allowed region, free from the $\dneff$ and the tensor-to-scalar ratio bounds. The three vertical lines labeled by specific $\Tre$ values divided the shaded regions into two. Above this temperature, the tensor-to-scalar ratio imposes a stricter constraint on the coupling parameter $\xi$ compared to the $\dneff$ bound. Three red-shaded regions indicate that these reheating temperatures are not allowed for the specified set of $\wre$ values as it violates $\dneff$ bound at high frequency near $\ke$ primarily sourced by PGWs.  }}\label{fig:xi_con}
\end{figure*}

\begin{itemize}
    \item 
    \underline{Constraining positive $\xi$ values for $\wre>1/3$ :}~~
We have discovered that for $\wre >1/3$ reheating scenarios, the scalar field undergoes a tachyonic instability beyond a threshold of the coupling constant $\xi$, particularly for the super-horizon modes, and generates large tensor fluctuations even at CMB scales. The current observational bound on the tensor-to-scalar ratio at CMB scales $r_{0.05} \leq 0.036$, \cite{Planck:2018vyg} can impose a stringent constraint on the coupling parameter $\xi$ through the following equation (assuming all contributions originate from secondary sources),
%%%%%%%%%%%%%%%%%%%%%%%%%%
\begin{align}\label{eq:r_con}
r_{0.05}&\simeq \frac{2\mathcal{A}^2\HI^4}{\pi^4M_{\rm pl}^4 A_s}\l\{ \frac{1}{2l(\delta-2)}+ \frac{1}{4l(1-l)-2l\delta}\r\}^2 \frac{8(1+2\nu_2)}{15(3-\nu_2)(4\nu_2-5)}\nn\\
    &\times \l( \frac{90 \He^2 M_{\rm pl}^2}{\pi^2 g_{\rm re}T_{\rm re}^4}\r)^{\frac{2(3\wre-1)}{3(1+\wre)}}\l( \frac{k_{\ast}}{\ke}\r)^{4(2-\nu_2)}\leq 0.036
\end{align}
%%%%%%%%%%%%%%%%%%%%%%%%
From Eq.(\ref{eq:r_con}), it is evident that the reheating temperature $T_{\rm re}$ and the average equation of state $\wre$, has a very sensitive dependence on the maximum coupling strength $\xi_{\rm max}$ to prevent the overproduction of tensor fluctuations at the CMB scale. In Fig.\ref{fig:gws}, coupling parameter has been chosen in a narrow range $2.85\leq\xi<3$ preventing the overproduction of large-scale(CMB scale) tensor fluctuations. For $\wre=3/5$ and $\Tre=10^{6}$ GeV, we have found the highest possible value of coupling strength $\xi_{\rm max}\approx2.9867$(See Table \ref{tab:param_est}). The SGW strength will be significant to overcome the PGW one for any $\xi$ close to this $\xi_{\rm max}$, $\xi\lesssim\xi_{\rm max}$. In the bottom left panel of Fig.\ref{fig:gws}, for any $\xi\geq 3$, total GW strength exceeds the maximum limit $\Omega_{\rm gw}h^2\gtrsim1.42\times 10^{-16}$ at the CMB scale, hence it is disfavored by the current observation. All the solid curves in the left panel of Figure (\ref{fig:xi_con}) depict maximum allowed values of $\xi_{\rm max}$ as a function of $\wre$ for different values of $\Tre$. And $\xi_{\rm max}$ represents the value of the coupling constant for a specific set of parameters exceeding which would violate the current bound on $r_{0.05}$ \cite{Planck:2018vyg}. For instance, when $\wre=0.5$ and $\Tre=10^{-2}$ GeV, the maximum value of the coupling constant is $\xi_{\rm max}\simeq 3.73$. 
On the other hand, if the reheating temperature is $\Tre=10^6$ GeV with the same EoS, the bound turns out to be $\xi_{\rm max}\simeq 3.99$.
The plot shows that $\xi_{\rm max}$ decreases with a reduction in the reheating temperature due to the prolonged duration of the reheating era
implying longer period of super-horizon growth.
We have listed the maximum allowed values of $\xi$ for different reheating temperatures and equation of state in Table \ref{tab:param_est}. It varies within $\sim (2,4)$. Maximum limit on $\xi$ is not sensitive to $r$ value. For example, future experiments such as LiteBIRD (\cite{LiteBIRD:2023iei, LiteBIRD:2022cnt, Drewes:2023bbs, Drewes:2022nhu}) will be able to detect $r< 0.002$ at 95$\%$ C.L (accounting for both statistical and systematic uncertainties)\cite{LiteBIRD:2022cnt}. Subject to this new bound we obtain $\xi_{\rm max}=(4.045,~3.089,~2.785,~2.640,~2.465,~2.277)$ corresponding to EoS $\wre=(1/2,~3/5,~2/3,~5/7,~4/5,~99/100)$ respectively for the reheating temperature $\Tre=10^6$GeV, which is not significantly different from that of the \textit{Planck} bound. We have already discussed the bound on $\xi$ for $r_{0.05}=10^{-4}$( see Table \ref{tab:param_est}) which anyway falls within the LiteBIRD limit. Henceforth, we shall stick to the current \textit{Planck} bound $r_{0.05}\lesssim 0.036$ while constraining the coupling for other EoS.\\
\item 

\underline{Constraining negative $\xi$ values for $\wre>1/3$:}
In this EoS range,
the dynamical equations (\ref{dynamicalinf}) and (\ref{dynamicalreh}) clearly show the presence of a strong inflationary IR instability but the absence of post-inflationary instability for any negative $\xi$ value. Subject to CMB bound, and considering the reheating parameters $(\Tre,~\He)=(10^6 {\rm GeV},~ 10^{-5} M_{\rm pl})$, we find $\xi_{\rm min}=(-0.15333,~-0.14079,~-0.13346,~-0.12866,~-0.12085,~-0.10659)$ for the reheating EoS $\wre=(1/2,~3/5,~2/3,~5/7,~4/5,~99/100)$ respectively. Therefore, any $|\xi|>|\xi_{\rm min}|$ is strictly prohibited for any $\wre>1/3$ based on the present study. We also further check that for a given EoS, the lower bound does not vary much with the reheating temperature. For instance, for $\wre=3/5$, we get $\xi_{\rm min}=(-0.11,~-0.12483,~-0.14079,~-0.158)$ for $\Tre=(10^{-2},~10^2,~10^6,~10^{10})$GeV respectively. \\\\

\item 
\underline{Constraining positive $\xi$ values for $\wre<1/3$ :}~~
In a similar fashion, for $\wre<1/3$, we calculate the tensor-to-scalar ratio at the CMB scale $r_{0.05}$ as
%%%%%%%%%%%%%%%%%%%%%%%%%%
\begin{align}\label{eq:r_con1}
r_{0.05}&\simeq \frac{2\mathcal{B}^2\HI^4}{\pi^4M_{\rm pl}^4 A_s}\l\{ \frac{1}{2l(\delta-2)}+ \frac{1}{4l(1-l)-2l\delta}\r\}^2 \frac{8(1+2(\nu_1+\nu_2))}{15(3-(\nu_1+\nu_2))(4(\nu_1+\nu_2)-5)}\nn\\
    &\times \l( \frac{90 \He^2 M_{\rm pl}^2}{\pi^2 g_{\rm re}T_{\rm re}^4}\r)^{\frac{2(3\wre-1)}{3(1+\wre)}}\l( \frac{k_{\ast}}{\ke}\r)^{4(2-\nu_1-\nu_2)}\leq 0.036
\end{align}
%%%%%%%%%%%%%%%%%%%%%%%%
The negative value of the index $(2-\nu_1-\nu_2)$ in the lower the value of $\xi <1/6$ gives significant GW strength for $\wre<1/3$. This fact sets the lower limit of $\xi_{\rm min}$ that satisfies the observational bound on $r_{0.05}$. For instance, for $\wre=0$, we obtain $\xi_{\rm min}\simeq (0.02036,~0.04179,~0.05935,~0.07394$), and for $\wre=0.1$, we obtain $\xi_{\rm min}\simeq (5.4\times 10^{-4},~0.01569,~0.02916,~0.04119$) for $\Tre=(10^{-2},~10^2,~10^{6},~10^{10}$) GeV respectively and for $\wre=0.2$, we obtain $\xi_{\rm min}\simeq (3.35\times 10^{-3},~0.01104$) for $\Tre=(~10^{6},~10^{10}$) GeV respectively. For $\wre=0.2$, no lower limit exists for the temperatures $\Tre=(10^{-2},~10^2$) GeV. In the right panel of Fig.\ref{fig:gws_1}, we have considered a narrow range of $0.06<\xi\leq 0.065$ for $\wre=0$ and $\Tre=10^{6}$ GeV, and shown how the GW spectrum changes near the CMB scale within the \textit{Planck} limit. The larger value of $\xi$ is unconstrained due to less GW production at the CMB scale ($k<<\ke$). Interestingly, we also find that as one gradually approaches $\wre=1/3$, the coupling remains unconstrained, for secondary GW is subdominant compared to the primary one. As an example, for $\wre=0.33$ and $\xi=0$, we find $r_{0.05}=(2.88\times 10^{-11},~ 3.46\times 10^{-11},~ 4.17\times 10^{-11},~ 5\times 10^{-11} )$ for $\Tre=(10^{-2},~10^2,~10^{6},~10^{10})$ GeV respectively.\\
\item
%%%%%%%%%%%%%%%%%%%%%%

\underline{Constraining negative $\xi$ values for $\wre<1/3$ :}
In this range, according to the dynamical equations (\ref{dynamicalinf}) and (\ref{dynamicalreh}), both inflationary and post-inflationary instability causes large GW production. For instance, for $\wre=0$ and $\Tre=10^6$ GeV, we find $r_{0.05}\approx(10^{26},~10^{24},~ 3.73\times10^{23},~ 7\times 10^{23} )$ for the coupling strength $\xi=(-0.01, -0.1, -1, -10)$ respectively. Therefore, the entire negative $\xi$ regime is found to be completely incompatible with CMB at least for $\wre=0$.
\end{itemize}
%%%%%%%%%%%%%%%%%%%%%%%%%%%%%%%%%%%%%%%%%
\begin{center}
\begin{table}[t]
\centering
   \renewcommand{\arraystretch}{1.6} \begin{tabular}{|c|c ||c| c| c| c| c|   }
    \hline
    \hline
    \multirow{5}{*}{$r_{0.05}=0.036$} & $\Tre$ (GeV) & $\wre=0$ 
    & $\wre=1/2$ &$\wre= 3/5$ &
    %$\wre=5/7$ & $\wre=3/4$ & $\wre=7/9$ 
     $\wre=4/5$ & $\wre=99/100$\\
    \cline{2-7}
     & $10^{-2}$ &0.02036& 3.7312 & 2.7649 &
     %2.4506 & 2.2967 & 2.2065 & 2.1473 &
     2.1056 & 1.8818\\
     & $ 10^2$ &0.04179& 3.8599 & 2.9034 &
     %2.5945 & 2.4451 & 2.3578 & 2.3008 & 
     2.2611 & 2.0529\\
    &  $10^6$ & 0.05935& 3.9955 & 2.9867 & 
    %2.7514 & 2.6076 & 2.5247 & 2.4714 & 
    2.4342 & 2.2482\\
    % &  $10^{14}$ & 0.08618& 4.288 & 3.3842 & 
    %2.7514 & 2.6076 & 2.5247 & 2.4714 & 
   % 2.8448 & 2.7317\\
  \cline{1-7}

   \multirow{4}{*}{$r_{0.05}=10^{-4}$}  & $10^{-2}$ & 0.01668& 3.7950 & 2.8136 &
   %2.4935 & 2.3375 & 2.2461 & 2.1859 &
   2.1443 &
1.9168\\
     & $ 10^2$ & 0.03903& 3.9289 & 2.9569 & 
     %2.6439 & 2.4922 & 2.4039 & 2.3463 & 
     2.3059 & 2.095\\
    &  $10^6$ & 0.05728 &4.0691 & 3.1111 &
    %2.8068 & 2.6613 & 2.5779 & 2.5242 & 
    2.4859 & 2.299\\
    %&  $10^{14}$ & 0.0850&4.3738 & 3.4581 &
    %2.8068 & 2.6613 & 2.5779 & 2.5242 & 
    %2.9161 & 2.807\\
      \hline
    
    \end{tabular}
         \caption{\textit{ %\textcolor{red}{Here give the values of both $\xi_{max/min}$, for $\wre =0,1/3,1/2, 3/5,1$. Do not need to give so many values of $\wre$. Also, quote the values of the highest $\Tre$ }
         In the above table, we have listed the minimum possible values $\xi_{\rm min}$ for $\wre=0$  and maximum possible values $\xi_{\rm max}$ for higher reheating EoS $\wre=1/2,~3/5,~4/5,~ 0.99$ for 
         %discrete reheating scenarios where we fixed the EoS $\wre$ and
         different reheating temperatures $\Tre$. The lower bound $\xi_{\rm min}$ for $\wre=0$ is derived from Eq. (\ref{eq:r_con1}) and the upper bound $\xi_{\rm max}$ for higher EoS $\wre>1/3$ is derived from Eq. (\ref{eq:r_con}) to avoid the overproduction of tensor perturbations at the CMB scale. Based on recent observations from Planck, We consider the upper bound on the tensor-to-scalar ratio to be $r_{0.05}= 0.036$. We also show the slight variation of both $\xi_{\rm min}$ and $\xi_{\rm max}$ with tensor-to-scalar ratio by considering another value at the CMB scale $r_{0.05}=10^{-4}$. }}
    \label{tab:param_est}
 \end{table}
\end{center}
%%%%%%%%%%%%%%%%%%%%%%%%%%%%%%

\paragraph{\underline{Constraining $\xi$ from the isocurvature bound:}}

Similar bound on the non-minimal coupling can also be obtained from the isocurvature constraint. The current constraints on the isocurvature power spectrum by \textit{Planck} are defined to be $\beta_{\rm iso}\equiv \Ps(k_{\ast})/\l(\Pr(k_{\ast})+\Ps(k_{\ast})\r)\lesssim 0.038$ at the $95\%$ C.L for the \textit{Planck} pivot scale $k_{\ast}$. The large-scale instability of the scalar field inevitably generates strong isocurvature perturbation at the CMB scale. In the recent literature \cite{Garcia:2023awt} and \cite{Garcia:2023qab}, authors have investigated such perturbation from the non-minimally coupled scalar field and background inflaton field. The pivot scale amplitude of curvature power spectrum $\Pr(k_{\ast})=2.1\times 10^{-9}$ gives the upper bound of the amplitude of isocurvature power spectrum at CMB scale $\Ps(k_{\ast})\lesssim 8.3\times 10^{-11}$\cite{Planck:2018vyg}. The second-order isocurvature power spectrum is evaluated by using the following expression as \cite{Garcia:2023awt, Garcia:2023qab, Chung:2004nh, Chung:2011xd, Ling:2021zlj, Kolb:2023ydq, Liddle:1999pr} 
%%%%%%%%%%%%%%%%%%%%%%%%%%%%%
\begin{equation}\label{isocurvpow}
  \Ps(k)= \frac{1}{\rho_{\chi}^2}\frac{k^3}{2\pi^2}\int d^3\vec{x}\langle\delta\rho_{\chi}(\vec{x})\delta\rho_{\chi}(0)\rangle e^{-i \vec{k}.\vec{x}}=\frac{k^3}{(2\pi)^5 \rho_{\chi}^2a^8}\int d^3\vec{p}~ P_{X}\l(p,|\vec{p}-\vec{k}|\r) 
\end{equation}
%%%%%%%%%%%%%%%%%%%%%%%%%%%%%
where $\rho_{\chi}$ and $\delta\rho_{\chi}$ are energy-density, and  
%%%%%%%%%%%%%%%%%%%%%%%%% 
\begin{equation}\label{PX1}
    P_{X}(p,q)=|X_p^{\prime}|^2|X_q^{\prime}|^2+a^4m_{\chi}^4|X_p|^2|X_q|^2
    %+a^2m_{\chi}^2 \l[(X_p X_p^{\prime\ast})(X_q X_q^{\prime \ast})+h.c\r]
\end{equation}
%%%%%%%%%%%%%%%%%%%%%%%%%%%%%%%%%
%As here we are mainly dealing with the massless scalar fluctuations, only non-vanishing contribution will come from the first term of the expression of $P_{X}(p,q)$ above and $\rho_{\chi}$ will be the energy-density of the massless scalar in the present context.
%%%%%%%%%%%%%%%%%%%
\begin{itemize}
    \item \underline{Isocurvature constraint for $\wre>1/3$:}
 With the help of the equations (\ref{Xreh9}), (\ref{isocurvpow}), and (\ref{PX1}) we compute the isocurvature amplitude at the CMB scale, and  we obtained the maximum bound on $\xi$ for $\wre>1/3$. The maximum values tunes out to be $\xi_{\rm max}=(3.750,~2.981,~ 2.545,~2.477)$ for EoS $\wre=(1/2,~3/5,~4/5,~ 99/100)$ respectively for the reheating temperature $\Tre=10^6$GeV and $\He=10^{-5}M_{\rm pl}$. For a given EoS, this maximum limit of $\xi$ does not vary much for a wide range of reheating temperatures. For example, for $\wre=1/2$, we get $\xi_{\rm max}=(3.761,~ 3.755,~3.750,~ 3.744 )$ for $\Tre=(10^{-2}, 10^2, 10^6, 10^{10})$ GeV respectively. Interestingly, these $\xi_{\rm max}$ values are close to the values we obtained from the the tensor-to-scalar ratio discussed earlier (See Table \ref{tab:param_est}).
\item 
\underline{Isocurvature constraint for $\wre<1/3$:} In \cite{Garcia:2023qab}, for $\wre=0$, authors have found a lower limit of $\xi\gtrsim 0.027$ for the massless non-minimally 
 coupled scalar fluctuations from the current isocurvature bound $\mathcal{P}_{S}(k_{\ast})<8.3\times 10^{-11}$. Interestingly, for $\wre=0$, this lower boundary of $\xi$ from isocurvature constraint is close to the prediction of CMB scale tensor-to-scalar ratio bound $r_{0.05}\leq 0.036$ as discussed above.
\end{itemize}
%%%%%%%%%%%%%%%%%%%%%%%
\paragraph{\underline{Constraining the coupling strength $\xi$ and reheating dynamics through $\Delta N_{\rm eff}$:}}

%\begin{itemize}
    %\item
    The total radiation density around the time of decoupling influences the cosmic microwave background (CMB). At that epoch, neutrinos comprised a significant fraction of the radiative energy, but additional radiation (such as dark radiation or primordial gravitational waves) may also impact the CMB spectrum. In this context, treating the $\chi$ field as dark radiation significantly affects the CMB spectra through the extra radiation component. The effective number of neutrino species, $\neff$, which represents the energy density stored in relativistic components (radiation), is defined as
%%%%%%%%%%%%%%%%%%%%%%%
\begin{align}
\rhora = \rhop + \rhon + \rho_{\rm x} = \left[1 + \frac{7}{8}\left(\frac{4}{11}\right)^{4/3} \neff \right]\rhop \label{eq:rhora}
\end{align}
where $\rhop$, $\rhon$, and $\rho_{\rm x}$ are the energy densities of photons, neutrinos, and extra radiation components (massless scalar field $(\rho_\chi)$ or primordial gravitational waves $(\rhogw)$), respectively. From Eq. (\ref{eq:rhora}), the excess radiation component can be defined as:
%%%%%%%%%%%%%%%%%%%%%%%%%
\begin{align}
\rho_{\rm \chi} = \frac{7}{8}\left(\frac{4}{11}\right)^{4/3}\rhop \Delta\neff \label{eq:rho_extra}
\end{align}
%%%%%%%%%%%%%%%%%%%%%%%%%%
where $\Delta\neff = \neff - N_\nu$ represents the extra relativistic degrees of freedom. Here, $\rho_{x}$ includes both dark radiation $\rho_\chi$ and primordial gravitational waves $\rhogw$, with $\neff$ being the observed total relativistic degrees of freedom and $N_\nu = 3.044$ representing the standard model neutrino degrees of freedom \cite{Bennett:2020zkv, Froustey:2020mcq,Akita:2020szl,Haque:2023yra,Chakraborty:2023ocr}.

To determine the contribution of the dark radiation, specifically the massless scalar field $\chi$, we can express Eq. (\ref{eq:rho_extra}) as:
%%%%%%%%%%%%%%%%%%%%%%%%%%%%%%
\begin{align}
\rho_\chi =\frac{1}{2\pi^2 a^4} \int_{k_{\rm re}}^{\ke} \frac{dk}{k}k^4 |\beta_k|^2=\frac{7}{8}\left(\frac{4}{11}\right)^{4/3}\rhop  \dneffc \label{eq:rho_chi}
\end{align}
%%%%%%%%%%%%%%%%%%%%%%%%%%%
where $\dneffc$ is the extra relativistic degree of freedom due to the dark radiation field $\chi$. This computation of $\rho_{\chi}$ in different $\xi$ ranges is detailed in the Appendix \ref{appenB}. For the sake of convenience, we introduce a new dimensionless variable $\omegac(\ere) = \rho_\chi(\ere) / \rhoc(\ere)$, where $\rhoc(\ere) = 3H_{\rm re}^2\mpl^2$ is the background energy density at the end of reheating. As both the background energy density $(\rho_c)$ and the massless scalar field (considered as dark radiation $\rho_\chi$) goes as $a^{-4}$ due the background expansion, the fractional energy density $\Omega_{\chi}(\eta>\ere<\eta_{\rm eq})$ remain conserved during radiation dominated era.
%%%%%%%%%%%%%%%%%%%%%%%%%%%%%%%%%%%%%%%%%%%
%\begin{center}
%\begin{table}[t]
%\centering
%    \begin{tabular}{c ||c c c c c c }
%    \hline
 %   \hline
%    $\Tre$ (GeV) & $\wre=1/2$ & $\wre=3/5$ &$\wre= 2/3$ & $\wre=5/7$ & $\wre=3/4$ & $\wre=7/9$  \\
%    \hline
%      $10^{-2}$ & 4.0408 & 0.235 &  &  &  &  \\
%      $ 10^2$ & 4.457 & 2.8376 & 0.863 & 0.192 &  &   \\
%      $10^6$ & 5.01 & 3.557 & 2.95 & 2.688 & 2.31 & 1.915 \\
%      \hline    
%    \end{tabular}
 %        \caption{In the above table, we have listed the $\xi_{\rm max}$ for discrete reheating scenarios where we fixed the EoS $\wre$ and reheating temperature $\Tre$. This bound is derived from Eq. (\ref{eq:chi_neff}) to avoid the overproduction of dark radiation that can hamper the current $\dneff$ bound.}
 %   \label{tab:param_est2}
% \end{table}
%\end{center}
%%%%%%%%%%%%%%%%%%%%%%%%%%%%%%%%%%%%%%%%%%%%
%Considering \(\chi\) as a massless scalar field, its energy density evolves as \(a^{-4}\) following the reheating phase due to the expansion of the universe.
The present-day fractional energy density of the \(\chi\) field can be expressed as
%%%%%%%%%%%%%%%%%%%%%%%%%%%%%%%
\begin{equation}
    \Omega_{\chi} h^2 \simeq \left( \frac{g_{r,0}}{g_{r,eq}} \right)^{1/3} \Omega_{\rm R} h^2 \, \Omega_{\chi}(\ere)
\end{equation}
%%%%%%%%%%%%%%%%%%%%%%%%%%%%%%
where \(\Omega_{\rm R} h^2 \simeq 4.3 \times 10^{-5}\) \cite{Planck:2018vyg} denotes the dimensionless energy density of radiation at the current epoch. Here, \(g_{r,eq}\) and \(g_{r,0}\) represent the number of relativistic degrees of freedom at the epochs of radiation-matter equality and the present day, respectively.
By expressing Eq. (\ref{eq:rho_chi}) in terms of this dimensionless variable, we obtain
%%%%%%%%%%%%%%%%%%%%%%%%%%%%%
\begin{align}
\omegac h^2 = \frac{7}{8}\left(\frac{4}{11}\right)^{4/3}\omegap h^2 \dneffc \
\simeq 1.6 \times 10^{-6}\left(\frac{\dneffc}{0.284}\right)\label{eq:chi_neff}
\end{align}
%%%%%%%%%%%%%%%%%%%%%%%%%%%%%%
where $\omegap h^2 = 2.47 \times 10^{-5}$ is the present-day photon energy density \cite{Planck:2018vyg}. The latest Planck data with Baryon Acoustic Oscillation (BAO) predicts $\dneff\leq 0.284$ (within $2\sigma$ range) \cite{Planck:2018vyg}. Using this bound as an upper limit, we can further constrain the coupling parameter $\xi$ through Eq.(\ref{eq:chi_neff}).\\

We find that coupling strength $\xi$ is further constrained through $\Delta N_{\rm eff}$ for $\wre>1/3$. In Fig.(\ref{fig:xi_con}), we have plotted the upper limit of the coupling parameter $\xi_{\rm max}$ as a function of the average equation of state $\wre>1/3$ for five different reheating temperatures. Lowering the reheating temperature for a fixed EoS tightens the constraint on $\xi$.
For example, if the EoS is $\wre=0.5$ and the reheating temperature of our universe is $\Tre=10^{-2}$ GeV, then to satisfy the current $\dneff$ bound, the maximum allowed value of the coupling parameter is $\xi_{\rm max}\simeq 4.02$. Note this is greater than the bound we obtained from the tensor-to-ratio ($\xi_{\rm max} \simeq 3.73)$ discussed before. On the other hand, if the reheating temperature is $\Tre=10^6$ GeV with the same EoS, the bound is $\xi_{\rm max}\simeq 4.99$, which is again higher than the bound from the tensor to ratio ($\xi_{\rm max} \simeq 3.99)$ discussed before.

It is to be noted that combining constraints from the tensor-to-scalar ratio and from $\dneff$ yields significant insights.
In the lower reheating temperature case as one reduces the equation state the tensor to scalar ratio tends (solid lines) to provide stronger constraints on $\xi_{\rm max}$ than the $\dneff$, and $\dneff$ leads to the maximum possible value of equation state $\wre$ (set by dashed lines). 
For instance, with $\wre = 0.6$ and a reheating temperature of $\Tre= 10^{-2}$ GeV, the $\dneff$ constraint predicts a maximum allowable value of the coupling constant $\xi_{\text{max}} \simeq 0.205$. In contrast, under the same reheating parameters, the tensor-to-scalar ratio constraint allows a higher upper bound for the coupling constant, $\xi_{\text{max}} \simeq 2.765$.
Therefore, we get an allowed region of $\xi$ bounded by solid and dashed lines which satisfy both the constraints.
%Specifically, we find that in such scenarios, the constraint from $\dneff$ imposes a stronger restriction on the coupling parameter $\xi$ compared to the tensor-to-scalar ratio constraint discussed in Eq. (\ref{eq:r_con}). 
This is indeed the case for red, blue, and magenta curves as depicted in the left panel of Fig.\ref{fig:xi_con} for $\Tre = (10^{-2}, 10^{2}, 10^{6})$ GeV accordingly. However, with the higher temperature, the constraint from \textit{Planck} on tensor to scalar ratio becomes increasingly important and tends to prove the entire bound on $\xi$. This indeed can be observed for cyan and brown curves with $\Tre = (10^{14}, 10^{10})$ GeV respectively.\\

%%%%%%%%%%%%%%%%%%%%%%%%%%%%%%%%%%%%%%%%%%
%%%%%%%%%%%%%%%%%%%%%%%%%%%%%%%%%%%%%%%%%%%
%\item 
%\underline{For $\wre<1/3$ :}

Unlike $\wre>1/3$, for $\wre<1/3$, we don't get any new constrain on $\xi$ through $\Delta N_{\rm eff}$ bound.  
% Here also we define the present-day fractional energy density of the $\chi$ field as
%%%%%%%%%%%%%%%%%%%%%%%%%%%%%%%
%\begin{align}
%\Omega_{\chi} h^2 \simeq \left( \frac{g_{r,0}}{g_{r,eq}} \right)^{1/3} \Omega_{\rm R} h^2 \, \Omega_{\chi}(\ere)\simeq 1.6 \times 10^{-6}\left(\frac{\dneffc}{0.284}\right) \nn
%\end{align}
%%%%%%%%%%%%%%%%%%%%%%%%%
%where the dimensionless variable at the end of reheating is defined as $\omegac(\ere) = \rho_\chi(\ere) / \rhoc(\ere)$.\\ 
As discussed earlier, for $\wre<1/3$, as enhancement effect is important in the range $\xi<1/6$, we find the lower limit of coupling $\xi_{\rm min}<\xi_{\rm cri}<1/6$ satisfying the tensor-to-scalar ratio bound $r_{0.05}\leq 0.036$. The energy spectrum remains blue-tilted in the entire range $\xi>\xi_{\rm cri}$. For example, for $\xi=0$, exploiting the Equations (\ref{eq:rho_chi})-(\ref{eq:chi_neff}) we obtain for $\wre=(0,~0.1,~0.2)$, $\Delta N_{\rm eff}=\l(2.88\times 10^{-13},~7\times 10^{-13},~1.77\times 10^{-12}\r)$ for a wide range of reheating temperatures $\Tre=\l(10^{-2}-10^{10}\r)$ GeV(Detail computations of $\rho_{\chi}$ for $\wre<1/3$ are given in Appendix \ref{appenB}.). These values are far below the maximum limit $\Delta N_{\rm eff}\leq 0.284$ for the lowest possible value of $\xi=0$ meaning no lower limit can be imposed on $\xi$ through $\Delta N_{\rm eff}$ for $\wre<1/3$. If we choose $\xi>\xi_{\rm cri}$, although the energy spectrum is blue-tilted, the secondary GW energy strength is too weak to overtake the primary strength around $\ke$(see Fig.\ref{fig:gws1}). Therefore, in this regime, the PGW contribution always dominates the total GW energy density $\Omega_{\rm gw}h^2$.\\

Similarly, primordial gravitational waves (PGWs) with frequencies $\geq 10^{-15}$ Hz may contribute significantly to the radiation density of the Universe during the decoupling of the cosmic microwave background (CMB). It can also be treated as an extra relativistic degree of freedom symbolized as $\dneffgw$. Hence, we can similarly express it as \cite{Caprini:2018mtu,Maiti:2024nhv}
%%%%%%%%%%%%%%%%%%%%%%%%%%%%%%
\begin{align}
    \ogwh=\frac{7}{8}\l( \frac{4}{11}\r)^{4/3}\omegap h^2 \dneffgw \simeq 1.6 \times 10^{-6}\left(\frac{\dneffgw}{0.284}\right)
\end{align}
%%%%%%%%%%%%%%%%%%%%%%%%%
where $\ogwh$ is the present-day dimensionless energy density of the gravitational waves produced in the early universe and it is defined as 
%%%%%%%%%%%%%%%%%%%%%%%%%%%%
\begin{align}
    \ogwh=\int_{k_{\rm min}}^{\ke} \frac{dk}{k}\ogw(k)h_0^2
\end{align}
%%%%%%%%%%%%%%%%%%%%%%
here $\ogw(k)h^2$ is the total contributions from both primary and secondary gravitational waves. To this end, we should point out that the GW is a secondary contribution 

%%%%%%%%%%%%%%%%%%%%%%%%%%
Assuming the present-day photon density parameter is $\Omega_{\gamma}h^2\simeq 2.47\times 10^{-5}$, a combination of the latest Planck-2018 and Baryon Acoustic Oscillation (BAO) data predicts $\Delta N_{\text{eff}}\simeq0.284$ (within a $2\sigma$ range) \cite{Planck:2018vyg}. Consequently, this prediction sets an upper limit on primordial gravitational waves, such that $\ogwh<1.6\times 10^{-6}$ \cite{Clarke:2020bil}. Using this result, we derive the following inequality \cite{Smith:2006nka, Clarke:2020bil,Maiti:2024nhv}
%%%%%%%%%%%%%%%%%%%%%%%
\begin{align}
\ogwh \leq 1.6 \times 10^{-6}\left(\frac{\dneffgw}{0.284}\right)\label{eq:rhogw_neff}
\end{align}
%%%%%%%%%%%%%%%%%%%%%%%%%%%%%%%%%%%%%%%%%%%
\begin{figure*}
\includegraphics[width=0.6\linewidth]{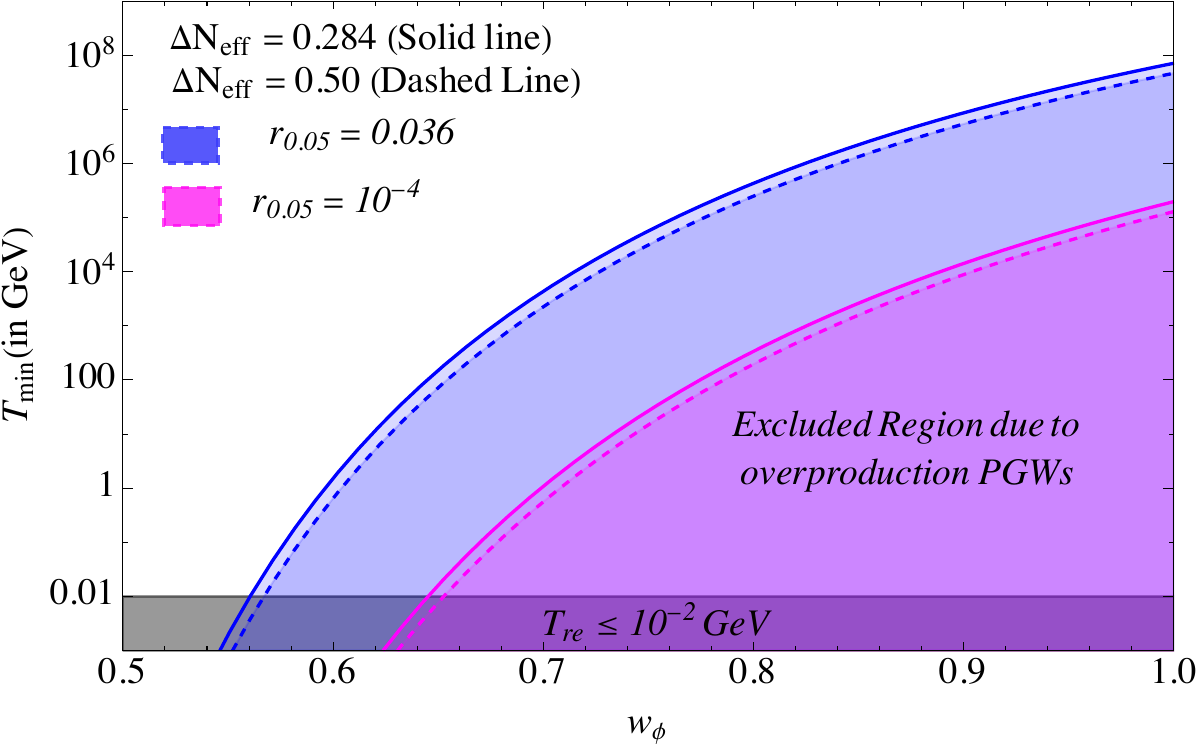}
\caption{\textit{The figure above shows the minimum allowed reheating temperature $T_{\text{min}}$ (in GeV) as a function of $\wre$. The blue lines correspond to $r_{0.05} = 0.36$, while the magenta lines represent $r_{0.05} = 10^{-4}$. The solid lines indicate $\dneff = 0.284$, and the dashed lines indicate $\dneff = 0.50$. The shaded regions in color mark areas were excluded due to the overproduction of primary gravitational waves at the high-frequency end $f = f_{\rm end}({k_e}/{2\pi})$. The gray region at the bottom is excluded due to the minimum reheating temperature required by Big Bang Nucleosynthesis (BBN), which is about $\Tre \simeq 10^{-2}$ GeV.}}\label{fig_tre}
\end{figure*}
%%%%%%%%%%%%%%%%%%%%%%%%%%%%%%%%%%%%%%%%%%%%%%%%%%%%%%%%%%%%%%%%%%
The constraint presented in Eq. (\ref{eq:rhogw_neff}) imposes a constraints on the reheating temperature;
%%%%%%%%%%%%%%%%%%%%%%
\begin{align}
T_{\text{min}} \geq \left(\frac{90\He^2M_{\rm pl}^2}{\pi^2g_{\text{re}}}\right)^{1/4}\beta^{\frac{3(1+\wre)}{4(3\wre-1)}}\left(\frac{0.284}{\dneffgw}\right)^{\frac{3(1+\wre)}{4(3\wre-1)}}\label{eq:tre_min}
\end{align}
%%%%%%%%%%%%%%%%%%%%%%%%%5
Here $T_{\text{min}}$ is the lowest possible reheating temperature to ensure the overproduction of the extra relativistic degree of freedom due to the PGWs. In the above we defined $\beta = 1.43 \times 10^{-11}\mathcal{D}_2(\He/10^{-5}M_{\rm pl})^2/(\nw)$. It is crucial to emphasize that this lower bound on the reheating temperature applies exclusively to the equation of state $\wre>1/3$.

%%%%%%%%%%%%%%%%%%%%%%%%%%%%%%%%%%%%%%%%%%%
\begin{figure*}
\includegraphics[width=0.48\linewidth]{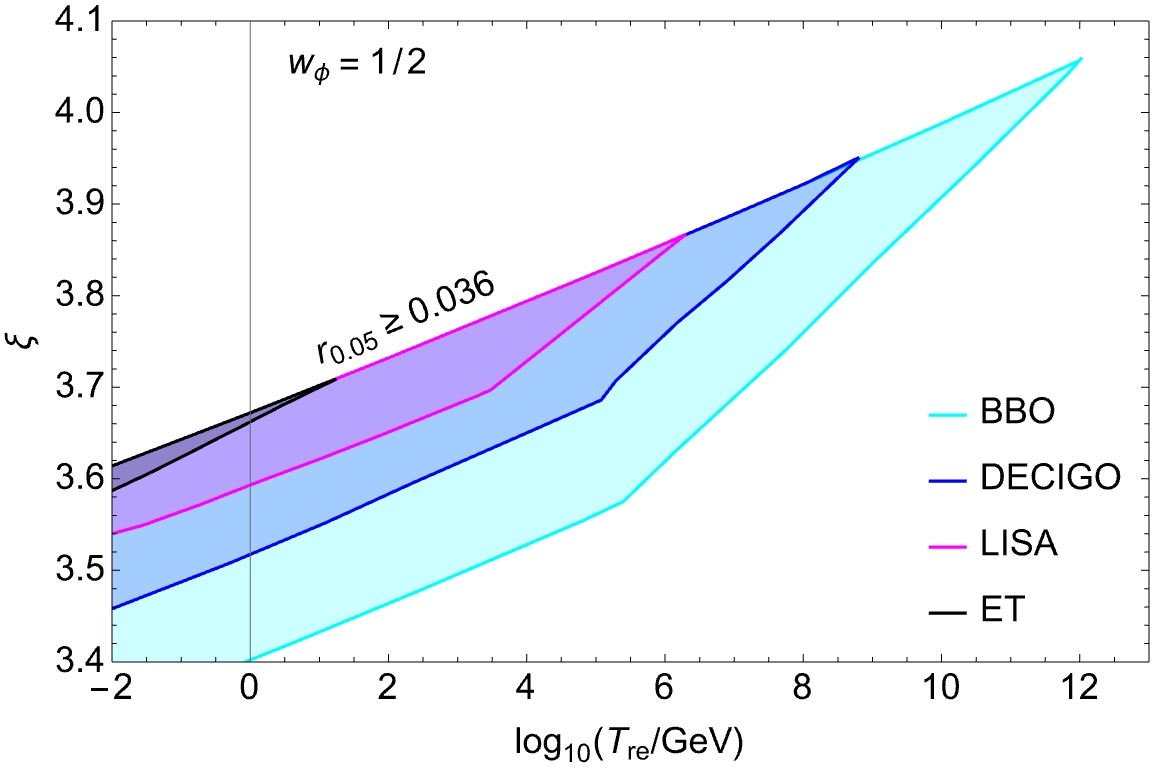}
\includegraphics[width=0.49\linewidth]{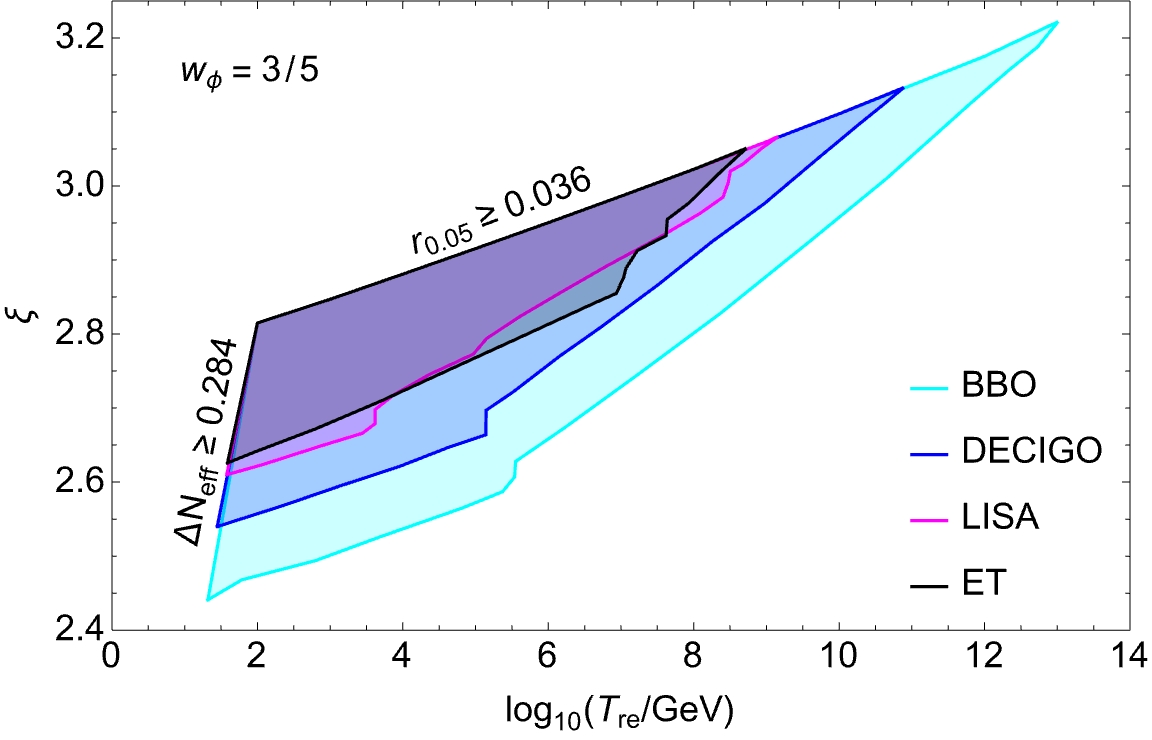}
\caption{\textit{In the figure above, we present the estimated parameter space of the coupling constant $\xi$ versus the reheating temperature $\Tre$ for two different equations of state (EoS): $\wre = 1/2$, and $\wre = 3/5$. These estimates can be tested by future gravitational wave (GW) detectors, including LISA (magenta), DECIGO (blue), BBO (cyan), and ET (gray). Within this parameter space, the signal generated by the scalar field $\chi$ can be detected by these GW experiments. In both plots, the upper bound comes from the tensor to scalar ratio at the CMB scale where in the right panel at a lower temperature the $\dneff$ puts a stronger constraint on the coupling parameter $\xi$.}}\label{xi_par}
\end{figure*}
%%%%%%%%%%%%%%%%%%%%%%%%%%%%%%%%%%%%%%%%%%%%%%%%%%%%%%%%%%%%%%%%%%
Using Eq. (\ref{eq:tre_min}), we have generated a parameter space plot of $T_{\text{min}}$ as a function of the average equation of state ($\wre$) in Fig. (\ref{fig_tre}), showing the minimum permissible reheating temperature as a function of $\wre$. In this figure, the blue lines correspond to $r_{0.05} = 0.036$, and the magenta lines to $r_{0.05} = 10^{-4}$. Solid lines represent $\dneff = 0.284$, while dashed lines correspond to $\dneff = 0.50$. The shaded regions indicate areas excluded due to the overproduction of gravitational waves at high frequencies. The gray shaded region at the bottom excludes reheating temperatures lower than the BBN bound, i.e., $\Tre^{\rm BBN} \simeq 10^{-2}$ GeV. While considering a stiff equation of state ($\wre > 1/3$), this bound must be taken into account. In all our plots, we have ensured that this bound is respected to avoid the overproduction of gravitational waves from primary quantum fluctuations during inflation.\\

\paragraph{\underline{Constraining Reheating Dynamics and Coupling Parameters $\xi$ via Future Gravitational Wave Experiments:}}
As depicted in Fig. (\ref{fig:gws}), for an appropriate set of reheating parameters and coupling constant $\xi$, the production of secondary gravitational waves (SGWs) by the scalar field is significant enough to pass through the sensitivity curves of several forthcoming gravitational wave (GW) detectors, including LISA, DECIGO, BBO, ET, and others. This observation suggests that future GW detectors could probe non-minimal coupling and associated dynamics in the early universe, potentially placing stringent constraints on the coupling parameter $\xi$ and the reheating dynamics. In this section, we point out the parameter space in which these detectors are expected to explore these dynamics.

Focusing on future GW experiments such as LISA, DECIGO, BBO, and ET, we utilize the proposed sensitivity curves to estimate the parameter ranges for $\xi$ and the reheating temperature $\Tre$ for two different equations of state, $\wre = 1/2$ and $\wre = 3/5$, as shown in Fig. (\ref{xi_par}). The shaded regions indicate the parameter space where the signal produced by the scalar field is expected to be detectable by these experiments, provided $\xi$ falls within the specified range for a given reheating temperature and equation of state. Notably, the upper bound on $\xi$ remains the same for all experiments, despite differing sensitivity thresholds. This is because the upper limit on $\xi$ is determined by the tensor-to-scalar ratio constraint by \textit{Planck} and the $\dneff$ bound (as illustrated in Fig. (\ref{fig:xi_con})). Consequently, for values of $\xi$ exceeding a certain threshold, the signal, although potentially detectable, is excluded due to the overproduction of tensor fluctuations and additional relativistic degrees of freedom.

In the right panel of Fig. (\ref{xi_par}), we observe that when the reheating temperature is relatively low and the scalar field is considered as dark radiation, $\dneff$ imposes stricter constraints on $\xi$, as discussed earlier. Conversely, in the case of $\wre = 0.5$ (see in the left panel of Fig.(\ref{xi_par})), the tensor-to-scalar ratio constraint is more restrictive than the $\dneff$ constraint. For all the plots the lower boundary is fixed by the lowest values of the gravitational wave strength (sensitivity line) that a particular experiment could measure.

To this end, we would like to point out that in the present study, we solely focus on the long-wavelength modes($k<\ke$) of the scalar fluctuations that suffer from tachyonic instability. The longer the wavelength, the stronger the instability and the larger the source field amplitude. Our study revealed that the presence of such instability causes large GW and isocurvature production, and gives tight constraints on $\xi$ which one should keep in mind in future studies. At this point let us mention the recent works \cite{Figueroa:2021iwm, Figueroa:2024yja, Garcia:2023qab}, where authors have explored the effect of $\xi$ on the dynamics of the scalar fluctuations, particularly for sub-horizon modes ($k\gtrsim \ke$). In order to obtain parametric resonance, the authors assumed a value of $\xi$ that is much larger than the bound we obtained. 
In our entire study, we have established that for higher EoS, the large-scale modes get enhanced even at a relatively lower $\xi$ value compared to the small-scale parametric resonance studied in \cite{Figueroa:2021iwm, Figueroa:2024yja, Garcia:2023qab}. It has also been found that beyond a certain $\xi$, the appearance of long-wavelength instability results in the development of a larger two-point correlation function $\langle\chi^2\rangle$(see the Appendix \ref{appenc}) which eventually generates a back-reaction effect on the inflaton background. 
%This back-reaction effect will slow down the production of fluctuations. Hence, it is likely to slightly extend the given parameter space of the coupling strength $\xi$.  
However, to address all these subtle issues 
%by taking into account the small-scale parametric resonance effect and the large-scale tachyonic instability together for larger $\xi$, 
we may resort to the lattice study which is beyond the scope of the present work and we leave it for our future endeavor.

%%%%%%%%%%%%%%%%%%%%%%%%%%%%%%%%%%%%%%%%%%%%%%%%%%%%%%%%%%%%%%%%%%%%%%%%%%%%%%%
\section{Conclusion}\label{sec4}

Over the past two decades, significant progress has been made in observational cosmology, enhancing our understanding of the universe and its evolutionary trajectories. However, the reheating epoch, a critical phase in cosmic history, remains poorly understood due to the lack of direct observational evidence. Reheating is a localized phenomenon, and information about its dynamics is obscured as the Standard Model (SM) plasma reaches local thermal equilibrium. Understanding how the universe achieves this thermal equilibrium is crucial for comprehending the physics of the early universe. Generally, this thermal bath is produced when the inflaton decays into SM particles, though the exact mechanisms of particle production at this stage are far from complete understanding. In this section, we shall now clearly highlight the key findings of this study.\\
%%%%%%%%%%%%%%%%%%%%%%%%%%%%
\begin{itemize}
    \item 
    \underline{Production of scalar fluctuations and infrared instability: }
    
    In this study, we investigate the production of a massless scalar field in the presence of a $\xi R\chi^2$ coupling during both inflation and reheating. 
We found that for $\xi<1/6$ and $\wre < 1/3$, the infrared(IR) instability helps scalar field fluctuation to grow both during and after inflation. Such growth turns out to be maximal for $\xi =0$. It potentially violates the \textit{Planck} bound on $r$ through its secondary GW production, thereby setting a tight constraint on the minimum value of $\xi$. On the contrary, for $\xi>1/6$, and $\wre > 1/3$ the infrared modes of the scalar field grow due to post-inflationary instability, and such instability grows with increasing $\xi$. Therefore, in this case, the associated induced gravitational waves put a tight constraint on the maximum value of $\xi$. Thus, in both cases, the infrared instability plays a pivotal role in constraining the non-minimal coupling strength. \\

\item 
\underline{Constraining $\xi$ through $\Delta N_{\rm eff}$, $r_{0.05}$, and isocurvature bound: }

To obtain the constraint on $\xi$, we mainly consider the observational bounds on the effective number of degrees of freedom $\Delta N_{\rm eff}$ at the time of BBN and \textit{Planck}'s upper limit on tensor to scalar ratio $r_{0.05}$ at the pivot scale $\l(k_{\ast}/a_0\r)=0.05~\mbox{Mpc}^{-1}$. The scalar field generated due to infrared instability can be treated as a possible candidate for dark radiation which contributes to $\Delta N_{\rm eff}$. On the other hand, associated secondary GW can affect both $\Delta N_{\rm eff}$ and in estimating the energy scale of inflation through $r$. Except for a few reheating parameters($\wre >1/3$, $N_{\rm re}, T_{\rm re}$), we have obtained bounds on maximum possible values of the coupling strength $\xi_{\rm max}$ as illustrated in Fig.(\ref{fig:xi_con}). In the higher reheating temperature regime, we find stronger constraints from the tensor-to-scalar ratio bound that shrinks the maximum allowed region of $\xi$ for given reheating parameters. Combining the latest \textit{Planck}-2018 data of both $r_{0.05} < 0.036$ and $\Delta N_{\rm eff} \simeq 0.284$ for $\wre\geq1/2$, we find that there exists an upper limit on  $\xi < \xi_{\rm max}(\simeq 4)$, and hence arbitrarily large values of $\xi$ are hardly acceptable. 
Likewise, for $\wre =0$, we find that there exists a lower limit on $\xi >\xi_{\rm min}(\simeq 0.02)$, and hence vanishing non-minimal coupling or in other words, the minimal scenario may not be tenable. We have also briefly discussed the constraints on $\xi$ based on current isocurvature bound at the CMB pivot scale, $\Ps(k_{\ast})\lesssim 8.3\times 10^{-11}$. Interestingly, we have found an almost similar prediction of both minimum(for $\wre=0$) and maximum bounds(for $\wre>1/3$) on $\xi$ as obtained through tensor-to-scalar ratio and isocurvature bound. \\

\item 
\underline{Constraining negative $\xi$ values:}

we have also estimated the possible constraint on negative $\xi$ values. Our study reveals that for $\wre=0$, the presence of strong inflationary and post-inflationary IR instability of the source field results in a heavily red-tilted energy-density spectrum, and this also results high value of $r_{0.05}$ which may violate the CMB bound. Hence, negative $\xi$ values are found to be completely forbidden for $\wre=0$. On the contrary, for $\wre>1/3$, we find lower limits of negative $\xi$ for different reheating parameters($\wre$,~$\Tre,~\He$), and for a given $\wre$, the typical values of negative $\xi$ don't vary much for a wide range of reheating temperatures as discussed in Section \ref{sec3}. This is opposite to the case of putting upper limits on positive $\xi$ for any $\wre>1/3$.

\item 
\underline{Detection prospect:}

Furthermore, we have found a distinctive gravitational wave spectrum for different parameters which could be  detectable by future GW detectors, allowing for more robust constraints on the coupling parameters and reheating dynamics in the near future GW experiments 
Towards the end, given the constraints from all the observations from \textit{Planck}, we derived the region of parameter space in $(\Tre,\xi)$ plane which can be probed by future experiments such as BBO, DECIGO, LISA, and ET, particularly for stiff inflaton equation of state. 
\end{itemize}
%%%%%%%%%%%%%%%%%%%%%%%%%%%%

%In connection with the aforementioned facts, our study upholds one important point: when the universe’s equation of state during reheating is $\wre>1/3$, arbitrarily large values of $\xi$ are hardly acceptable. Furthermore, specific parameter sets can produce distinctive gravitational wave signals detectable by future GW detectors, allowing for more robust constraints on the coupling parameters and reheating dynamics in the near future GW experiments.\\

\section{Acknowledgments}
AC would like to thank the Ministry of Human Resource Development, Government of India (GoI), for financial assistance. SM gratefully acknowledges the financial support provided by the Council of Scientific and Industrial Research (CSIR), Ministry of Science and Technology, Government of India (GoI), through the Senior Research Fellowship (File No. 09/731(0192)/2021-EMR-I). DM wishes to acknowledge support from the Science and Engineering Research Board~(SERB), Department of Science and Technology~(DST), Government of India~(GoI), through the Core Research Grant CRG/2020/003664. The authors are also thankful to the anonymous referee for his/her valuable suggestions for the improvement of the quality of this work. 
%We thank the anonymous referee for his valuable comments and suggestions which helped us improving the paper. 

%\newpage
\appendix

\section{Analyzing the behavior of the suppression factor $\l(1-{a^2\xi\langle\chi^2\rangle}/{M_{\rm pl}^2}\r)^{-2}$ associated with the two-point correlation function $\langle\chi^2\rangle$}  \label{appenc}

%The presence of $\l(1-{a^2\xi\langle\chi^2\rangle}/{M_{\rm pl}^2}\r)^{-1}$ term in the dynamical equation(\ref{eq3}) causes the appearance of a suppression factor $\l(1-\frac{a^2\xi\langle\chi^2\rangle}{M_{\rm pl}^2}\r)^{-2}$ in the expression of the secondary tensor power spectrum(\ref{eq:ptra_chi}) and this again appears in the energy-density spectrum(\ref{eq:gws}).
This factor  might play a very crucial role in estimating the strength of the secondary GW spectrum for larger values of the two-point correlation function $\langle\chi^2\rangle$. Therefore, the typical values of this suppression factor in the different ranges of $\xi$ must be analyzed very carefully. With the help of the equations (\ref{Xreh4}),(\ref{Xreh9}) and (\ref{chisquarevev}) we can compute the typical values of the two-point correlation function $\langle\chi^2\rangle$ with varying theory parameters.

\begin{figure*}
\includegraphics[width=0.8\linewidth]{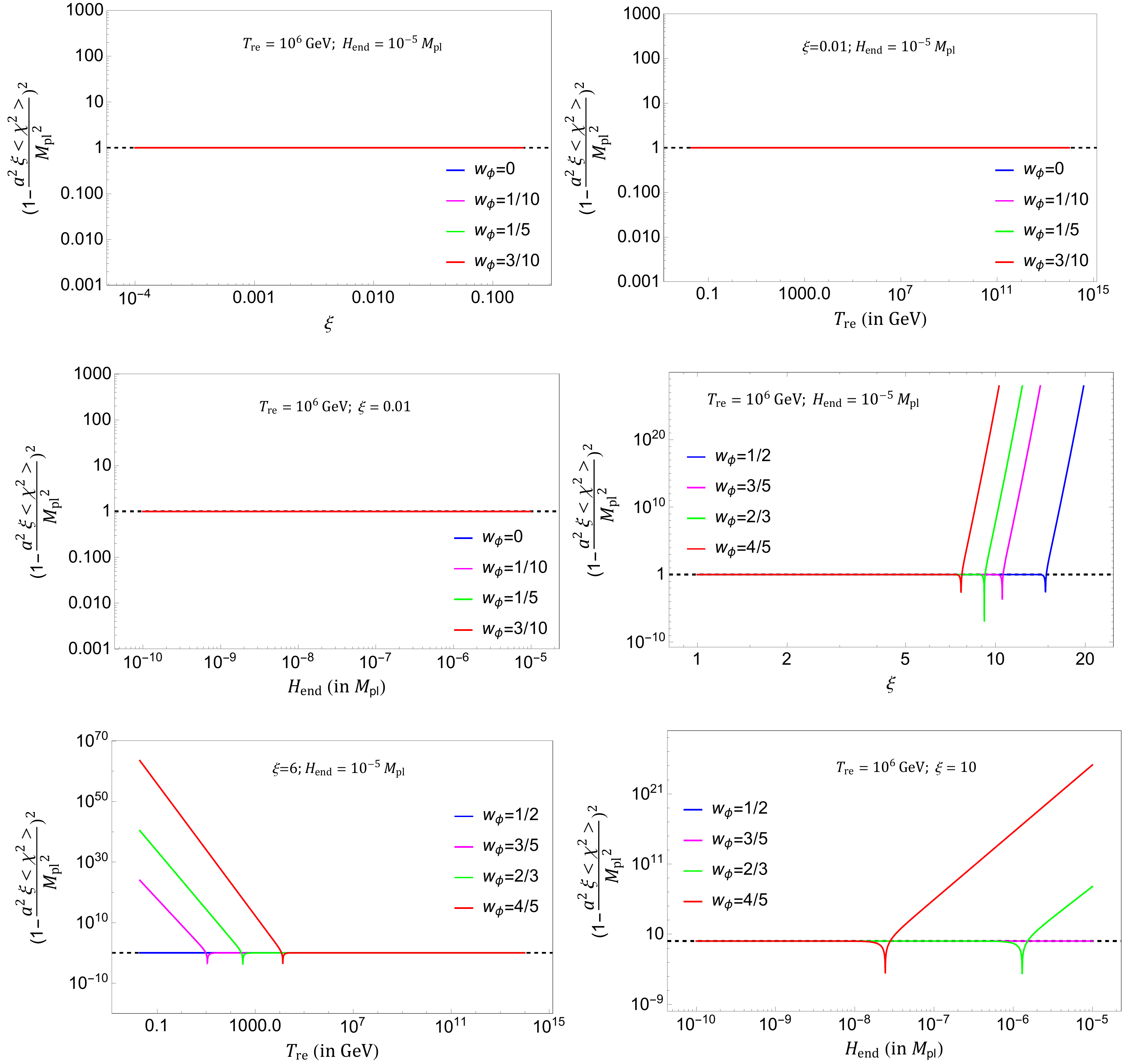}
\caption{\textit{
The figure presents the behavior of the suppression factor with the variation of three parameters $\xi, \Tre, \He$ for different reheating EoS in the regimes $\wre<1/3$ and $\wre>1/3$. In all the plots, the black dotted line depicts $\l(1-\frac{a^2\xi\langle\chi^2\rangle}{M_{\rm pl}^2}\r)^{2}=1$}}\label{fig: suppression factor}
\end{figure*}

The behavior of $\l(1-{a^2\xi\langle\chi^2\rangle}/{M_{\rm pl}^2}\r)^{2}$ is depicted in Fig.(\ref{fig: suppression factor}) for different reheating EoS in two regimes $\wre<1/3$ and $\wre>1/3$. 
 %We have already argued in the earlier discussion that for $\wre<1/3$, the instability effect is significant for $\xi<3/16$, for $\wre>1/3$, significant instability occurs for $\xi>3/16$. In Fig.(\ref{fig: suppression factor}), we have chosen the values of the coupling strength accordingly for both $\wre<1/3$ and $\wre>1/3$. Obviously, for any $\wre<1/3$, the suppression factor always remains unity irrespective of the choice of the reheating parameters $\Tre, \He, \xi$ in a wide range. This, on the other hand, proves the condition $a^2\xi\langle\chi^2\rangle<<M_{\rm pl}^2$ true. It is clearly explained in Section \ref{sec3} that for $\wre<1/3$, $\xi<1/6$, the presence of strong IR divergence in the energy-density spectrum of the source field imposes a lower bound on the coupling $\xi_{\rm min}$ from the current bound on tensor-to-scalar $r_{0.05}\lesssim 0.036$(See Eq.(\ref{eq:r_con1})). And for any $\wre<1/3$, the larger the values of $\xi$, the feebler the source field instability. This in turn causes a very low strength of SGW spectra as obvious in Fig.(\ref{fig:gws1}). Hence, the larger $\xi>3/16$ regime is uninteresting for the lower EoS $\wre<1/3$. \\ 
 In the higher EoS $\wre>1/3$, and $\xi>3/16$ regime, we
 observed that it can become larger than unity as evident in Fig.(\ref{fig: suppression factor}).  %$\langle\chi^2\rangle$
 For instance, for $\Tre=10^6$ GeV and $\He=10^{-5} M_{\rm pl}$, we get $\xi=(14.94,~ 10.75,~9.27,~ 7.77)$ for $\wre=(1/2, 3/5, 2/3, 4/5)$ respectively, and beyond these typical values, the suppression factor starts diverging($a^2\xi\langle\chi^2\rangle>>M^2_{\rm pl}$) with further increase of the coupling strength as shown in Fig.(\ref{fig: suppression factor})(middle row, right panel). The last two bottom plots of Fig.(\ref{fig: suppression factor}) represent the variation of the said quantity in terms of $\Tre$ and $\He$ for different EoS with a fixed $\xi$. We see that for $\xi=6, \He=10^{-5} M_{\rm pl}$, the factor grows rapidly with the decrease of $\Tre$(bottom row, left panel), and for $\xi=10, \Tre=10^6$ GeV, the suppression term grows with $\He$ for higher reheating EoS $\wre=2/3, 4/5$(bottom row, right panel). 
However, our analysis has already shown well below the above values of $\xi$, we have found either large GW or isocurvature modes which already violate the \textit{Planck} bound.   
%
%important to realize that the $\xi$ values required to have Although the two-point correlation function $\langle\chi^2\rangle$ is vital beyond a certain coupling $\xi$, our study reveals that the current \textit{Planck} bound on tensor-to-scalar ratio already imposes an upper boundary on coupling $\xi_{\rm max}$(See Eq.(\ref{eq:r_con})) which is lower than the previous cut-off for any $\wre>1/3$. Therefore, in our entire study, we get rid of the effect of strong suppression of the SGW amplitude and we can safely assume $\l(1-\frac{a^2\xi\langle\chi^2\rangle}{M_{\rm pl}^2}\r)^{-2}\approx 1$. 

%%%%%%%%%%%%%%%%%%%%%%%%
%%%%%%%%%%%%%%%%%%%%%%%%%%%%

\section{Computing the Secondary Tensor power spectrum:}\label{appenA}
As we are mainly interested in the higher post-inflationary EoS regime, $\wre>1/3$, so, we shall first share a detailed calculation of the tensor power spectrum during reheating for a generalized power law type matter power spectrum in stiff EoS regime $\wre>1/3$. The computation of the tensor power spectrum for $\wre<1/3$ will follow the same procedure.\\

\subsection{For $\wre> 1/3$ :}
For $\wre>1/3$ and $\xi>3/16$,  combining Eqs. (\ref{Xreh9}) and (\ref{powerspec2}) we can write the matter power spectrum in terms of the rescaled field $X_k$ for long-wavelength modes as follows.
%%%%%%%%%%%%%%%%%%%%%%%%%%%
\begin{align}
    \mathcal{P}_X(k,\eta)=\frac{k^3}{2\pi^2}\l| X^{\rm long}_k(k,\eta)\r|^2=\frac{\mathcal{A}_3}{\pi^2}~\ke^2~\mI(k,\eta)\l( \frac{k}{\ke}\r)^{2(1-\nu_2)}\label{eq:pchi_a}
\end{align}
%%%%%%%%%%%%%%%%%%%%%%%%%%%
In the above equation, $\mathcal{A}_3$ is a constant part depending on the initial parameters as given in (\ref{comovingenergy3}) and $\mI=\text{cos}^2(k\eta)$ is the time-dependent part of the matter power spectrum.

Now we can write the tensor power spectrum in terms of the matter power spectrum as
%%%%%%%%%%%%%%%%%%%%%%%%%%%%%%%%%%%%
\begin{align}
    \Pt^{\rm sec}(k,\ere)=\frac{8}{M_{\rm pl}^4}\l( \int_{\xe}^{\xre} dx_1 \frac{\Gkre(\xre,x_1)}{a^2(x_1)}\r)^2\times \int_{\kmin}^{\ke}\frac{dq}{k} \int_{-1}^1d\gamma (1-\gamma^2)^2\frac{(q/k)^3\mathcal{P}_X(q,\eta_1)\mathcal{P}_X(|\vk-\vq|\eta_1)}{|1-q/k|^3}\label{eq:pt_a}
\end{align}
Now using Eq.(\ref{eq:pchi_a}) in the above Eq.(\ref{eq:pt_a}) we get the following form.
%%%%%%%%%%%%%%%%%%%%%%%%%%%
\begin{align}
    \Pt^{\rm sec}(k,\ere)=&\frac{8\mathcal{A}_3^2\ke^4}{\pi^4M_{\rm pl}^4}\l( \int_{\xe}^{\xre}dx_1 \frac{\Gkre(\xre,x_1)}{a^2(x_1)}\mI(x_1)\r)^2\nonumber\\
    &\times \int_{\kmin}^{\ke}\frac{dq}{k}\int_{-1}^1d\gamma (1-\gamma^2)^2\frac{(q/k)^3(q/\ke)^{2(1-\nu_2)}(|\vk-\vq|/\ke)^{2(1-\nu_2)}}{|1-q/k|^3}\label{eq:pt_a2}
\end{align}
%%%%%%%%%%%%%%%%%%%%%%%%%%%%%%%%
During the reheating era, the scale factor can be approximately written as $a^2(\eta)\approx\ae^2(x/\xe)^{\delta}$, we recall that $\delta(\wre)=4/(1+3\wre)$. We obtain the following expression by substituting this approximate form of the scale factor to the above Eq.(\ref{eq:pt_a2}).
%%%%%%%%%%%%%%%%%%%%%%%%%%
\begin{align}\label{eq:pt_a3}
    \Pt^{\rm sec}(k,\ere) &=\frac{8\mathcal{A}_3^2\He^4}{\pi^4 M_{\rm pl}^4}\l( \frac{k}{\ke}\r)^{4+2\delta-4\nu_2}\l( \int_{\xe}^{\xre}dx_1x_1^{-\delta}\Gkre(\xre,x_1)\mI(x_1)\r)^2\nonumber\\
    &\times\int_{\kmin}^{\ke}\frac{dq}{k}\int_{-1}^1d\gamma(1-\gamma^2)^2 (q/k)^{5-2\nu_2}(|1-q/k|)^{-(1+2\nu_2)}\nonumber\\
    %=\frac{2\mathcal{A}^2k^4}{\mpl^4\ae^4}\l( \frac{k}{\ke}\r)^{2\delta+2\alpha_1}\l( \int_{\xe}^{\xre}dx_1x_1^{-\delta}\Gkre(\xre,x_1)\mI(x_1)\r)^2 \mathcal{F}(k)\\
    &=\frac{8\mathcal{A}_3^2\He^4}{\pi^4\mpl^4}\l( \frac{k}{\ke}\r)^{4+2\delta-4\nu_2}\l(\underbrace{ \int_{\xe}^{\xre}dx_1x_1^{-\delta}\Gkre(\xre,x_1)\mI(x_1)}_{\text{Time Integral}}\r)^2\mathcal{F}(k)
\end{align}
%%%%%%%%%%%%%%%%%%%%%%%%%%
where we define $\mathcal{F}(k)$ as
%%%%%%%%%%%%%%%%%%%%%%%%%%
\begin{align}\label{Fk}
    \mathcal{F}(k)=\int_{\umin}^{\umax}du\int_{-1}^1 d\gamma (1-\gamma^2)^2u^{5-2\nu_2}(1+u^2-2u\gamma)^{-(\nu_2+1/2)}
\end{align}
%%%%%%%%%%%%%%%%%%%%%%%%%%%%
\subsubsection{Computation of  $\mathcal{F}(k)$ :}
Here we define another variable $u=q/k$. In order to perform the momentum integral, we have to break it into two separate limits, i.e. $\umin\leq u<1$ and $1<u\leq\umax$, where $\umin=k_{\rm min}/k$ and $\umax=\kf/k$ where $k_{\rm min}$ represents the largest observable scale as of today(~CMB scale,  $k_{\ast}$) and $\ke$ represent the largest mode that leaves the horizon at the end of inflation.

  Now in the following range $\umin\leq u<1$ the above integral (\ref{Fk}) boils down to
%%%%%%%%%%%%%%%%%%%%%%%%%%%%%%%%%%%%%%%%%%%%%%
  \begin{align}
\mF_1\simeq\int_{\umin}^1du u^{5-2\nu_2}\int_{-1}^1d\gamma(1-\gamma^2)^2= \frac{16}{15(6-2\nu_2)}\l\{1-\l(\frac{k_{\rm min}}{k}\r)^{6-2\nu_2}\r\}\label{eq:fk1}
  \end{align}
%%%%%%%%%%%%%%%%%%%%%%%%%%
 where we have used $(1+u^2-2\gamma u)\simeq 1$. On the other hand for $1<u\leq\umax$ range the above integral Eq.(\ref{Fk}) boils down to the following one.
 %%%%%%%%%%%%%%%%%%%%%%%%%%%%%
  \begin{align}
      \mF_2&\simeq\int_{1}^{\umax}du \int_{-1}^1d\gamma(1-\gamma^2)^2 u^{5-2\nu_2}u^{-(2\nu_2+1)}=\frac{16}{15(5-4\nu_2)}\l\{ \l(\frac{\ke}{k}\r)^{5-4\nu_2}-1\r\}\nonumber\\
     & =\frac{16}{15(5-4\nu_2)}\l(\frac{\ke}{k}\r)^{5-4\nu_2}\l\{ 1-\l(\frac{k}{\ke}\r)^{5-4\nu_2}\r\}\label{eq:fk2}
  \end{align}
%%%%%%%%%%%%%%%%%%%%%%%%%%%%
  where we have used $(1+u^2-2\gamma u)\simeq u^2$.\\

  Now combining Eq.(\ref{eq:fk1}) and Eq.(\ref{eq:fk2}) we get
  \begin{align}
      \mF(k)=(\mF_1+\mF_2)\simeq \frac{16}{15} \l\{ \frac{1}{6-2\nu_2}\l( 1-\l(\frac{\kmin}{k}\r)^{6-2\nu_2}\r) +\frac{1}{5-4\nu_2}\l( \l(\frac{\ke}{k}\r)^{5-4\nu_2}-1\r)\r\}
  \end{align}
%%%%%%%%%%%%%%%%%%%%%%%%%%%%%%
\subsubsection{Simplification of the Time Integral :}
  During reheating, the Green's function of Eq.(\ref{eq:hij_2}) is 
  \begin{align}
      \mGk^{\rm re}(x,x_1)=\frac{\pi x^l x_1^{1-l}}{2\sin(l\pi)}(J_l(x)J_{-l}(x_1)-J_{-l}(x)J_l(x_1))\label{eq:gkre}
  \end{align}
We recall again $l(\wre)=3(\wre-1)/2(1+3\wre)$.
  Now using Eq.(\ref{eq:gkre}) we are going to perform the time integral part of Eq.(\ref{eq:pt_a3})
  %%%%%%%%%%%%%%%%%%%%%%%
\begin{align}\label{Lt1}
    \mI_t(\xre,\xe)=\int_{\xe}^{\xre} dx_1 x_1^{-\delta}\mI(x_1)\Gkre(\xre,x_1)=\frac{\pi}{2\sin(l\pi)} \frac{x_{\rm re}^lx_1^{2-2l-\delta}}{2^{1+l}}\l( -x_1^{2l}J_{-l}(x_{\rm re})\Gamma\l[1-\frac{\delta}{2} \r]\r.\nonumber\\
    \l.\text{HypergeometricPFQRegularized}\l[\l\{1-\frac{\delta}{2}\r\},\l\{1+l,2-\frac{\delta}{2}\r\},-\frac{x_1^2}{4} \r]\r.\nonumber\\
   \l. \l.+4^l J_l(x_{\rm re})\Gamma\l[1-l-\frac{\delta}{2} \r]\text{HypergeometricPFQRegularized}\l[ \l\{1-l-\frac{\delta}{2} \r\},\l\{1-l,2-l-\frac{\delta}{2}\r\},-\frac{x_1^2}{4} \r] \r)\r|_{\xe}^{\xre}
\end{align}
%%%%%%%%%%%%%%%%%%
\begin{align}
    \mI_t(\xre,\xe)=(\mI_t(\xre,\xre)-\mI_t(\xre,\xe))\simeq\mI_t(\xre,\xre)
\end{align}
For the sub-horizon limit i.e.
%%%%%%%%%%%%%%%%%% $\xre>>1$, the above integral boils down to the simplified form below.
%%%%%%%%%%%%%%%%%%%%%%%%%%%
\begin{align}
    \mI_t(\xre,\xe) &\simeq -\frac{\pi}{2\sin(l\pi)}\frac{2^{1-l-\delta}\pi}{\Gamma(l+\delta/2)\Gamma(\delta/2)}\xre^l\l[ \csc(\pi\delta/2)J_{-l}(\xre)-\csc\l[\pi(2l+\delta)/2\r]J_l(\xre)\r ]\nonumber\\
    &\simeq  -\frac{2^{1-l-\delta}\pi \Gamma(1-l)\Gamma(l)}{2\Gamma(l+\delta/2)\Gamma(\delta/2)}\xre^l \sqrt{\frac{2}{\pi \xre}}\l\{ \csc(\pi\delta/2)\cos\l[ \xre-\pi(1-2l)/4\r]- \csc\l[\pi(2l+\delta)/2\r]\cos\l[\xre-\pi(1+2l)/4\r] \r\}\nonumber\\
    &\simeq  -\frac{2^{1-l-\delta}\pi \Gamma(1-l)\Gamma(l)}{2\Gamma(l+\delta/2)\Gamma(\delta/2)}\xre^l \sqrt{\frac{2}{\pi \xre}}\label{eq:Itre2}
\end{align}
%%%%%%%%%%%%%%%%%%%%%
Plugging the above Eq.(\ref{eq:Itre2}) into the Eq.(\ref{eq:pt_a3})we have the following expression.
%%%%%%%%%%%%%%%%%%%%%%%%%%%%%%%
\begin{align}
    \lim_{k>>\kre}\Pt^{\rm sec}(k,\ere)\simeq \frac{2\mathcal{A}_3^2\HI^4}{\pi^4\mpl^4}\frac{2^{1-2l-2\delta}\pi\Gamma^2(1-l)\Gamma^2(l)}{\Gamma^2(l+ \frac{\delta}{2})\Gamma^2\l( \frac{\delta}{2}\r)}  \frac{8(1+2\nu_2)}{15(3-\nu_2)(4\nu_2-5)} \l(\frac{\kre}{\ke}\r)^{\delta}\l( \frac{k}{\ke}\r)^{4+\delta-4\nu_2}\label{eq:ptre_ap2}
\end{align}
%%%%%%%%%%%%%%%%%%%%%%%%%%%%%%%%%

Similarly for the super-horizon limit i.e. $\xre<<1$ limit, the above integral boils down to the simplified form below.
%%%%%%%%%%%%%%%%%%%%%%%%%%%%%%%%
\begin{align}
    \lim_{k<<\kre}\mI_t(\xre,\xe)\simeq \xre^{2-\delta}\l\{ \frac{1}{2l(\delta-2)}+ \frac{1}{4l(1-l)-2l\delta}\r\}
\end{align}
%%%%%%%%%%%%%%%%%%%%%%%%%%%%%
Now utilizing this expression in Eq.(\ref{eq:pt_a3}) we obtain the tensor power spectrum at the super-horizon scale behaving as
%%%%%%%%%%%%%%%%%%%%%%%%%%%
\begin{align}
    \lim_{k<<\kre}\Pt^{\rm sec}(k,\ere)\simeq \frac{2\mathcal{A}_3^2\HI^4}{\pi^4\mpl^4}\l\{ \frac{1}{2l(\delta-2)}+ \frac{1}{4l(1-l)-2l\delta}\r\}^2\mathcal{F}(k)\l( \frac{\ke}{\kre}\r)^{4-2\delta}\l(\frac{k}{\ke}\r)^{4(2-\nu_2)}\nn\\
    \simeq \frac{2\mathcal{A}_3^2\HI^4}{\pi^4\mpl^4}\l\{ \frac{1}{2l(\delta-2)}+ \frac{1}{4l(1-l)-2l\delta}\r\}^2 \frac{8(1+2\nu_2)}{15(3-\nu_2)(4\nu_2-5)}\l( \frac{\ke}{\kre}\r)^{4-2\delta}\l( \frac{k}{\ke}\r)^{4(2-\nu_2)}\label{eq:ptre_ap1}
\end{align}
%%%%%%%%%%%%%%%%%%%%%%%%%%%%%%%%%
~~
\paragraph{GWs spectral behavior for extreme limit $k<<\kre$ and $k>>\kre$:}
During the radiation-dominated era, the spectral energy density can be written as
%%%%%%%%%%%%%%%%%%%%%%%%%%%%%%%
\begin{align}
    \ogw(k,\eta) = \frac{\rhogw(k,\eta)}{\rho_c(\eta)} = \frac{1}{12} \frac{k^2 \Ptra(k,\eta)}{a^2(\eta) H^2(\eta)}\label{eq:ogw_a}
\end{align}
%%%%%%%%%%%%%%%%%%%%%%%%%%%%%%%
where $\Ptra(k,\eta)$ is the tensor power spectrum during radiation dominated era and it can be written in terms of $\Pt(k,\ere)$ as
%%%%%%%%%%%%%%%%%%%%%%%%
\begin{align}
  \Ptra(k,\eta)=  \left(1 + \frac{k^2}{\kre^2}\right) \frac{\mathcal{P}_{\rm T}(k,\ere)}{2k^2 \eta^2} \label{eq:ptra_a}
\end{align}
%%%%%%%%%%%%%%%%%%%%%%%
Now utilizing Eq.(\ref{eq:ptra_a}) and  (\ref{eq:ptre_ap1}) in Eq.(\ref{eq:ogw_a}) we obtain the GW spectrum behaving in super-horizon limit as
%%%%%%%%%%%%%%%%%%%%%%%%%%
\begin{align}\label{gw1A}
    \lim_{k<<\kre} \ogw(k,\eta)\simeq \frac{2\mathcal{A}_3^2\HI^4}{24\pi^4\mpl^4}\l\{ \frac{1}{2l(\delta-2)}+ \frac{1}{4l(1-l)-2l\delta}\r\}^2 \frac{8(1+2\nu_2)}{15(3-\nu_2)(4\nu_2-5)}\l( \frac{\ke}{\kre}\r)^{4-2\delta}\l( \frac{k}{\ke}\r)^{2(4-2\nu_2)}
\end{align}
%%%%%%%%%%%%%%%%%%%%%%%%
Similarly, for sub-horizon modes, using Eq.(\ref{eq:ptre_ap2}) and (\ref{eq:ptra_a}) in Eq.(\ref{eq:ogw_a}), the GW spectral energy density can be written as
%%%%%%%%%%%%%%%%%%%%%%%%%%
\begin{align}\label{gw2A}
    \lim_{k>>\kre}\ogw(k,\eta)\simeq \frac{2\mathcal{A}_3^2\HI^4}{24\pi^4\mpl^4} \frac{2^{1-2l-2\delta}\pi\Gamma^2(1-l)\Gamma^2(l)}{\Gamma^2(l+ \frac{\delta}{2})\Gamma^2\l( \frac{\delta}{2}\r)}  \frac{8(1+2\nu_2)}{15(3-\nu_2)(4\nu_2-5)} \l(\frac{\ke}{\kre}\r)^{2-\delta}\l( \frac{k}{\ke}\r)^{6+\delta-4\nu_2}
\end{align}
%%%%%%%%%%%%%%%%%%%%%%%%%%%%%%%%%%%%%%%%%%%%%%%%%%%%%%%%%%

\subsection{For $\wre<1/3$ :}

It is already discussed in Section \ref{sec2}, for $\wre<1/3$, the system experiences greater enhancement in long-wavelength regime in the range $\xi<1/6$.   Combining Eqs. (\ref{Xreh4}) and (\ref{powerspec1}) we obtain the expression of the field power spectrum as
%%%%%%%%%%%%%%%%%%%%%%%%
\begin{equation}
 \mathcal{P}_X(k,\eta)=\frac{k^3}{2\pi^2}\l| X^{\rm long}_k(k,\eta)\r|^2=\frac{\mathcal{A}_1}{\pi^2}~\ke^2~\mI(k,\eta)\l( \frac{k}{\ke}\r)^{2(1-\nu_1-\nu_2)}\label{eq:pchi_a1}
\end{equation}
%%%%%%%%%%%%%%%%%%%%%%%%%%%
In the above equation, $\mathcal{A}_1$ is a constant part depending on the initial parameters as given in (\ref{comovingenergy1}) and $\mI=\text{cos}^2(k\eta)$ is the time-dependent part of the matter power spectrum as it was for $\wre>1/3$. Subject to this field power spectrum, following the same steps as followed in the previous case, we obtain the secondary tensor power spectrum as
%%%%%%%%%%%%%%%%%%%%%%%%%%%
\begin{align}
    \Pt^{\rm sec}(k,\ere)=&\frac{8\mathcal{A}_1^2\ke^4}{\pi^4\mpl^4}\l( \int_{\xe}^{\xre}dx_1 \frac{\Gkre(\xre,x_1)}{a^2(x_1)}\mI(x_1)\r)^2\nonumber\\
    &\times \int_{\kmin}^{\ke}\frac{dq}{k}\int_{-1}^1d\gamma (1-\gamma^2)^2\frac{(q/k)^3(q/\ke)^{2(1-\nu_1-\nu_2)}(|\vk-\vq|/\ke)^{2(1-\nu_1-\nu_2)}}{|1-q/k|^3}\label{eq:pt_a22}
\end{align}
%%%%%%%%%%%%%%%%%%%%%%%%%%%%%%%%
Comparing Eqs. (\ref{eq:pt_a2}) and (\ref{eq:pt_a22}), apart from the time-independent amplitude part($\mathcal{A}_1$ and $\mathcal{A}_3$), we notice the only difference in the index of $(q/\ke)$ and $\l(|\vk-\vq|/\ke\r)$ ratios. This fact confirms that the substitution of $(\nu_1+\nu_2)$ in place of $\nu_2$ will essentially reproduce all the results up to Eq.(\ref{gw2A}) for $\wre<1/3$ in a similar fashion. Simplification of the following time-integral in (\ref{eq:pt_a22})gives the same expression as in (\ref{Lt1}). Following the same procedure, the above momentum-integral is simplified as
%%%%%%%%%%%%%%%%%%%%%%%%%%%
\begin{align}
\mE(k)\simeq \frac{16}{15} \l\{ \frac{1}{6-2(\nu_1+\nu_2)}\l( 1-\l(\frac{\kmin}{k}\r)^{6-2(\nu_1+\nu_2)}\r) +\frac{1}{5-4(\nu_1+\nu_2)}\l( \l(\frac{\ke}{k}\r)^{5-4(\nu_1+\nu_2)}-1\r)\r\}
  \end{align}
%%%%%%%%%%%%%%%%%%%%%%%%%%%%%%
Having followed the same methodology as outlined in the previous case for $\wre>1/3$, we finally reach the GW spectrums in both super-horizon and sub-horizon limits as follows:\\

\paragraph{GW spectrum in super-horizon($k<<\kre$) limit:}
Likewise (\ref{gw1A}), we compute the GW energy density spectrum for super-horizon modes, $k<<\kre$ as
%%%%%%%%%%%%%%%%%%%%%%%
\begin{align}\label{gw11A}
\lim_{k<<\kre} \ogw(k,\eta)\simeq& \frac{2\mathcal{A}_1^2\HI^4}{24\pi^4\mpl^4}\l\{ \frac{1}{2l(\delta-2)}+ \frac{1}{4l(1-l)-2l\delta}\r\}^2 \nonumber \\&\times \frac{8(1+2(\nu_1+\nu_2))}{15(3-(\nu_1+\nu_2))(4(\nu_1+\nu_2)-5)} \l( \frac{\ke}{\kre}\r)^{4-2\delta}\l( \frac{k}{\ke}\r)^{2(4-2(\nu_1+\nu_2))}
\end{align}
%%%%%%%%%%%%%%%%%%%%%%%%

\paragraph{GW spectrum in sub-horizon($k>>\kre$) limit:} 
 
 Likewise (\ref{gw2A}), we compute the GW energy density spectrum for sub-horizon modes, $k>>\kre$ as
%%%%%%%%%%%%%%%%%%%%%%%
\begin{align}\label{gw22A}
\lim_{k>>\kre}\ogw(k,\eta)\simeq& \frac{2\mathcal{A}_1^2\HI^4}{24\pi^4\mpl^4} \frac{2^{1-2l-2\delta}\pi\Gamma^2(1-l)\Gamma^2(l)}{\Gamma^2(l+ \frac{\delta}{2})\Gamma^2\l( \frac{\delta}{2}\r)} \nonumber\\ &\times \frac{8(1+2(\nu_1+\nu_2))}{15(3-(\nu_1+\nu_2))(4(\nu_1+\nu_2)-5)} \l(\frac{\ke}{\kre}\r)^{2-\delta}\l( \frac{k}{\ke}\r)^{6+\delta-4(\nu_1+\nu_2)}
\end{align}
%%%%%%%%%%%%%%%%%%%%%%%%%%%%%%%%%%%%%%%%%%%%%%%%

\section{Energy density of Massless scalar field $\chi$ as a dark radiation component for $0\leq\wre\leq1$ :}\label{appenB}

In this section, employing the Eq.(\ref{eq:rho_chi}) with the knowledge of $\beta_k$ as calculated in Section \ref{sec2}, we shall calculate the energy density of the dark radiation component in different ranges of $\xi$ values for $0\leq\wre\leq 1$.\\

\underline{For $1/3\leq\wre\leq1$ :}

\subsection{For $0\leq\xi<3/16$ }

In the given range $0\leq\xi<3/16$, the total energy density at the reheating end is computed to be
%%%%%%%%%%%%%%%%%%%%%%%%%%%%%%%%
\begin{align}\label{comovingenergy1}
    &\rho_{\chi}\left(\frac{a}{a_{\rm end}}\right)^4=\frac{\mathcal{A}_1 \He^4}{2\pi^2}\int_{\kre/\ke}^1 d\left(k/\ke\right)\left(k/\ke\right)^{3-2(\nu_1+\nu_2)}\nonumber\\
    & \Rightarrow \rho_{\chi}\approx \frac{\mathcal{A}_1 \He^4}{4\pi^2\left(2-(\nu_1+\nu_2)\right)} \text{exp}(-4 N_{\rm re}) 
\end{align}
%%%%%%%%%%%%%%%%%%%%%%%%%%%%%%%%%%%%%%%%%%
where $\mathcal{A}_1=\left(\frac{\Gamma(\nu_1)\Gamma(\nu_2)2^{\nu_1}}{8\pi}\left(\frac{2}{3\mu-1}\right)^{\nu_2}\left(\frac{3\mu(1-2\nu_1)+2(\nu_1-\nu_2)}{\sqrt{(3\mu-1)}}\right)\right)^2$. For $\wre>1/3$, in this specified range of $\xi$, we always have $(4-2(\nu_1+\nu_2))>0$. So, the maximum contribution to energy is coming from $\ke$. This property of the blue-tilted spectrum is used to reach the final expression of $\rho^{\rm com}_{\chi}$ in Eq.(\ref{comovingenergy1}).

\subsection{For $\xi=3/16$ :}

Likewise the previous case, the total energy density at the reheating end for $\xi=3/16$ is evaluated to be
%%%%%%%%%%%%%%%%%%%%%%%%%%%%%%%%%%%%
\begin{equation}\label{comovingenergy2}
   \rho_{\chi}\approx \frac{\mathcal{A}_2 \He^4}{4\pi^2\left(2-\nu_2\right)}\text{exp}(-4 N_{\rm re}) 
\end{equation}
%%%%%%%%%%%%%%%%%%%%%%%%%%%%%%%%
where $\mathcal{A}_2=\left(\frac{\Gamma(\nu_2)}{2}\left(\frac{2}{3\mu-1}\right)^{\nu_2}\left|\frac{3\mu-2\nu_2}{4\sqrt{(3\mu-1)}}+\frac{i\sqrt{3\mu-1}}{\pi}\right|\right)^2$\\

\subsection{For $\xi>3/16$ :}
In this range, the existence of the critical coupling strength $\xi_{\rm cri}$ $\left(\text{a function of}~ \wre, ~\xi_{\rm cri}=\frac{(9 \wre +7) (15 \wre+1)}{48 (3 \wre-1)}\right)$ of the field energy density spectrum, $\rho_{\chi_k}$ or $k^4|\beta_k|^2$,  distinguishes two regions with a junction that is at $\xi=\xi_{\rm cri}$.\\ 

\subsubsection{For $3/16<\xi<\xi_{\rm cri}$ :}

For $3/16<\xi<\xi_{\rm cri}$, the energy spectrum being blue-tilted again satisfies the same expressions of reheating parameters as in the range $0\leq\xi\leq3/16$. The total energy density at the reheating end is computed as
%%%%%%%%%%%%%%%%%%%%%%%%%%%%%%%%%%%%
\begin{equation}\label{comovingenergy3}
     \rho_{\chi}\approx \frac{\mathcal{A}_3 \He^4}{4\pi^2\left(2-\nu_2\right)}\text{exp}(-4 N_{\rm re}) 
\end{equation}
%%%%%%%%%%%%%%%%%%%%%%%%%%%%%%%5
where $\mathcal{A}_3\approx\left(\frac{\Gamma(\nu_2)\text{exp}(-\pi \tilde{\nu}_1/2)}{4}\left(\frac{2}{3\mu-1}\right)^{\nu_2}\sqrt{3\mu-1}\left|\frac{\left(\pi+i\text{cosh}(\pi\tilde{\nu_1})\Gamma(1-i\tilde{\nu_1})\Gamma(i\tilde{\nu_1})\right)}{\pi\Gamma(i\tilde{\nu_1})}\right|\right)^2$\\

\subsubsection{For $\xi=\xi_{\rm cri}$ :}

At this junction point, the total energy density at the reheating end is computed as
%%%%%%%%%%%%%%%%%%%%%%%%%%%%%%%
\begin{align}\label{comovingenergy4}
    &\rho_{\chi}= \frac{\mathcal{A}_3 \He^4}{2\pi^2}\text{ln}\left(\frac{\ke}{\kre}\right)\text{exp}(-4 N_{\rm re})\nonumber\\
    &\Rightarrow \rho_{\chi}=\frac{\mathcal{A}_3 \He^4(1+3\wre)N_{\rm re}}{4\pi^2}\text{exp}(-4 N_{\rm re})
\end{align}
%%%%%%%%%%%%%%%%%%%%%%%%%%%%%%%%%

\subsubsection{For $\xi>\xi_{\rm cri}$ :}

After crossing the critical coupling or the junction $\xi_{\rm cri}$, the energy spectrum turns out to be red-tilted. The total energy density at the reheating  end is now computed as
%%%%%%%%%%%%%%%%%%%%%%%%%%%%%%%%%%%%%%
\begin{align}\label{comovingenergy5}
   &\rho_{\chi}= \frac{\mathcal{A}_3 \He^4}{4\pi^2(\nu_2-2)}\left(\frac{\ke}{\kre}\right)^{2\nu_2-4}\text{exp}(-4 N_{\rm re})\nonumber\\
   &\Rightarrow \rho_{\chi}=\frac{\mathcal{A}_3 \He^4}{4\pi^2(\nu_2-2)}\text{exp}\Big(\big((1+3\wre)(\nu_2-2)-4\big)N_{\rm re}\Big)
    \end{align}
%%%%%%%%%%%%%%%%%%%%%%%%%%%%%%%%

\underline{For $0\leq\wre<1/3$ :}

\subsection{For $0\leq\xi<3/16$ }

According to the energy density spectrum described in Section \ref{sec2}, for $0\leq\wre<1/3$, we find the existence of a critical coupling $\xi_{\rm cri}$ lying in the range 
$0<\xi_{\rm cri}<1/6$, below which the energy spectrum is red-tilted and above which the spectrum is blue-tilted. In the entire range $\xi>\xi_{\rm cri}$, the energy spectrum $\rho_{\chi_k}$ remains blue-tilted.

\subsubsection{For $0\leq\xi<\xi_{\rm cri}$ :}

In the given range, energy spectrum is red-tilted. The total energy density at the reheating end is calculated to be

%%%%%%%%%%%%%%%%%%%%%%%%%%%%%%%
\begin{align}\label{comovingenergy6}
    &\rho_{\chi}\left(\frac{a}{a_{\rm end}}\right)^4=\frac{\mathcal{A}_1 \He^4}{2\pi^2}\int_{\kre/\ke}^1 d\left(k/\ke\right)\left(k/\ke\right)^{3-2(\nu_1+\nu_2)}\nonumber\\
    & \Rightarrow \rho_{\chi}\approx \frac{\mathcal{A}_1 \He^4}{4\pi^2\left((\nu_1+\nu_2)-2\right)} \text{exp}\Big( N_{\rm re}\big((1+3\wre)(\nu_1+\nu_2-2)-4\big)\Big) 
\end{align}
%%%%%%%%%%%%%%%%%%%%%%%%%%%%%%%%%%%%%%%%%%

\subsubsection{For $\xi_{\rm cri}<\xi<3/16$ :}

In the given range, energy spectrum is blue-tilted. The total energy density at the reheating end is calculated to be

%%%%%%%%%%%%%%%%%%%%%%%%%%%%
\begin{align}\label{comovingenergy7}
    &\rho_{\chi}\left(\frac{a}{a_{\rm end}}\right)^4=\frac{\mathcal{A}_1 \He^4}{2\pi^2}\int_{\kre/\ke}^1 d\left(k/\ke\right)\left(k/\ke\right)^{3-2(\nu_1+\nu_2)}\nonumber\\
    & \Rightarrow \rho_{\chi}\approx \frac{\mathcal{A}_1 \He^4}{4\pi^2\left(2-(\nu_1+\nu_2)\right)} \text{exp}(-4 N_{\rm re}) 
\end{align}
%%%%%%%%%%%%%%%%%%%%%%%%%%%%%%%%%%%%%%%%%%

\bibliographystyle{apsrev4-1}
\bibliography{references}

\end{document}